\documentclass{article}

\usepackage{amsfonts}

\def\noheaderplainsetup{ 

\topmargin=0pt \headheight=0pt \headsep=0pt  \oddsidemargin=0pt \evensidemargin=0pt  \textheight=9truein \textwidth=6.5truein}   

\noheaderplainsetup

\newcommand{\valid}{\mbox{$\vdash\hspace{-4pt}\vdash$}}

\newcommand{\invalid}{\mbox{$\vdash\hspace{-4pt}\not\vdash$}}

\newcommand{\chand}{\sqcap} 

\newcommand{\pand}{\wedge} 

\newcommand{\sand}{\hspace{2pt}\mbox{\small \raisebox{0.0cm}{$\bigtriangleup$}}\hspace{2pt}} 

\newcommand{\tand}{\mbox{\hspace{2pt}$\wedge$\hspace{-1.29mm}\raisebox{0.02mm}{\rule{0.13mm}{2mm}}}\hspace{5pt}}    

\newcommand{\chor}{\sqcup} 

\newcommand{\por}{\vee} 

\newcommand{\sor}{\hspace{2pt}\mbox{\small \raisebox{0.049cm}{$\bigtriangledown$}}\hspace{2pt}} 

\newcommand{\tor}{\mbox{\hspace{2pt}$\vee$\hspace{-1.29mm}\raisebox{0.1mm}{\rule{0.13mm}{2mm}}\hspace{5pt}}}    

\newcommand{\blall}{\mbox{\large $\forall$}} 

\newcommand{\chall}{\hspace{1pt}\mbox{\Large $\sqcap$}} 

\newcommand{\pall}{\hspace{1pt}\mbox{{\Large $\wedge$}}\hspace{1pt}} 

\newcommand{\sall}{\mbox{\large \raisebox{0.0cm}{$\bigtriangleup$}}} 

\newcommand{\tall}{\mbox{\hspace{1pt}\Large $\wedge$\hspace{-1.84mm}\raisebox{0.02mm}{\rule{0.13mm}{3.0mm}}\hspace{6pt}}}   

\newcommand{\blexists}{\mbox{\large $\exists$}} 

\newcommand{\chexists}{\hspace{1pt}\mbox{\Large $\sqcup$}} 

\newcommand{\pexists}{\hspace{0pt}\mbox{{\Large $\vee$}}\hspace{0pt}} 

\newcommand{\sexists}{\mbox{\large \raisebox{0.07cm}{$\bigtriangledown$}}} 

\newcommand{\texists}{\hspace{1pt}\mbox{\Large $\vee$\hspace{-1.84mm}\raisebox{0.1mm}{\rule{0.13mm}{3.0mm}}\hspace{5pt}}} 

\newcommand{\chimplication}{\sqsupset} 

\newcommand{\pimplication}{\rightarrow} 

\newcommand{\simplication}{\hspace{3pt}\mbox{\Large $\triangleright$}\hspace{3pt}} 

\newcommand{\timplication}{>\hspace{-11pt}-\hspace{2pt}} 

\newcommand{\precurrence}{\hspace{1pt}\mbox{\raisebox{-0.01cm}{\scriptsize $\wedge$}\hspace{-4pt}\raisebox{0.16cm}{\tiny $\mid$}}\hspace{2pt}} 

\newcommand{\srecurrence}{\mbox{\raisebox{-0.07cm}{\scriptsize $-$}\hspace{-5.9pt}\mbox{\raisebox{-0.01cm}{\scriptsize $\wedge$}\hspace{-4pt}\raisebox{0.16cm}{\tiny $\mid$}}}\hspace{2pt}} 

\newcommand{\trecurrence}{\mbox{\raisebox{-0.01cm}{\scriptsize $\wedge$}\hspace{-3.95pt}\raisebox{0.06cm}{\small $\mid$}\hspace{2pt}}}  

\newcommand{\brecurrence}{\mbox{\raisebox{-0.05cm}{$\circ$}\hspace{-0.13cm}\raisebox{3.9pt}{\tiny $\mid$}}\hspace{1.5pt}} 

\newcommand{\coprecurrence}{\hspace{1pt}\mbox{\raisebox{0.12cm}{\scriptsize $\vee$}\hspace{-3.8pt}\raisebox{0.02cm}{\tiny $\mid$}}\hspace{2pt}} 

\newcommand{\cosrecurrence}{\mbox{\raisebox{0.20cm}{\scriptsize $-$}\hspace{-5.9pt}\mbox{\raisebox{0.12cm}{\scriptsize $\vee$}\hspace{-3.8pt}\raisebox{0.02cm}{\tiny $\mid$}}}\hspace{2pt}} 

\newcommand{\cotrecurrence}{\mbox{\raisebox{0.12cm}{\scriptsize $\vee$}\hspace{-3.95pt}\raisebox{0.04cm}{\small $\mid$}\hspace{2pt}}}  

\newcommand{\cobrecurrence}{\mbox{\raisebox{0.12cm}{$\circ$}\hspace{-0.13cm}\raisebox{0.02cm}{\tiny $\mid$}}\hspace{1.5pt}} 

\newcommand{\primplication}{\hspace{2pt}\mbox{\raisebox{0.033cm}{\tiny $>$}\hspace{-0.18cm} \raisebox{-0.043cm}{\large --}}\hspace{2pt}} 

\newcommand{\srimplication}{\hspace{3pt}\mbox{\mbox{\raisebox{0.1pt}{\small $\triangleright$}}\hspace{-4pt}  \raisebox{-0.8pt}{\large --}}\hspace{3pt}} 

\newcommand{\trimplication}{\mbox{\hspace{2pt}\raisebox{0.033cm}{\tiny $>$}\hspace{-0.28cm} \raisebox{-2.4pt}{\LARGE --}\hspace{2pt}}}

\newcommand{\brimplication}{\hspace{3pt}\mbox{$\circ$\hspace{-0.14cm} \raisebox{-0.043cm}{\Large --}}\hspace{3pt}} 

\newcommand{\prepudiation}{\mbox{\raisebox{1.1pt}{\tiny $>$}\hspace{-1.6pt}{\scriptsize $\neg$}}} 

\newcommand{\srepudiation}{\hspace{1pt}\mbox{\raisebox{0.1pt}{\small $\triangleright$}\hspace{-1pt}{\scriptsize $\neg$}}} 

\newcommand{\trepudiation}{\mbox{\raisebox{0.033cm}{\tiny $>$}\hspace{-0.28cm} \raisebox{-0.5pt}{\large -}\hspace{-1pt}{\scriptsize $\neg$}}} 

\newcommand{\brepudiation}{\mbox{$\circ$\hspace{-0.14cm} \hspace{-0.8pt}\raisebox{0.2pt}{\scriptsize $\neg$}}} 

\newcommand{\sti}{\mbox{\raisebox{-0.02cm}{\scriptsize $\circ$}\hspace{-0.121cm}\raisebox{0.08cm}{\tiny $.$}\hspace{-0.079cm}\raisebox{0.10cm}{\tiny $.$}\hspace{-0.079cm}\raisebox{0.12cm}{\tiny $.$}\hspace{-0.085cm}\raisebox{0.14cm}{\tiny $.$}\hspace{-0.079cm}\raisebox{0.16cm}{\tiny $.$}\hspace{1pt}}}

\newcommand{\costi}{\mbox{\raisebox{0.08cm}{\scriptsize $\circ$}\hspace{-0.121cm}\raisebox{-0.01cm}{\tiny $.$}\hspace{-0.079cm}\raisebox{0.01cm}{\tiny $.$}\hspace{-0.079cm}\raisebox{0.03cm}{\tiny $.$}\hspace{-0.085cm}\raisebox{0.05cm}{\tiny $.$}\hspace{-0.079cm}\raisebox{0.07cm}{\tiny $.$}\hspace{1pt}}}

\newcommand{\seq}[1]{\langle #1 \rangle}           
\newcommand{\lp}[2]{\mbox{\bf Lp}^{#1}_{#2}} 
\newcommand{\legal}[2]{\mbox{\bf Lr}^{#1}_{#2}} 
\newcommand{\win}[2]{\mbox{\bf Wn}^{#1}_{#2}} 
\newcommand{\chess}{\mbox{\em Chess}}

\newtheorem{theoremm}{Theorem}[section]
\newtheorem{thesiss}[theoremm]{Thesis}
\newtheorem{conjecturee}[theoremm]{Conjecture}
\newtheorem{exercisee}[theoremm]{Exercise}
\newtheorem{definitionn}[theoremm]{Definition}
\newtheorem{lemmaa}[theoremm]{Lemma}
\newtheorem{propositionn}[theoremm]{Proposition}
\newtheorem{conventionn}[theoremm]{Convention}
\newtheorem{examplee}[theoremm]{Example}
\newtheorem{remarkk}[theoremm]{Remark}
\newtheorem{factt}[theoremm]{Fact}
\newtheorem{notationn}[theoremm]{Notation}
\newtheorem{openproblemm}[theoremm]{Open Problem}
\newtheorem{terminologyy}[theoremm]{Terminology}

\newenvironment{definition}{\begin{definitionn} \em}{ \end{definitionn}}
\newenvironment{theorem}{\begin{theoremm}}{\end{theoremm}}

\newenvironment{convention}{\begin{conventionn} \em}{\end{conventionn}}

\newenvironment{example}{\begin{examplee} \em}{\end{examplee}}
\newenvironment{exercise}{\begin{exercisee} \em}{\end{exercisee}}

\newenvironment{thesis}{\begin{thesiss} \em}{ \end{thesiss}}
\newenvironment{notation}{\begin{notationn} \em}{ \end{notationn}}
\newenvironment{openproblem}{\begin{openproblemm} \em}{ \end{openproblemm}}
\newenvironment{terminology}{\begin{terminologyy} \em}{ \end{terminologyy}}

\makeindex

\begin{document}

\title{Fundamentals of computability logic 2020}
 \author{Giorgi Japaridze \\
 \\  
 Villanova University  and \\  Institute of Philosophy, Russian Academy of Sciences}

\date{}

\maketitle \           
\tableofcontents 

\section{Introduction}\label{intro}

 Not to be confused with the generic term ``computational logic'', computability logic (CoL) is the proper name of a philosophical  platform and mathematical framework for developing ever more expressive computationally meaningful extensions of traditional logic. The main pursuit of this  ongoing long-term  project is to offer a convenient language  for specifying computational tasks and  relations between them in a systematic way, and to provide a deductive apparatus for systematically telling what can be computed and how. This line of research was officially introduced in \cite{Jap03} and developed in a series \cite{bauerTOCL}-\cite{bauerLMCS},\cite{Jap03}-\cite{Jap20},\cite{kwon},\cite{Ver2010}-\cite{Qu},\cite{xuIGPL}-\cite{xu2016}  of subsequent papers.

Under the approach of CoL, formulas represent computational problems, logical operators stand for operations on such entities, and   ``truth''  is seen as computability. Computational problems, in turn, are understood in their most general, interactive sense, and are mathematically construed as games played by a machine against its environment, with computability meaning existence of a machine (algorithmic strategy) that always wins.     
 
CoL understands propositions or predicates of traditional logic as games with no moves, automatically won by the machine when true and lost when false. This naturally makes the classical concept of truth a special case of computability ---  computability by doing nothing. Further, all operators of classical logic are conservatively generalized from moveless games to all games, which eventually makes classical logic a conservative fragment of the otherwise much more expressive CoL. Based on the overall philosophy and intuitions associated with intuitionistic and linear logics, the latter can also be seen as special fragments of CoL, even though, unlike classical logic, ``not quite'' conservative ones. 

A long list of related or unrelated game semantics can be found in the literature proposed by various authors. Out of those, Blass's \cite{Bla92} game semantics, which in turn is a 
refinement of Lorenzen's \cite{Lor59} dialogue semantics, is the closest precursor of the semantics of CoL, alongside with Hintikka's \cite{Hin73} game-theoretic semantics. 
More often than not, the motivation for studying games in logic has been to achieve a better understanding of some already existing systems, such as intuitionistic (\cite{Bla72,Fel85,Lor59}), classical (\cite{Hin73}) or linear (\cite{Abr94,Bla92}) logics. In contrast, CoL's motto is  that logic should serve games rather than the other way around. For  logic is meant to be the most general and universal intellectual tool for navigating real life; and it is games that offer the most adequate mathematical models for the very essence of all ``navigational''  activities of agents: their interactions with the surrounding world. An agent and its environment  translate into game-theoretic terms as two players; their actions as moves; situations arising in the course of interaction as  positions; and successes or failures as wins or losses.

This chapter is a semitutorial-style introduction to the basics of CoL, containing  many definitions, illustrations,  claims and even exercises  but no technical proofs whatsoever. 
It is primarily focused on the language of CoL and its semantics, paying considerably less attention to the associated proof theory or applications. A more detailed and continuously updated survey of the subject is maintained online at \cite{CL}.

\section{Games}\label{sgames}
 Computability  is a property of {\em computational problems} and,  before attempting to talk about the former, we  need to agree on the precise meaning of the latter.  According to the mainstream understanding going back to Church \cite{church} and Turing \cite{Tur36},  a computational problem is a {\em function}---more precisely, the task of systematically generating  the values of that function at different arguments. Such a view, however, as more and more researchers have been acknowledging     \cite{goldin}, is too narrow. Most tasks performed by computers are {\em interactive}, far from being as simple as functional transformations from inputs to outputs. Think of the work of a  network server for instance, where the task is to maintain a certain infinite process, with incoming (``input'') and outgoing (``output'') signals interleaved in some complex and probably unregulated fashion, depending on not only immediately preceding signals but  also various events taken place in the past.  In an attempt to advocate for the conventional view of computational problems, one might suggest to understand an interactive computational task as the task of repeatedly computing the value of a function whose  argument is not just the latest input but the whole preceding interaction. This is hardly a good solution though, which becomes especially evident with computational complexity considerations in mind. If the task performed by your personal computer was like that, then you would have noticed its performance worsening after every use due to the need to read the ever longer history of interaction with you. 

Instead, CoL postulates that a computational problem is a {\em game} between two agents: a machine and its environment, symbolically named $\top$ and $\bot$, respectively. $\top$ is a mechanical device only capable of following algorithmic strategies, while there are no similar assumptions about $\bot$ whose behavior can be arbitrary. Computational tasks in the traditional sense now become special cases of games with only two moves, where   the first move (``input'') is by $\bot$ and the second move (``output'') by $\top$. 

The following notational and terminological conventions are adopted. A {\bf move} is any finite string over the standard keyboard alphabet. A {\bf labeled move} is a move prefixed with $\top$ or $\bot$, with such a prefix ({\bf label}) indicating which player is the author of the move. We will not always be very strict about differentiating between moves and labeled moves, sometimes saying ``move'' where, strictly speaking, ``labeled move'' is meant. A {\bf run} is a (finite or infinite) sequence of labeled moves, and a {\bf position} is a finite run.  We usually use lowercase Greek letters as metavariables for moves, and uppercase Greek letters for runs. We will be writing runs and positions as
$\seq{\alpha,\beta,\gamma}$, $\seq{\Theta,\Gamma}$, $\seq{\Theta,\alpha,\Gamma}$, etc. 
 The meanings of such expressions should be clear. For instance, $\seq{\Theta,\alpha,\Gamma}$ is the run consisting of the (labeled) moves of the position $\Theta$,  followed   by the move $\alpha$, and then by the moves of the  run $\Gamma$. 

A set $S$ of runs is said to be {\bf prefix-closed} iff, whenever a run is in $ S$, so are all of its initial segments.  The {\bf limit-closure} of a  set $S$ of runs is the result of adding to $S$ every infinite run $\Gamma$ such that all finite initial segments of $\Gamma$ are in $S$.

\begin{definition}\label{game}
A {\bf game} is a pair  $G=(\lp{G}{},\win{G}{})$, where:

1.   $\lp{G}{}$  is a nonempty, prefix-closed set of positions. We write $\legal{G}{}$ for  
the limit-closure of $\lp{G}{}$. 

2.      
$\win{G}{}$ is a mapping from $\legal{G}{}$ to $\{\top,\bot\}$. 
\end{definition}

Intuitively, in the context of a given game $G$, $\lp{G}{}$ is the set of {\bf legal positions} and $\legal{G}{}$ is the set of  {\bf legal runs}. 
  Note  that, since $\lp{G}{}$ is required to be nonempty and prefix-closed, the {\bf empty position} $\seq{}$, being an initial segment of all runs,
 is always legal.  With $\wp$ here and elsewhere standing for either player and $\overline{\wp}$ for its adversary 
$\wp\not= \overline{\wp}\in\{\top,\bot\}$, 
a  {\bf legal move} by  $\wp$  in a position $\Theta$ is a move $\alpha$ such that $\seq{\Theta,\wp\alpha}\in \lp{G}{}$. We say that a run $\Gamma$ 
is {\bf $\wp$-legal}  iff    either $\Gamma$ is legal, or else, where $\Theta$ is the shortest illegal initial segment of $\Gamma$, the last move 
of $\Theta$ is $\overline{\wp}$-labeled. Intuitively, such a $\Gamma$ is a run where $\wp$ has not made any illegal moves unless its 
adversary $\overline{\wp}$ has done so first. 
In all cases, we shall say ``{\bf illegal}'' for ``not legal'' and  ``lost'' for ``not won''. 
For each legal run, $\win{G}{}$ tells us which of the two players $\wp\in\{\top,\bot\}$ has   won  the run. The following definition extends this 
meaning of the word ``won''  from legal runs to all runs by stipulating that an illegal run is always lost by the player that has made the first illegal move:

\begin{definition}\label{newdef}
For a game $G$, run $\Gamma$ and player $\wp$, we say that $\Gamma$ is a {\bf $\wp$-won} (or won by $\wp$) run of $G$ iff $\Gamma$ is 

1. either  a legal run of $G$ with $\win{G}{}\seq{\Gamma}=\wp$, or

2. a $\overline{\wp}$-illegal run of $G$. 

\end{definition}

Games---at least when they are finite---can be visualized as trees in the style of Figure 1. Each complete or incomplete branch of such a tree represents a legal run, namely, the sequence of the labels of the edges of the branch. The nodes represent positions, where the label $\top$ or $\bot$ of a node indicates which player is the winner if the play ends in the corresponding position.  
  
\begin{center}
\begin{picture}(322,150)

\put(159,132){\circle{16}}
\put(155,128){$\bot$}
\put(151,132){\line(-4,-1){88}}
\put(95,122){{\tiny $\top$}{\footnotesize $\alpha$}}
\put(159,124){\line(0,-1){14}}
\put(160,116){{\tiny $\bot$}{\footnotesize $\beta$}}
\put(167,132){\line(4,-1){88}}
\put(212,122){{\tiny $\bot$}{\footnotesize $\gamma$}}

\put(62,102){\circle{16}}
\put(58,98){$\top$}
\put(55,98){\line(-5,-3){31}}
\put(26,88){{\tiny $\bot$}{\footnotesize $\beta$}}
\put(69,98){\line(5,-3){31}}
\put(88,88){{\tiny $\bot$}{\footnotesize $\gamma$}}
\put(159,102){\circle{16}}
\put(155,98){$\top$}
\put(159,94){\line(0,-1){14}}
\put(161,84){{\tiny $\top$}{\footnotesize $\alpha$}}
\put(256,102){\circle{16}}
\put(252,98){$\bot$}
\put(249,97){\line(-5,-3){31}}
\put(221,88){{\tiny $\top$}{\footnotesize $\alpha$}}
\put(256,94){\line(0,-1){14}}
\put(257,84){{\tiny $\top$}{\footnotesize $\beta$}}
\put(262,97){\line(5,-3){31}}
\put(279,88){{\tiny $\top$}{\footnotesize $\gamma$}}

\put(20,72){\circle{16}}
\put(16,68){$\top$}
\put(104,72){\circle{16}}
\put(100,68){$\bot$}
\put(102,64){\line(-2,-3){9}}
\put(85,56){{\tiny $\top$}{\footnotesize $\beta$}}
\put(107,64){\line(2,-3){9}}
\put(113,56){{\tiny $\top$}{\footnotesize $\gamma$}}
\put(159,72){\circle{16}}
\put(155,68){$\top$}
\put(214,72){\circle{16}}
\put(210,68){$\bot$}
\put(211,64){\line(-2,-3){9}}
\put(194,56){{\tiny $\top$}{\footnotesize $\beta$}}
\put(222,56){{\tiny $\top$}{\footnotesize $\gamma$}}
\put(217,64){\line(2,-3){9}}
\put(256,72){\circle{16}}
\put(252,68){$\top$}
\put(256,64){\line(0,-1){14}}
\put(257,56){{\tiny $\top$}{\footnotesize $\alpha$}}
\put(298,72){\circle{16}}
\put(294,68){$\bot$}
\put(298,64){\line(0,-1){14}}
\put(299,56){{\tiny $\top$}{\footnotesize $\alpha$}}

\put(92,42){\circle{16}}
\put(88,38){$\top$}
\put(116,42){\circle{16}}
\put(112,38){$\bot$}
\put(202,42){\circle{16}}
\put(198,38){$\top$}
\put(226,42){\circle{16}}
\put(222,38){$\bot$}
\put(256,42){\circle{16}}
\put(252,38){$\top$}
\put(298,42){\circle{16}}
\put(294,38){$\bot$}

\put(120,10){{\bf Figure 1:} A game}
\end{picture}
\end{center}

A distinguishing feature of CoL games is the absence of rules governing the order in which (legal) moves can or should be made. In some situations, such as in the root position of the game of   Figure 1, both players may have legal moves, and which  if any  player moves first depends on which one wants or can act faster. Imagine a simultaneous play of chess on two boards, where you play white on both boards. At the beginning, only you have  legal moves. But once you make an opening move---say, on board \#1---the situation changes. Now both you and your environment have legal moves: the environment may respond on board \#1, while you can make another opening move on board \#2. It would be unnatural to impose rules determining the next player to move in this case, especially if your environment consists of two independent and non-communicating adversaries. The relaxed nature of our games makes them more direct and adequate tools for modeling real-life interactions like this and beyond than stricter games would be. 

But how are such loose  games played and, most importantly, what does an algorithmic winning strategy mean? Below is an example of such a strategy. It is left to the reader to convince himself or herself that following it guarantees $\top$ a win in the game of Figure 1: 

\begin{quote} {\em Regardless of what the adversary is doing or has done, go ahead and make move $\alpha$; make $\beta$ as your second move if and when you see that the adversary has made move $\gamma$, no matter whether this happened before or after your first move. } 
\end{quote}

Formally, $\top$'s algorithmic strategies  can be understood as what CoL, for historical reasons,  calls  {\bf HPM}s (``hard-play machines''). An HPM  is a Turing machine with the capability of making moves. This is just like the capability of generating an output, with the only difference that, while an ordinary Turing machine halts after generating an output, an HPM generally does not halt after making a move, so it can continue its work and make more moves later. Also, an HPM is equipped with an additional, read-only tape called the {\em run tape}, initially empty. Every time the HPM makes a move $\alpha$, the string $\top\alpha$ is automatically appended to the content of this tape. At any time, any $\bot$-labeled move $\bot \beta$ may also be nondeterministically appended to the content of the run tape. This event is interpreted as that the environment has just made move $\beta$. This way, at any step of the process, the run tape spells the current position of the play. It is hardly necessary to define HPMs in full detail here, for the Church-Turing thesis extends from ordinary Turing machines to HPMs, according to which HPMs adequately correspond to what we intuitively perceive as algorithmic strategies. So, rather than attempting to formally describe an HPM playing a given game, we can simply describe its work in relaxed, informal terms in the style of the earlier-displayed strategy for the game of Figure 1.  
There is also no need to anyhow define $\bot$'s strategies: all possible behaviors by $\bot$ are accounted for  by the above-mentioned different nondeterministic updates of the run tape, including $\bot$'s relative speed because there are no restrictions on when or with what frequency the updates can take place.  

Depending on what nondeterministic events ($\bot$'s moves) occur in the course of the work of an HPM $\cal M$ and when, different runs will be eventually (in the limit) spelled on $\cal M$'s run tape. We call any such run a {\bf run generated by $\cal M$}.


\begin{definition}\label{solution}
We say that an HPM $\cal M$ {\bf computes} a game $G$ iff every run  generated by $\cal M$ is $\top$-won run of $G$. Such an $\cal M$ is said to be a 
{\bf solution}  (or an {\bf algorithmic winning strategy} for) $G$.
\end{definition}

By the {\bf depth} of a game we mean the (possibly infinite) length of its longest legal run.  Computational problems in the traditional sense, i.e. functions, are  games of depth 2 of the kind seen in \mbox{Figure 2}. In such a game, the upper level edges represent possible inputs provided by the environment. This explains why their labels are $\bot$-prefixed.  The lower level edges represent possible outputs generated by the machine, so their labels are $\top$-prefixed. The root is $\top$-labeled because it corresponds to the situation where nothing happened, namely, no input was provided by the environment. The machine has nothing to answer for in this case, so it wins. The middle level nodes are $\bot$-labeled because they correspond to situations where there was an input but the machine failed to generate an output, so the machine loses. Each group of the bottom level nodes has exactly one $\top$-labeled node, because a function has exactly one (correct) value at each argument. 

\begin{center}
\begin{picture}(322,160)

\put(175,142){\circle{16}}
\put(-4,109){\scriptsize Input}
\put(171,138){$\top$}
\put(175,134){\line(-3,-1){106}}
\put(83,109){\tiny $\bot 0$}
\put(175,134){\line(0,-1){34}}
\put(163,109){\tiny $\bot 1$}
\put(175,134){\line(3,-1){106}}
\put(226,109){\tiny $\bot 2$}
\put(175,134){\line(6,-1){119}}
\put(285,109){\Huge ...}

\put(-4,59){\scriptsize Output}
\put(65,92){\circle{16}}
\put(61,88){$\bot$}
\put(30,59){\tiny $\top 0$}
\put(65,84){\line(-1,-3){11}}
\put(46,59){\tiny $\top 1$}
\put(65,84){\line(-1,-1){34}}
\put(62,59){\tiny $\top 2$}
\put(65,84){\line(1,-3){11}}
\put(77,59){\tiny $\top 3$}
\put(65,84){\line(1,-1){34}}
\put(95,59){\large ...}
\put(65,84){\line(3,-2){34}}

\put(29,42){\circle{16}}
\put(25,38){$\bot$}
\put(53,42){\circle{16}}
\put(49,38){$\top$}
\put(77,42){\circle{16}}
\put(73,38){$\bot$}
\put(101,42){\circle{16}}
\put(97,38){$\bot$}

\put(175,92){\circle{16}}
\put(171,88){$\bot$}
\put(140,59){\tiny $\top 0$}
\put(175,84){\line(-1,-3){11}}
\put(156,59){\tiny $\top 1$}
\put(175,84){\line(-1,-1){34}}
\put(172,59){\tiny $\top 2$}
\put(175,84){\line(1,-3){11}}
\put(187,59){\tiny $\top 3$}
\put(175,84){\line(1,-1){34}}
\put(205,59){\large ...}
\put(175,84){\line(3,-2){34}}

\put(139,42){\circle{16}}
\put(135,38){$\bot$}
\put(163,42){\circle{16}}
\put(159,38){$\bot$}
\put(187,42){\circle{16}}
\put(183,38){$\top$}
\put(211,42){\circle{16}}
\put(207,38){$\bot$}

\put(285,92){\circle{16}}
\put(281,88){$\bot$}
\put(250,59){\tiny $\top 0$}
\put(285,84){\line(-1,-3){11}}
\put(266,59){\tiny $\top 1$}
\put(285,84){\line(-1,-1){34}}
\put(282,59){\tiny $\top 2$}
\put(285,84){\line(1,-3){11}}
\put(297,59){\tiny $\top 3$}
\put(285,84){\line(1,-1){34}}
\put(315,59){\large ...}
\put(285,84){\line(3,-2){34}}

\put(249,42){\circle{16}}
\put(245,38){$\bot$}
\put(273,42){\circle{16}}
\put(269,38){$\bot$}
\put(297,42){\circle{16}}
\put(293,38){$\bot$}
\put(321,42){\circle{16}}
\put(317,38){$\top$}

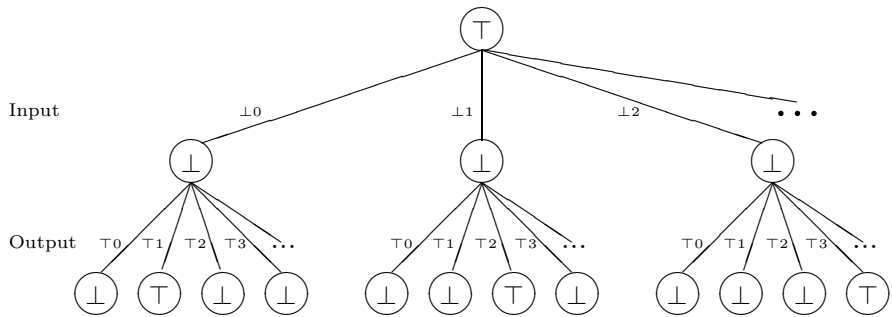
\put(65,10){{\bf Figure 2:} The successor function as a game}
\end{picture}
\end{center}

But why limit ourselves only to trees of the above sort?    
First of all, we may want to allow branches to be longer than $2$, or even infinite to be able to model long or infinite tasks performed by computers.  And why not allow any other sorts of arrangements of $\top$ and $\bot$ in nodes or on edges? For instance, consider the task of computing the function $5/x$. It would be natural to make the node to which the input $0$ takes us not $\bot$-labeled, but $\top$-labeled. For the function is not defined at $0$, and the machine cannot be held responsible for failing  to generate an output on such an input.  

It makes sense to generalize computational problems not only in the direction of increasing their depths, but also decreasing. 
Games of depth $0$, i.e., games that have no nonempty legal runs,  are said to be {\bf elementary}. 
There are exactly two elementary games, for which we use the same symbols $\top,\bot$ as for the two players.  Namely, $\top$ is the elementary game $G$ with  
$\win{G}{}\seq{}=\top$, and $\bot$ is the elementary game $G$ with  
$\win{G}{}\seq{}=\bot$. Intuitively, $\top$ and $\bot$ are moveless games,  with (the only legal run $\seq{}$ of)   $\top$ automatically won by the machine and   $\bot$ won by the environment. While the game $\bot$ has no solution, the ``do nothing'' strategy is a solution of $\top$.   

Extensionally, true propositions  of classical logic such as ``snow is white'' are understood in CoL as the game $\top$, and false propositions such as  ``$2+2=5$'' as the game $\bot$. Propositions are thus  special---elementary---cases of our games. This allows us to say that games are generalized propositions.  

\section{Gameframes}\label{frames}
This section is devoted to the basic concepts   necessary for lifting CoL from the propositional level to the first-order level. What we call {\em gameframes}  are generalized predicates in the same sense as games are generalized propositions.   

We fix an infinite set $\mbox{\em Variables}  =  \{var_1, var_2, var_3, \cdots\}$ of {\bf variables}. As usual, lowercase letters near the end of the Latin alphabet will be used as metavariables for variables. We further fix the set $\mbox{\em Constants} =  \{0,1,2,3,\cdots\}$ of decimal numerals, and call its elements  {\bf constants}.

A {\bf universe} (of discourse) is a pair $U = (Dm, Dn)$, where $Dm$, called the {\bf domain} of $U$, is a nonempty set, and $Dn$, called the {\bf denotator} of $U$,  is a total function of the type $\mbox{\em Constants}\rightarrow  Dm$.  Elements of $Dm$ will be referred to as   {\bf individuals}. The intuitive meaning of $d=Dn(c)$ is that the individual $d$ is the {\bf denotat} of the constant $c$ and thus $c$ is a {\bf name}  of $d$. 

A nice natural example of a universe is the {\bf arithmetical universe},  whose domain is the set of natural numbers and whose denotator is the bijective function that sends each constant to the number it represents in standard decimal notation.   
Generally, however, the denotator is required to be neither injective nor surjective, meaning that some individuals may have multiple names, and some no names at all. For instance, in the  informal  universe of astronomy, most individuals---celestial bodies---have no names  while some  have several names (Morning Star = Evening Star = Venus). A natural example of a mathematical universe with an intrinsically non-surjective denotator would be  one whose domain is the set of real numbers. Even if the set of constants was not fixed, the denotator here could not be surjective for the simple reason that, while there are uncountably many real numbers,  there can only be countably many names. This is so because names, by their very nature and purpose, have to be finite objects. 

Many properties of common interest, such as computability or decidability, are  sensitive with respect to how objects (individuals) are named, as they deal with the names of those objects rather than the objects themselves. For instance, strictly speaking, computing a function $f(x)$ means the ability to tell, after seeing a (the) name of an arbitrary individual $a$,  a (the) name of the individual $b$ with $b=f(a)$.  Similarly, an algorithm deciding a predicate $p(x)$ on a set $S$, strictly speaking, takes as inputs not elements of $S$---which may be abstract objects such as numbers or graphs---but rather names of those elements, such as decimal  codes. It is not hard to come up with a nonstandard naming of the natural numbers via decimal numerals where the predicate ``$x$ is even'' is undecidable.  On the other hand, for any undecidable arithmetical predicate $p(x)$, one can come up with a naming  such that $p(x)$ becomes decidable---for instance, one that assigns even-length names to all $a$ satisfying $ p(a)$ and assigns odd-length names to all $a$ with $\neg p(a)$. Classical logic exclusively deals with individuals of a universe without a need to also consider names for them, as it is not concerned with decidability or computability. CoL, on the other hand, with its computational semantics, inherently calls for being more careful about differentiating between individuals and their names, and hence for explicitly considering universes in the form $(Dm, Dn)$ rather than just $Dm$ as classical logic does.

For a set $Vr$  of variables and a domain $Dm$,   by a {\bf $(Vr,Dm)$-valuation} we mean a total function $e$ of the  type $Vr\rightarrow Dm$. When $Vr$ is finite,    such a valuation $e$ 
can be simply written as an $n$-tuple $(a_1,\cdots, a_n)$ of individuals, meaning that $e(x_1)=a_1,\cdots,e(x_n)=a_n$, where  $x_1,\cdots, x_n$ are the variables of $Vr$ listed lexicographically.

\begin{definition}\label{gameframe}
Let $n$ be a natural number. An  $n$-ary {\bf gameframe} is a quadruple $(Dm,Dn,Vr,G)$, where $(Dm,Dn)$ is a universe, $Vr$ is a set of $n$ distinct variables, and $G$ is a mapping that sends every $(Vr,Dm)$-valuation $e$ to a  game $G(e)$.
\end{definition}

Given a gameframe ${\cal G}=(Dm,Dn,Vr,G)$, we refer to $Dm$ as the {\em domain  of ${\cal G}$}, to $Dn$ as the {\em denotator  of ${\cal G}$}, to $(Dm,Dn)$ as the {\em universe of ${\cal G}$}, to the elements of $Vr$ as the {\em variables on which ${\cal G}$ depends} (or simply the {\em variables of} ${\cal G}$), and   to $G$ as the {\bf extension  of} ${\cal G}$.  
For a gameframe $(Dm,Dn,Vr,G)$ we customarily use the same name $G$ as for its extension. This never causes ambiguity, as it is usually clear from the context whether $G$ refers to the gameframe itself or just its extension. In informal contexts where a universe  is either fixed or irrelevant, we think of games as  special---nullary---cases of gameframes. Namely, a nullary gameframe $(Dm,Dn,\emptyset,G)$ will be understood as the game $G()$, usually simply written as $G$. 
  
In classical logic, under an intensional (variable-sensitive) understanding, the definition of the concept of an $n$-ary predicate would look exactly like our definition of an $n$-ary gameframe after omitting the redundant denotator component, with the only difference that there the extension function would return propositions rather than games. And, just like propositions are nothing but $0$-ary predicates, games are nothing but $0$-ary gameframes. Thus, gameframes generalize games in the same way as predicates generalize propositions.

In formal contexts, we choose a similar intensional approach to functions. The definition of a function given below is literally the same as our definition of a gameframe, with the only difference that the extension component now  maps valuations to individuals   rather than games.  

\begin{definition}\label{function}
Let $n$ be a natural number. An $n$-ary {\bf function} is a tuple $(Dm,Dn,Vr,f)$, where $(Dm,Dn)$ is a universe, $Vr$ is a set of $n$ distinct variables, and $f$ is a mapping that sends every $(Vr,Dm)$-valuation to an element $f(e)$ of $Dm$. 
\end{definition}

Just as in the case of gameframes, we customarily use the same name $f$ for a function $(Dm,Dn,Vr,f)$ as for its last component.  We refer to the elements of $Vr$ as the variables on which the function $f$ depends, refer to $Dm$ as the domain of $f$, etc.

Given a gameframe $(Dm,Dn,Vr,G)$, a set $X$ of variables with  $Vr\subseteq X$ and an  $(X,Dm)$-valuation $e$, we write $G(e)$ to mean the game $G(e')$, where $e'$ is the restriction of $e$ to $Vr$ (i.e.,  the $(Vr,Dm)$-valuation that agrees with $e$ on all variables from $Vr$). Such a  game $G(e)$ is said to be an {\bf instance} of $G$, and the operation that generates $G(e)$ from $G$ and $e$ is said to be the {\bf instantiation} operation.  For readability, we usually write $\lp{G}{e}$,  $\legal{G}{e}$ and  $\win{G}{e}$  instead of $\lp{G(e)}{ }$,  $\legal{G(e)}{ }$ and  $\win{G(e)}{ }$.  Similarly, given a function $(Dm,Dn,Vr,f)$,  a set $X$ of variables with  $Vr\subseteq X$ and an $(X,Dm)$-valuation $e$, we write $f(e)$ to denote the individual $f(e')$ to which $f$ maps $e'$, where $ e'$ is the restriction of $e$ to $Vr$.

We say that a gameframe is {\bf elementary} iff so are all of its instances. Thus, gameframes generalize elementary gameframes in the same sense as   games generalize  elementary games. In turn,  elementary gameframes generalize  elementary games in the same sense as predicates generalize propositions in classical logic. So, just as we identify   elementary games with propositions, we will identify elementary gameframes with predicates.  Specifically, in the context of a given universe $(Dm,Dn)$, we understand a predicate $p$ on $Dm$ as the elementary gameframe $(Dm,Dn,Vr,G)$, where $Vr$ is the set of variables on which $p$ depends, and $G$ is such that, for any $(Vr,Dm)$-valuation $e$, $\win{G}{e}\seq{}=\top$ iff $p$ is true at $e$. And vice versa: an elementary gameframe $G$ will be understood as the predicate $p$ that depends on the same variables as $G$ does  and  is true at a given valuation $e$ iff $\win{G}{e}\seq{}=\top$.

\begin{convention}\label{cuxu}
Assume $U=(Dm,Dn)$ is a universe, $a\in Dm$, $c\in\mbox{\em Constants}$, and $x\in\mbox{\em Variables}$. We shall write $a^{\mbox{\tiny{\it U}}}$ to mean the nullary (constant) function $(Dm,Dn,\emptyset,f)$ such that  $f()=a$. We shall write $c^{\mbox{\tiny{\it U}}}$ to mean the nullary  function $(Dm,Dn,\emptyset,f)$ such that  $f()=Dn(c)$. And we shall write $x^{\mbox{\tiny{\it U}}}$ to mean the unary function $(Dm,Dn,\{x\},f)$  such that, for any $a\in Dm$,  $f(a)=a$.
\end{convention}

\begin{convention}\label{cuxuu}
Assume $K=(Dm,Dn,Vr,K)$  is a function (resp. gameframe). Following the standard readability-improving practice established in the literature for functions and predicates, we may fix a tuple $(x_1,\cdots,x_n)$ of pairwise distinct variables for $K$ when first mentioning it, and write  $K$ as $K(x_1,\cdots,x_n)$. When doing so, we do not necessarily mean that $\{x_1,\cdots,x_n\}=Vr$. Representing $K$ as $K(x_1,\cdots,x_n)$ sets a context in which, for whatever functions $f_1=(Dm,Dn,Vr_1,f_1), \cdots, f_n=(Dm,Dn,Vr_n,f_n)$, we can write $K(f_1,\cdots,f_n)$ to mean the function (resp. gameframe) $(Dm,Dn,Vr',K')$ such that:\vspace{-7pt}  
\begin{itemize}
  \item $Vr'=(Vr -\{x_1,\cdots,x_n\})\cup Vr_1\cup\cdots\cup Vr_n$.\vspace{-7pt}
  \item For any $(Vr',Dm)$-valuation $e'$, $K'(e')=K(e)$, where $e$ is the $(Vr,Dm)$-valuation such that $e(x_1)=f_1(e'),  \cdots, e(x_n)=f_n(e')$ and $e$ agrees with $e'$ on all other variables from $Vr$.\vspace{-7pt}
\end{itemize}
Further, we allow for any of the above   $f_i$ to  be (written as) just an individual $a$, just a constant $c$ or just a variable $x$. In such   cases, $f_i$  should be correspondingly understood as the function $a^{\mbox{\tiny{\it U}}}$, $c^{\mbox{\tiny{\it U}}}$  or $x^{\mbox{\tiny{\it U}}}$, where $U=(Dm,Dn)$. So, for instance, $K(0,x)$ is our lazy way to write $K(0^{\mbox{\tiny{\it U}}},x^{\mbox{\tiny{\it U}}})$.
\end{convention}

\section{The operator zoo of computability logic}\label{zoo}
Logical operators in CoL stand for operations on gameframes. With games seen as nullary gameframes, such operations are automatically also operations on games.   There is an open-ended pool of operations of potential interest, and which of those to study may depend on particular needs and taste. Below is an incomplete list of  the operations that have been officially introduced so far.\vspace{-5pt}

\begin{itemize}
  \item Negation:  $\neg$.\vspace{-5pt}
  \item Conjunctions:  $\pand$ (parallel);    $\chand$ (choice);   $\sand$ (sequential);  $\tand$ (toggling).\vspace{-5pt}
  \item Disjunctions:       $\por$ (parallel);     $\chor$ (choice);   $\sor$ (sequential);    $\tor$ (toggling).\vspace{-5pt}
  \item Implications:    $\pimplication$ (parallel);     $\chimplication$ (choice);    $\simplication$ (sequential);    $\timplication$ (toggling).\vspace{-5pt}
  \item Universal quantifiers:   $\blall$ (blind);  $\pall$ (parallel);     $\chall$ (choice);   $\sall$ (sequential);    $\tall$ (toggling).  \vspace{-5pt}    
  \item Existential quantifiers:  $\blexists$ (blind);    $\pexists$  (parallel);     $\chexists $ (choice);   $ \sexists$ (sequential);   $\texists$ (toggling).\vspace{-5pt}      
  \item Recurrences:  $\brecurrence$ (branching);    $\precurrence$  (parallel);      $\srecurrence$ (sequential);   $\trecurrence$ (toggling).\vspace{-5pt}    
  \item Corecurrences:    $\cobrecurrence$ (branching);    $\coprecurrence$  (parallel);      $\cosrecurrence$ (sequential);   $\cotrecurrence$ (toggling).\vspace{-5pt}
  \item Rimplications:       $\brimplication$ (branching);    $\primplication$ (parallel);      $\srimplication$ (sequential);   
 $\trimplication$ (toggling).\vspace{-5pt}   
  \item Repudiations:         $\brepudiation$ (branching);    $\prepudiation$ (parallel);      $\srepudiation$ (sequential);    $\trepudiation$ (toggling).\vspace{-5pt} 
\end{itemize}

Among the symbolic names for the above  operations we see all operators of classical logic, and our choice of the classical notation for them is no accident: classical first-order logic is nothing but the result of discarding all other operators in CoL and forbidding all but elementary gameframes.  
Indeed, after analyzing the relevant definitions, each of the classically-shaped operators, when restricted to elementary gameframes, can be easily seen to be virtually the same as the corresponding operator of classical logic. For instance, if $A$ and $B$ are elementary games or gameframes, then so 
is $A\pand B$, and the latter is exactly the classical conjunction of $A$ and $B$ understood as  propositions or predicates. In the nonelementary case, however, the logical behavior of $\neg$, $\pand$, $\por$, $\pimplication$  becomes more reminiscent of---yet not the same as---that of the corresponding 
operators of multiplicative linear logic.

This section contains formal definitions of all of the above-listed operations. We agree that, throughout those definitions, $\Phi$ ranges over positions, $\Gamma$   over runs and $e$ over $(Vr,Dm)$-valuations, where $Dm$ is the domain of the gameframe $G$ that is being defined and $Vr$ is the set of variables on which that gameframe depends. All such metavariables should be considered universally quantified in the corresponding clause(s) of the definition unless otherwise implied by the context.  Each definition has two clauses, one defining $\lp{
G}{}$ 
 and the other $\win{G}{}$.  The second clause, telling us who wins a given run   of $G(e)$, always implicitly assumes that such a run  is in $\legal{G}{e}$.

This section also contains many examples and informal explanations.  For clarity  let us agree that
in all such cases, unless otherwise implied by the context, we have the arithmetical universe (cf. Section \ref{frames}) in mind. This is so even if we  talk about seemingly non-number individuals such as  people, Turing machines, etc.   The latter should simply be understood as the  natural numbers  that encode the corresponding objects in some fixed 
encoding, and the (non-numeral) names of such objects understood as the corresponding decimal numerals.   
Fixing the universe allows us to understand games as nullary gameframes as explained in \mbox{Section \ref{frames}.} The informal discussions found in this section sometimes use the word ``valid'', which, intuitively, should be understood as ``always computable''. The precise meaning(s) of this concept will only be defined later in Section \ref{s6}. When describing machine's winning strategies, we usually assume implicitly that the environment never makes illegal moves, for, if it does, the machine automatically wins regardless of what happens afterwards. 

 From our formal definitions of propositional (non-quantifier) operations it can be seen immediately that   instantiation commutes with all such operations: $(\neg A)(e)=\neg(A(e))$, $(A\pand B)(e)=A(e)\pand B(e)$, etc. So, in order to understand the meanings of the propositional operations, it would be sufficient to just understand how they modify nullary gameframes. For this reason, in the corresponding informal explanations we always implicitly assume that the gameframes that we talk about are nullary, and call them simply ``games''. For similar reasons, when informally explaining the meaning of 
$\mathbb{Q}xA$ where $\mathbb{Q}$ is one of our quantifiers, it will be implicitly assumed that $A$ is a unary gameframe that only depends on $x$ and (hence) $\mathbb{Q}xA$ is nullary.  

When omitting parentheses in compound expressions, we assume that all unary operators (negation, repudiations, recurrences, corecurrences and quantifiers) take precedence over all binary operators (conjunctions, disjunctions, implications, rimplications), among which implications and rimplications have the lowest precedence. So, for instance, $A\pimplication \neg B\por C$ should be understood as $A\pimplication \bigl((\neg B)\por C\bigr)$.

\subsection{Prefixation and negation}

Unlike the operations listed in the preceding outline, the operation of prefixation is not meant here to act as a logical operator in the formal language of CoL. Yet, it is very useful in characterizing and analyzing games, and we want to start our tour of the zoo with it.

\begin{definition} Assume $A = (Dm, Dn, Vr, A)$ is a gameframe and $\Psi$ is a legal position of every instance of $A$ (otherwise the operation is undefined). The {\bf $\Psi$-prefixation} of $A$, denoted $\seq{\Psi}A$, is defined as the gameframe $G=(Dm, Dn, Vr, G)$ such that:
\begin{itemize}
  \item $\lp{G}{e}=\{\Phi\ |\ \seq{\Psi,\Phi}\in\lp{A}{e}\}$;
  \item $\win{G}{e}\seq{\Gamma}=\win{A}{e}\seq{\Psi,\Gamma}$.
\end{itemize}
\end{definition}

Intuitively, $\seq{\Psi}A$   is a game playing which means playing $A$ starting (continuing) from position $\Psi$. That is, $\seq{\Psi}A$  is the game to which $A$ {\bf evolves} (is  {\bf brought down}) after the moves of $\Psi$ have been made. Visualized as a tree, $\seq{\Psi}A$  is nothing but the subtree of $A$ rooted at the node corresponding to position $\Psi$.  

To define the {\bf negation} operation $\neg$, read as ``{\em not}\hspace{1pt}'', let us agree that, for a run $\Gamma$,  $\overline{\Gamma}$ means the result of  changing the label  $\top$ to $\bot$ and vice versa in each move of $\Gamma$.  

\begin{definition} Assume $A = (Dm, Dn, Vr, A)$ is a gameframe. $\neg A$ is defined as the gameframe $G=(Dm, Dn, Vr, G)$ such that:  
\begin{itemize}
  \item $\lp{G}{e}=\{\overline{\Phi}\ |\ \Phi\in\lp{A}{e}\}$;
  \item $\win{G}{e}\seq{\Gamma}=\top$ iff $\win{A}{e}\seq{\overline{\Gamma}}=\bot$.
\end{itemize}
\end{definition} 

Intuitively, $\neg A$ is $A$ with the roles of the two players interchanged: $\top$'s (legal) moves and wins become $\bot$'s moves and wins, and vice versa.  
Let \chess, here and later, be the game of chess from the point of view of White, with draws ruled out (say, by declaring them to be wins for Black). Then $\neg\chess$ is the same game but as seen by Black.

Obviously the   double negation principle  $\neg\neg A = A$  holds: interchanging the players' roles twice restores the original roles of the players. It is also easy to see that we always have  $\neg \seq{\Psi}A=\seq{\overline{\Psi}}\neg A$. So, for instance, if $\alpha$ is $\top$'s legal move in the empty position of $A$ that brings $A$ down to $B$, then the same  $\alpha$ is $\bot$'s  legal move in the empty position of  $\neg A$, and it brings  $\neg A$ down to  $\neg B$.

\subsection{Choice operations}
This group of operations consists of $\chand$ ({\bf choice conjunction}, read as ``{\em chand}\hspace{1pt}''), $\chor$ ({\bf choice disjunction}, read as ``{\em chor}\hspace{1pt}''), $\chimplication$ ({\bf choice implication}, read as ``{\em chimplication}\hspace{1pt}''), $\chall$ ({\bf choice universal quantifier}, read as ``{\em chall}\hspace{1pt}'') and $\chexists$ ({\bf choice existential quantifier}, read as ``{\em chexists}\hspace{1pt}'').

$A\chand B$ is a game where, in the initial (empty) position, only the environment has legal moves. Such a move should be either ``$0$'' or ``$1$''.  If the environment moves $0$, the game continues as $A$, meaning that $\seq{\bot 0}(A\chand B) = A$; if it moves $1$, then the game continues as $B$, meaning that$\seq{\bot 1}(A\chand B) = B$; and if it fails to make either move (``choice''), then it loses. $A\chor B$ is similar, with the difference that here it is the machine who has initial moves and who loses if no such move is made. Formally, we have:

\begin{definition}
Assume $A_0=(Dm, Dn, Vr_0, A_0)$ and $A_1 =(Dm, Dn, Vr_1, A_1)$ are gameframes.
\begin{description}
  \item[(a)] 
$A_0\chand A_1$ is  defined as the gameframe $G=(Dm, Dn, Vr_0\cup Vr_1, G)$ such that:
\begin{itemize}
  \item $\lp{G}{e}=\{\seq{}\}\cup \{\seq{\bot i,\Phi}\ |\ i\in\{0,1\}, \ \Phi\in\lp{A_i}{e}    \}$.
  \item $\win{G}{e}\seq{}=\top$; $\win{G}{e}\seq{\bot  i,\Gamma}=\win{A_i}{e}\seq{\Gamma}$.
\end{itemize}
\item[(b)]
 $A_0\chor A_1$ is  defined as the gameframe $G=(Dm, Dn, Vr_0\cup Vr_1, G)$ such that:
\begin{itemize}
  \item $\lp{G}{e}=\{\seq{}\}\cup \{\seq{\top i,\Phi}\ |\ i\in\{0,1\}, \ \Phi\in\lp{A_i}{e}    \}$.
  \item $\win{G}{e}\seq{}=\bot$; $\win{G}{e}\seq{\top  i,\Gamma}=\win{A_i}{e}\seq{\Gamma}$.
\end{itemize}
\item[(c)] $A_0\chimplication A_1\ =_{def}\ \neg A_0\chor A_1$.
\end{description}
\end{definition}

The symbol $\chimplication$ is seldom used in this chapter:  instead of $A\chimplication B$, we often prefer to write the intuitively more transparent $\neg A\chor B$.

Note the perfect symmetry between the first two clauses of the above definition: clause (b) is nothing but clause (a) with $\top$ and $\bot$ interchanged everywhere, and vice versa. Such symmetry is called {\em duality}:

\begin{terminology} We say that a concept $\mathbb{B}$ is {\bf dual} to a concept $\mathbb{A}$ iff the definition of $\mathbb{B}$ can be obtained from the definition of $\mathbb{A}$ by interchanging $\top$ and $\bot$.  For instance, $\chor$ (or $A\chor B$) is dual to $\chand$ (or $A\chand B$), and vice versa. 
\end{terminology}

It is not hard to see that, due to duality,  the De Morgan laws go through for $\chand,\chor$: we always have  $\neg(A\chand B) = \neg A\chor \neg B$  and $\neg(A\chor B) = \neg A\chand \neg B$. Together with the earlier observed double negation principle, this means that 
$A\chor B =  \neg(\neg A\chand\neg B)$ and $A\chand B =  \neg(\neg A\chor \neg B)$. Similarly for the quantifier counterparts $\chall$  and $\chexists$   of $\chand$ and  $\chor$. And similarly for all other sorts of conjunctions, disjunctions, recurrences, corecurrences  and  quantifiers defined in this section. 

$\chall xA(x)$ can be understood as the infinite conjunction $A(0)\chand A(1)\chand A(2)\chand\cdots$, and $\chexists xA(x)$  as the infinite disjunction $A(0)\chor A(1)\chor A(2)\chor\cdots$. Specifically, $\chall xA(x)$ is a game where, in the initial position, only the environment has legal moves, and such a move should be one of the constants. If the environment moves $c$, then the game continues as $A(c)$, and if the environment fails to make an initial move/choice, then it loses.  $\chexists xA(x)$ is similar, with the difference that here it is the machine who has initial moves and who loses if no such move is made.  So, we always have  
$\seq{\bot c}\chall xA(x)=A(c)$ and $\seq{\top c}\chexists xA(x)=A(c)$. Below is a formal definition of the choice quantifiers:

\begin{definition}
Assume $A(x) =(Dm, Dn, Vr, A)$ is a gameframe.
\begin{description}
  \item[(a)] 
$\chall xA(x)$ is  defined as the gameframe $G=(Dm, Dn, Vr-\{x\}, G)$ such that:
\begin{itemize}
  \item $\lp{G}{e}=\{\seq{}\}\cup \{\seq{\bot c,\Phi}\ |\ c\in\mbox{\em Constants}, \ \Phi\in\lp{A(c)}{e}    \}$.
  \item $\win{G}{e}\seq{}=\top$; $\win{G}{e}\seq{\bot  c,\Gamma}=\win{A(c)}{e}\seq{\Gamma}$.
\end{itemize}
\item[(b)]
$\chexists xA(x)$ is dual to $\chall xA(x)$.  
\end{description}
\end{definition}

With choice operators we can easily express the most common sorts of computational problems, such as the problem of computing a function $f$ or the problem of deciding a predicate $p$. The former can be written as $\chall x\chexists y\bigl(y=f(x)\bigr)$, and the latter as 
$\chall x\bigl(p(x)\chimplication p(x)\bigr)$. That is, $f$ is computable in the standard sense iff $\chall x\chexists y\bigl(y=f(x)\bigr)$ is computable in our sense, and $p$ is decidable in the standard sense iff $\chall x\bigl(p(x)\chimplication p(x)\bigr)$ is computable in our sense. 
So,   the  game of Figure 2 is nothing but   $\chall x\chexists y (y=x+1)$. Every run of this game can be seen as a short dialogue between the machine and its environment. The first move---say, $2$---is by $\bot$, and intuitively it amounts to asking  ``what is the successor of $2$?''. It brings the game down to  $\chexists y (y=2+1)$.  In order win, $\top$ has to make the move $3$, amounting to saying that $3$ is the successor of $2$. Any other move, or no move at all, would be a loss for $\top$. 

Classical logic has been repeatedly criticized for its operators not being constructive. Consider, for example, $\blall x\blexists y\bigl(y=f(x)\bigr)$. It is always true in the classical sense (as long as $f$ is a total function). Yet its truth has no practical import, for ``$\blexists y$''  merely signifies existence of $y$, without implying that such a $y$ can actually be found. And, indeed, if $f$ is an incomputable function, there is no method for finding $y$. On the other hand, the choice operations of CoL are constructive. Computability (``truth'') of  $\chall x\chexists y\bigl(y=f(x)\bigr)$ means more than just existence of $y$; it means the possibility to actually find (compute, construct) the corresponding $y$ for every $x$.

Similarly, let $\mbox{\em Halts}(x,y)$ be the predicate ``Turing machine $x$ halts on input $y$''. Consider the statement $\blall x\blall y\bigl(\neg \mbox{\em Halts}(x,y)\por  \mbox{\em Halts}(x,y ) \bigr) $.   
 It is true in  classical logic, yet not in a constructive sense. Its truth means that, for all $x$ and $y$, either $\neg \mbox{\em Halts}(x,y)$  or $\mbox{\em Halts}(x,y)$ is true, but it does not 
imply existence of an actual way to tell which of these two is true after all. And such a way does not really exist, as the halting problem is undecidable. This means that $\chall x\chall y(\neg \mbox{\em Halts}(x,y) \chor \mbox{\em Halts}(x,y)\bigr)$ is not computable. Generally,  the law of   excluded middle  $\neg A\mbox{ OR } A$, validated by classical logic and causing the indignation of the constructivistically-minded, is not valid in computability logic with OR understood as choice disjunction. The following is an example of a  game of the form $\neg A\chor  A$ with no algorithmic solution (why, by the way?):
\[\neg \chall x\chall y(\neg \mbox{\em Halts}(x,y) \chor \mbox{\em Halts}(x,y)\bigr)\chor \chall x\chall y(\neg \mbox{\em Halts}(x,y) \chor \mbox{\em Halts}(x,y)\bigr).\]

\subsection{Parallel operations}\label{sp}
This group of operations consists of $\pand$ ({\bf parallel conjunction}, read as ``{\em pand}\hspace{1pt}''), $\por$ ({\bf parallel disjunction}, read as ``{\em por}\hspace{1pt}''), $\pimplication$ ({\bf parallel implication}, read as ``{\em pimplication}\hspace{1pt}''), $\pall$ ({\bf parallel universal quantifier}, read as ``{\em pall}\hspace{1pt}''), $\pexists$ ({\bf parallel existential quantifier}, read as ``{\em pexists}\hspace{1pt}''), $\precurrence$ ({\bf parallel recurrence}, read as ``{\em precurrence}\hspace{1pt}''), $\coprecurrence$ ({\bf parallel corecurrence}, read as ``{\em coprecurrence}\hspace{1pt}''), $\primplication$ ({\bf parallel rimplication}, read as ``{\em primplication}\hspace{1pt}'') and $\prepudiation$ ({\bf parallel repudiation}, read as ``{\em prepudiation}\hspace{1pt}'').

$A\pand B$ and $A\por  B$ are games playing which means playing the two games simultaneously. In order to win in $A\pand B$ (resp. $A\por B$),  $\top$ needs to win in both (resp. at least one) of the components $A,B$. For instance, $\neg \chess\por \chess$ is a two-board game, where $\top$ plays black on the left board and white on the right board, and where it needs to win in at least one of the two parallel sessions of chess. A win can be easily achieved here by just mimicking in $\chess$ the moves that the adversary is making in $\neg\chess$, and vice versa. This {\bf copycat strategy} guarantees that the positions on the two boards always remain symmetric (``synchronized''),  and thus $\top$ eventually loses on one board but wins on the other. This is very different from 
$\neg \chess\chor  \chess$. In the latter $\top$ needs to choose between the two components and then win the chosen one-board game, which makes $\neg \chess\chor  \chess$ essentially as hard to win as either $\neg \chess $ or $  \chess$. A game of the form $A\por B$ is generally easier (at least, not harder) to win than $A\chor B$, the latter is easier to win than $A\chand B$, and the latter in turn is easier to win than $A\pand B$. 
 
Technically, a move $\alpha$ in the left (resp. right) $\pand$-conjunct or $\por$-disjunct is made by prefixing $\alpha$ with ``$0.$'' (resp. ``$1.$''). For instance, in the initial position of $(A\chor B)\por (C\chand D)$, the move ``$1.0$'' is legal for $\bot$, meaning choosing the left $\chand$-conjunct in the right $\por$-disjunct of the game. If such a move is made, the game continues as 
$(A\chor B)\por  C $. The player $\top$, too, has initial legal moves in $(A\chor B)\por (C\chand D)$, which are ``$0.0$''  and ``$0.1$''. 

The rest of this chapter will rely on the following important notational convention:
\begin{notation} For a run $\Gamma$ and string $\alpha$, $\Gamma^\alpha$  means  the result of removing from $\Gamma$  all moves except those of the form $\alpha\beta$,  and then deleting the prefix ``$\alpha$'' in the remaining moves. For instance, $\seq{\top 0.1, \bot 3.1, \bot 0.0}^{0.}=\seq{\top 1, \bot 0}$.  
\end{notation}

\begin{definition}
Assume $A_0=(Dm, Dn, Vr_0, A_0)$ and $A_1 =(Dm, Dn, Vr_1, A_1)$ are gameframes.
\begin{description}
  \item[(a)] 
$A_0\pand A_1$ is  defined as the gameframe $G=(Dm, Dn, Vr_0\cup Vr_1, G)$ such that:
\begin{itemize}
  \item $\Phi\in \lp{G}{e}$ iff every move of $\Phi$ has the prefix ``$0.$'' or ``$1.$'' and, for both $i\in\{0.1\}$, $\Phi^{i.}\in 
 \lp{A_i}{e}$.
  \item $\win{G}{e}\seq{\Gamma}=\top$ iff, for both $i\in\{0,1\}$, $\win{A_i}{e}\seq{ \Gamma^{i.}}=\top$.
\end{itemize}
\item[(b)]
 $A_0\por A_1$ is dual to  $A_0\pand A_1$.
\item[(c)]
 $A_0\pimplication A_1\ =_{def}\ \neg A_0\por A_1$. 
\end{description}
\end{definition}

\begin{example}\label{sqr}
  $\Gamma=\seq{\bot 1.5, \top 0.5, \bot 0.25, \top 1.25}$ is a legal run of the game $A = \chexists x\chall y(y\not=x^2) \por \chall x\chexists y(y=x^2)$. It induces the following what we call {\em evolution sequence}, showing how things evolve as $\Gamma$ runs, i.e., how the moves of $\Gamma$  affect/modify the game that is being  played:
\[
\begin{array}{ll}
    \chexists x\chall y(y\not=x^2) \por \chall x\chexists y(y=x^2)  &   \mbox{ i.e. } A\\
   \chexists x\chall y(y\not=x^2) \por  \chexists y(y=5^2)  & \mbox{ i.e. } \seq{\bot 1.5}A\\
  \chall y(y\not=5^2) \por  \chexists y(y=5^2) & \mbox{ i.e. } \seq{\bot 1.5,\top 0.5}A\\ 
  25\not=5^2  \por  \chexists y(y=5^2) & \mbox{ i.e. } \seq{\bot 1.5,\top 0.5,\bot 0.25}A\\ 
    25\not=5^2  \por  25=5^2  & \mbox{ i.e. }  \seq{\bot 1.5,\top 0.5,\bot 0.25,\top 1.25}A
\end{array}\]
The run hits the true proposition $25\not=5^2\por  25=5^2$, and hence is won by the machine.
\end{example}

As one may guess, $\pall xA(x)$ is nothing but $A(0)\pand A(1)\pand A(2)\pand\cdots$,  and  $\pexists xA(x)$ is nothing but $A(0)\por A(1)\por A(2)\por\cdots$. Formally these two quantifiers are defined as follows:

\begin{definition}
Assume $A(x) =(Dm, Dn, Vr, A)$ is a gameframe.
\begin{description}
  \item[(a)] 
$\pall xA(x)$ is  defined as the gameframe $G=(Dm, Dn, Vr-\{x\}, G)$ such that:
\begin{itemize}
  \item $\Phi\in \lp{G}{e}$ iff every move of $\Phi$ has the prefix ``$c.$'' for some  $c\in\mbox{\em Constants}$ and, for all such $c$, $\Phi^{c.}\in 
 \lp{A(c)}{e}$.
  \item $\win{G}{e}\seq{\Gamma}=\top$ iff, for all  $c\in\mbox{\em Constants}$, $\win{A(c)}{e}\seq{ \Gamma^{c.}}=\top$.
\end{itemize}
\item[(b)]
 $\pexists xA(x)$ is dual to $\pall xA(x)$.
\end{description}
\end{definition}

The next group of parallel operators are $\precurrence$ and its dual $\coprecurrence$. Intuitively, playing  $\precurrence A$  means simultaneously playing in infinitely many  ``copies'' of $A$, and  $\top$ is the winner iff it wins $A$ in all copies. $\coprecurrence A$  is similar, with the only difference that here winning in just one copy is sufficient.  So, $\precurrence A$   is nothing but the infinite parallel conjunction $A\pand A\pand A\pand \cdots$, and $\coprecurrence A$   is  $A\por A\pand A\por \cdots$. Equivalently, $\precurrence A$  and $\coprecurrence A$    can be respectively understood as $\pall xA$ and $\pexists xA$, where $x$ is a dummy variable on which $A$ does not depend. The following definition formalizes these intuitions:

\begin{definition}
Assume $A =(Dm, Dn, Vr, A)$ is a gameframe.
\begin{description}
  \item[(a)] 
$\precurrence A$ is  defined as the gameframe $G=(Dm, Dn, Vr, G)$ such that:
\begin{itemize}
  \item $\Phi\in \lp{G}{e}$ iff every move of $\Phi$ has the prefix ``$c.$'' for some  $c\in\mbox{\em Constants}$ and, for all such $c$, $\Phi^{c.}\in \lp{A}{e}$.
  \item $\win{G}{e}\seq{\Gamma}=\top$ iff, for all  $c\in\mbox{\em Constants}$, $\win{A}{e}\seq{ \Gamma^{c.}}=\top$.
\end{itemize}
\item[(b)]
 $\coprecurrence A$ is  dual to $\precurrence A$.
\end{description}
\end{definition}

The prefix ``r'' in the qualification ``rimplication'' stands for ``recurrence''. Generally, a rimplication  of one or another sort is a weak (recurrence-based) implication, and a repudiation is a  weak negation. The parallel versions 
$\primplication$ and $\prepudiation$ of such operations are defined as follows. 
\begin{definition} {\bf (a)}  $A\primplication B\ =_{def}\ \precurrence A\pimplication B$. \ 
{\bf (b)} $\prepudiation A\ =_{def}\ \coprecurrence \neg A$.
\end{definition}

Just like negation and unlike choice operations, parallel operations preserve the elementary property of games. When restricted to elementary games, the meanings of $\pand$, $\por$ and $\pimplication$ coincide with those of classical conjunction, disjunction and implication. Further, as long as all individuals of the universe have naming constants,  the meanings of  $\pall$ and $\pexists$   coincide with those of classical universal quantifier and existential quantifier. The same conservation of classical meaning (but without any conditions on the universe) is going to be the case with the blind quantifiers $\blall,\blexists$ defined later; so, at the elementary level, when all individuals of the universe have naming constants, $\pall$ $\pexists$ and  are indistinguishable from  and $\blall$ and $\blexists$, respectively. As for the parallel recurrence and corecurrence, for an elementary $A$ we simply have $A=\precurrence A=\coprecurrence A$. 

While all classical tautologies automatically remain valid when parallel operators are applied to elementary games, in the general case the class of valid (in the strict sense of either sort of validity defined in Section \ref{s6}) principles shrinks. For example, $P\pimplication P\pand P$, i.e. $\neg P\por (P\pand P)$, is not valid.  Back to our chess example, one can see that the earlier copycat strategy successful for $\neg \chess\por\chess$  would be inapplicable to $\neg \chess\por (\chess\pand \chess)$. The best that $\top$ can do in this three-board game is to  synchronize $\neg \chess$  with one of the two conjuncts of $\chess\pand \chess$. It is possible that then $\neg\chess$ and the unmatched $\chess$ are both lost, in which case the whole game will be lost as well.

The principle $P\pimplication P\pand P$ is valid in classical logic because the latter sees no difference between $P$ and $P\pand P$. On the other hand, in virtue of its semantics, CoL is resource-conscious, and in it $P$ is by no means the same as $P\pand P$ or $P\por P$.  Unlike $P\pimplication P\pand P$,  $P\primplication  P\pand P$ is a valid principle. Here, in the antecedent, we have infinitely many ``copies'' of $P$. Pick any two copies and, via copycat, synchronize them with the two conjuncts of the consequent. A win is guaranteed. The principle $P\pimplication P\chand P$ can also be seen to be  valid. 

This talk about resource-consciousness immediately reminds us of linear logic \cite{Gir87}. The latter, for instance, also rejects $P\pimplication P\pand P$ while accepting both $P\pimplication P\chand P$ and $\precurrence P\pimplication P\pand P$ with $\pand,\por,\pimplication$ understood as multiplicatives, $\chand,\chor$ as additives  and $\precurrence,\coprecurrence$ as exponentials. Together with similarities, there are also considerable discrepancies though. The class of principles provable in linear logic or even its extension known as affine logic forms a proper subclass of the principles validated by the semantics of CoL. An example of a purely multiplicative formula separating the two classes is Blass's \cite{Bla92} principle
\[(P\pand P)\por (P\pand P)\pimplication (P\por P)\pand (P\por P).\] 
Other examples include $\coprecurrence \precurrence P\pimplication \precurrence \coprecurrence P$. On the other hand, it is believed (but never has been officially proven) that the class of CoL's valid principles in the signature $\{\neg,\pand,\por,\chand,\chor,\chall,\chexists\}$ is indistinguishable from the class of principles validated by Blass's \cite{Bla92} game semantics. This, however, stops being the case if $\precurrence$---or any later-defined sort of recurrence for that matter---is added to the signature as a purported   counterpart of Blass's {\em repetition} operator. For instance, $\precurrence (P\chor Q)\pimplication \precurrence P\chor \precurrence Q$ is valid in Blass's sense but it is not a valid principle of CoL; on the other hand, CoL validates $P\pand \precurrence (P\pimplication Q\pand P)\pimplication \precurrence P$ (cf. \cite{separating}) which is not  valid in Blass's sense. 

\subsection{Reduction}\label{sreduc}
The operator $\pimplication$   deserves a separate subsection. The intuition associated with $A\pimplication B$ is that this is the problem of {\em reducing} $B$ to $A$: solving $A\pimplication B$ means solving $B$ while having $A$ as a computational resource. Specifically,
 $\top$  may observe how $A$ is being solved by its adversary, and utilize this information in its own solving $B$. Resources 
are symmetric to problems: what is a problem to solve for one player is a resource that the other player can use, and vice versa. Since $A$ is negated in $A\pimplication B=\neg A\por B$ and negation means switching the players' roles, $A$ (as opposed to $\neg A$) comes as a resource rather than problem to $\top$ in $A\pimplication B$.  Our copycat strategy for $\neg\chess\por\chess$ was an example of reducing 
$\chess$ to $\chess$. The same strategy was underlying Example \ref{sqr}, where $\chall x\chexists y(y=x^2)$  was reduced to itself.

Let us look at a more meaningful example: reducing the acceptance problem to the halting problem. The former, as a decision problem, will be written as $\chall x\chall y \bigl(\neg \mbox{\em Accepts}(x,y)\chor  \mbox{\em Accepts}(x,y)\bigr)$, where $\mbox{\em Accepts}(x,y)$ is the predicate ``Turing machine $x$ accepts input $y$''. Similarly, 
as we already agreed, the halting problem is written as $\chall x\chall y \bigl(\neg \mbox{\em Halts}(x,y)\chor  \mbox{\em Halts}(x,y)\bigr)$. Neither problem has an algorithmic solution, yet the following implication does:
\begin{equation}\label{horse}
\chall x\chall y \bigl(\neg \mbox{\em Halts}(x,y)\chor\mbox{\em Halts}(x,y)\bigr)\pimplication \chall x\chall y \bigl(\neg \mbox{\em Accepts}(x,y)\chor \mbox{\em Accepts}(x,y)\bigr).\end{equation}

Here is $\top$'s winning strategy for (\ref{horse}). Wait till $\bot$ makes the moves $1.m$ and $1.n$ for some $m$ and $n$. Making these moves essentially means asking the question ``Does machine $m$ accept input $n$?''.  If such moves are never made, you (the machine) win. Otherwise, the moves bring the game down to
 \[\chall x\chall y \bigl(\neg \mbox{\em Halts}(x,y)\chor\mbox{\em Halts}(x,y)\bigr)\pimplication  \neg \mbox{\em Accepts}(m,n)\chor \mbox{\em Accepts}(m,n) .\]
Make the moves $0.m$ and $0.n$, thus asking the counterquestion ``Does machine $m$ halt on input $n$?''.  Your moves further bring the game down to  \[\neg \mbox{\em Halts}(m,n)\chor\mbox{\em Halts}(m,n)\pimplication  \neg \mbox{\em Accepts}(m,n)\chor \mbox{\em Accepts}(m,n).\]
$\bot$ will have to answer this counterquestion, or else it loses (why?). If it answers by   $0.0$ (``No, $m$ does not halt on $n$''), you make the move $1.0$ (say ``$m$ does not accept $n$''). The game will be brought down to \(\neg \mbox{\em Halts}(m,n)\pimplication  \neg \mbox{\em Accepts}(m,n)\). You win, because this is a true proposition: if $m$ does not halt on $n$, then it does not accept $n$, either. Otherwise, if $\bot$ answers by  $0.1$ (``Yes, $m$ halts on $n$''), start simulating $m$ on $n$ until $m$ halts. If you see that $m$ accepted $n$, make the move $1.1$ (say ``$m$ accepts $n$''); otherwise make the move $1.0$ (say ``$m$ does not accept $n$''). Of course, it is a possibility that this simulation goes on forever. But then $\bot$ has lied when saying ``$m$ halts on $n$''; in other words, the antecedent is false, and you win regardless of what happens in the consequent.   Note that what the machine did when following this strategy was indeed reducing the acceptance problem to the halting problem: it solved the former using an external (environment-provided) solution of the latter.

There are many natural concepts of reduction, and a strong case can be made that  {\bf pimplicative reduction}, i.e. the reduction captured by $\pimplication$, is the most basic one. For this reason  we agree that, if we simply say ``reduction'', it always 
means pimplicative reduction. A great variety of other reasonable concepts of reduction is expressible 
in terms of $\pimplication$.  Among those is Turing reduction. Remember that a predicate $q(x)$ is said to be 
{\em Turing reducible} to a predicate $p(x)$ if $q(x)$ can be decided by a Turing machine equipped with an oracle for $p(x)$.  
For a positive integer $n$, {\em $n$-bounded Turing reducibility} is defined the same way, with the only difference that here 
the oracle is allowed to be used only $n$ times. It turns out that $\primplication$  is a conservative generalization of 
Turing reduction. Namely, when $p(x)$ and $q(x)$ are elementary games (i.e. predicates), $q(x)$ is Turing reducible to $p(x)$ 
if and only if the problem $\chall x\bigl(\neg p(x)\chor p(x)\bigr)\primplication \chall x\bigl(\neg q(x)\chor  q(x)\bigr)$ has 
an algorithmic solution. If here we change $\primplication$ back to $\pimplication$, we get the same result for 1-bounded
 Turing reducibility. More generally, as one might guess, $n$-bounded Turing reduction will be captured by
\[\chall x_1\bigr(\neg p(x_1)\chor  p(x_1)\bigr)\pand \cdots \pand \chall x_n\bigr(\neg p(x_n)\chor  p(x_n)\bigr)\pimplication  \chall x\bigl(\neg q(x)\chor q(x)\bigr).\]
If, instead, we write 
\[\chall x_1\cdots\chall x_n\Bigl(\bigr(\neg p(x_1)\chor  p(x_1)\bigr)\pand \cdots \pand  \bigr(\neg p(x_n)\chor  p(x_n)\bigr)\Bigr)\pimplication  \chall x\bigl(\neg q(x)\chor q(x)\bigr),\]
then we get what is called {\em $n$-bounded weak truth-table reduction}. The latter differs from $n$-bounded Turing reduction in that here all $n$ oracle queries should be made at once, before seeing responses to any of those queries. What is called {\em mapping} (or {\em many-one}) {\em reducibility} of $q(x)$ to $p(x)$ is nothing but computability of $\chall x\chexists y\bigl(q(x)\leftrightarrow p(y)\bigr)$, where $A\leftrightarrow B$ abbreviates $(A\pimplication B)\pand (B\pimplication A)$.  One  could go on and on with this list.

And yet many other natural concepts of reduction expressible in the language of CoL  may have no established names in the literature. For example, from the previous discussion it can be seen that a certain reducibility-style relation holds between the predicates $\mbox{\em Accepts}(x,y)$ and $\mbox{\em Halts}(x,y)$ in an even stronger sense than computability of (\ref{horse}). 
In fact, not only (\ref{horse}) has an algorithmic solution, but also the generally harder-to-solve problem
\[
\chall x\chall y \bigl( \neg \mbox{\em Halts}(x,y)\chor\mbox{\em Halts}(x,y) \pimplication    \neg \mbox{\em Accepts}(x,y)\chor \mbox{\em Accepts}(x,y) \bigr).\]
Among the merits of CoL is that it offers a formalism and deductive machinery for systematically expressing and studying computation-theoretic relations such as reducibility, decidability, enumerability, etc., and all kinds of variations of such concepts.

Back to reducibility, while the standard approaches only allow us to talk about (a whatever sort of) reducibility as a {\em relation} between problems, in our approach reduction becomes an {\em operation} on problems, with reducibility as a relation simply meaning computability of the corresponding combination  of games, such as $A\pimplication B$ for pimplicative reducibility. Similarly for other relations or properties such as the property of {\em decidability}. The latter becomes the operation of {\em deciding} if we define the problem of deciding a predicate $p(x)$ as the game $\chall x\bigl(\neg p(x)\chor  p(x)\bigr)$. So, now we can meaningfully ask questions such as ``{\em Is the reduction of the problem of deciding $q(x)$ to the problem of deciding $p(x)$ always reducible to the mapping reduction of $q(x)$ to $p(x)$?}\hspace{1pt}''. This question would be equivalent to whether the following formula is valid in CoL:
\begin{equation}\label{e1}\chall x\chexists y\bigl(q(x)\leftrightarrow p(y)\bigr)\pimplication \Bigl(\chall x\bigl(\neg p(x)\chor p(x)\bigr)\pimplication    \chall x\bigl(\neg q(x)\chor q(x)\bigr)\Bigr).\end{equation}
The answer turns out to be ``Yes'', meaning that mapping reduction is at least as strong as pimplicative reduction. Here is a strategy that wins this game no matter what particular predicates $p(x)$ and $q(x)$ are. At first, wait till, for some $m$, the environment brings the game down to  
\[\chall x\chexists y\bigl(q(x)\leftrightarrow p(y)\bigr)\pimplication \Bigl(\chall x\bigl(\neg p(x)\chor p(x)\bigr)\pimplication    \neg q(m)\chor q(m)\Bigr).\]
Respond by bringing the game down to  
\[\chexists y\bigl(q(m)\leftrightarrow p(y)\bigr)\pimplication \Bigl(\chall x\bigl(\neg p(x)\chor p(x)\bigr)\pimplication    \neg q(m)\chor q(m)\Bigr).\]
Wait again till, for some $n$, the environment further brings the above game down to  
\[\bigl(q(m)\leftrightarrow p(n)\bigr)\pimplication \Bigl(\chall x\bigl(\neg p(x)\chor p(x)\bigr)\pimplication    \neg q(m)\chor q(m)\Bigr).\]
Bring this game down to  
\(\bigl(q(m)\leftrightarrow p(n)\bigr)\pimplication \bigl(\neg p(n)\chor p(n)\pimplication    \neg q(m)\chor q(m)\bigr),\)
after which wait till the environment further brings the game down to either \(\bigl(q(m)\leftrightarrow p(n)\bigr)\pimplication \bigl( \neg  p(n)\pimplication    \neg q(m)\chor q(m)\bigr)\)  or \(\bigl(q(m)\leftrightarrow p(n)\bigr)\pimplication \bigl(  p(n)\pimplication    \neg q(m)\chor q(m)\bigr)\). In the former case, bring the game  down to \(\bigl(q(m)\leftrightarrow p(n)\bigr)\pimplication \bigl( \neg  p(n)\pimplication    \neg q(m)\bigr)\), and you have won; 
in the latter case, bring the game down to \(\bigl(q(m)\leftrightarrow p(n)\bigr)\pimplication \bigl(p(n) \pimplication q(m)\bigr)\), and you have won, again.

One could also ask: ``{\em Is the mapping reduction of $q(x)$ to $p(x)$ always reducible to the reduction of the problem of deciding $q(x)$ to the problem of deciding $p(x)$?}\hspace{1pt}''. This question would be equivalent to whether the following formula is valid:
\begin{equation}\label{e2}
\Bigl(\chall x\bigl(\neg p(x)\chor p(x)\bigr)\pimplication \chall x\bigl(\neg q(x)\chor q(x)\bigr)\Bigr)\pimplication \chall x\chexists y\bigl(q(x)\leftrightarrow p(y)\bigr).\end{equation}
The answer here turns out to be ``No'', meaning that mapping reduction is properly stronger than pimplicative reduction. This negative answer can be  obtained by showing that the above formula is not provable in one of the   sound and complete deductive systems for  CoL whose language allows to write (\ref{e2}), such as system {\bf CL12} found later in Section \ref{scl12}. Similarly, 
 had our ad hoc attempt to come up with a strategy for (\ref{e1}) failed, its validity could have been easily established by finding a proof of it in such a system.

To summarize, CoL offers not only a convenient language for specifying computational problems and relations or operations on them, but also a systematic tool for asking and answering questions in the above style and beyond.

\subsection{Blind operations} 
This group only includes  $\blall$ ({\bf blind universal quantifier}, read as ``{\em blall}\hspace{1pt}'') and $\blexists$ ({\bf blind existential quantifier}, read as ``{\em blexists}\hspace{1pt}''), with no propositional counterparts. Our definition of $\blall xA(x)$ and $\blexists xA(x)$ below assumes that the gameframe $A(x)$ is ``unistructural'' in $x$.  Intuitively, unistructurality in $x$ means that the $\lp{}{}$ component of the gameframe does not depend on the (value of the) variable $x$. Formally, we say that a gameframe $A(x)=(Dm,Dn,Vr,A)$  is {\bf unistructural in $x$} iff, for any $(Vr,Dm)$-valuation $e$ and any  $a,b\in Dm$, we have $\lp{A(a)}{e} = \lp{A(b)}{e}$.  All nullary or elementary gameframes are unistructural in (whatever variable) $x$. And all operations of CoL are known to preserve this property. 

\begin{definition}
 Assume $A(x)  = (Dm, Dn, Vr, A)$ is a gameframe unistructural in   $x$.

\begin{description}
  \item[(a)] 
$\blall xA(x)$ is  defined as the gameframe $G=(Dm, Dn, Vr-\{x\}, G)$ such that:
\begin{itemize}
  \item $\lp{G}{e}=\lp{A(x)}{e}$. 
  \item $\win{G}{e}\seq{\Gamma}=\top$ iff, for all  $a\in Dm$, $\win{A(a)}{e}\seq{ \Gamma}=\top$.
\end{itemize}
\item[(b)]
$\blexists xA(x)$ is dual to $\blall xA(x)$.
\end{description}
\end{definition}

Intuitively, playing $\blall xA(x)$ or $\blexists xA(x)$ means playing $A(x)$ ``blindly'', without knowing the value of $x$. In $\blall xA(x)$, the machine wins iff the play it generates is successful for every possible value of $x$ from the domain, while in $\blexists xA(x)$ being successful for just one value is sufficient. When applied to elementary games, the blind quantifiers act exactly like the corresponding quantifiers of classical logic.

Unlike $\pall xA(x)$ which is a game on infinitely many boards, both $\blall xA(x)$ and $\chall xA(x)$  are one-board games. Yet, they are very different from each other. To see this difference, compare the problems $\chall x \bigl(\mbox{\em Even}(x)\chor \mbox{\em Odd}(x)\bigr)$ and $\blall x \bigl(\mbox{\em Even}(x)\chor \mbox{\em Odd}(x)\bigr)$. The former is an easily winnable game of depth 2: the environment selects a number, and the machine tells whether that number is even or odd. The latter, on the other hand, is a game which is impossible to win. This is a game of depth 1, where the value of $x$ is not specified by either player, and only the machine moves---tells whether (the unknown) $x$ is even or odd. Whatever the machine says, it loses, because there is always a value for $x$ that makes the answer wrong.

This should not suggest that nontrivial $\blall$-games can never be won. For instance, the problem 
\[\blall x\Bigl(\mbox{\em Even}(x) \chor \mbox{\em Odd}(x)\pimplication \chall y\bigl(\mbox{\em Even}(x+y)\chor \mbox{\em Odd}(x+y)\bigr)\Bigr)\] has an easy solution. The idea of a winning strategy here is that, for any given $y$, in order to tell the parity of $x+y$, it is not really necessary to know the value of $x$. Rather, just knowing the parity of $x$ is sufficient. And such knowledge can be obtained from the antecedent. In other words, for any known $y$ and unknown $x$, the problem of telling whether $x+y$ is even or odd is reducible to the problem of telling whether $x$ is even or odd. Specifically, if both $x$ and $y$ are even or both are odd, then $x+y$ is even; otherwise $x+y$ is odd. Below is the evolution sequence (cf. Exercise \ref{sqr}) induced by the run $\seq{\bot 1.7, \bot 0.0, \top 1.1}$ where the machine has used such a strategy.
\[\begin{array}{l}
    \blall x\Bigl(\mbox{\em Even}(x)\chor \mbox{\em Odd}(x)\pimplication \chall   y\bigl(\mbox{\em Even}(x+y)\chor  \mbox{\em Odd}(x+y)\bigr)\Bigr)\\
   \blall x\bigl(\mbox{\em Even}(x)\chor  \mbox{\em Odd}(x)\pimplication   \mbox{\em Even}(x+7)\chor  \mbox{\em Odd}(x+7)\bigr)\\
    \blall x\bigl(\mbox{\em Even}(x)\pimplication \mbox{\em Even}(x+7) \chor \mbox{\em Odd}(x+7)\bigr)\\
    \blall x\bigl(\mbox{\em Even}(x) \pimplication    \mbox{\em Odd}(x+7)\bigr)
\end{array}\]
The machine  won because the play hit the true $\blall x\bigl(\mbox{\em Even}(x) \pimplication    \mbox{\em Odd}(x+7)\bigr)$. Notice how $\blall x$ persisted throughout the sequence. Generally, the $(\blall,\blexists)$-structure of a game will remain unchanged in such sequences. The same is the case with   parallel operations such as $\pimplication$ in the present case.

To help us appreciate the contrast between the logical behaviors of $\blall$, $\chall$ and $\pall$,  the following list shows some valid ($\valid$) and invalid ($\invalid$) principles of CoL, where validity (``always computability'') can be understood in either sense defined later in Section \ref{s6}.\vspace{5pt}
  
1. $\valid\ \blall xP(x)\pimplication  \chall xP(x)$   

2. $\invalid\ \chall xP(x) \pimplication  \blall xP(x)$   

3.  $\invalid\ \blall xP(x)\pimplication  \pall xP(x)$   

4. $\invalid\ \pall xP(x)\pimplication  \blall xP(x)$ 

5. $\valid\ \pall xP(x)\pimplication  \chall xP(x)$

6. $\invalid\ \chall xP(x)\pimplication  \pall xP(x)$

7. $\valid\ \mathbb{Q} x P(x)\pand \mathbb{Q} x R(x)\pimplication \mathbb{Q} x\bigl( P(x)\pand R(x)\bigr)$ for all three $\mathbb{Q}\in\{\blall,\chall,\pall\}$ 

8. $\valid\ \blall x\bigl(P(x)\pand R(x)\bigr)\pimplication  \blall xP(x)\pand \blall xR(x)$

9. $\invalid\ \chall x\bigl(P(x)\pand R(x)\bigr)\pimplication  \chall xP(x)\pand \chall xR(x)$

10. $\valid\ \pall x\bigl(P(x)\pand R(x)\bigr)\pimplication  \pall xP(x)\pand \pall xR(x)$

\subsection{Branching operations}
This group consists of $\brecurrence$ ({\bf branching recurrence}, read as ``{\em brecurrence}\hspace{1pt}''), $\cobrecurrence$ ({\bf branching corecurrence}, read as ``{\em cobrecurrence}\hspace{1pt}''), $\brimplication$ ({\bf branching rimplication}, read as ``{\em brimplication}\hspace{1pt}'') and $\brepudiation$ ({\bf branching repudiation}, read as ``{\em brepudiation}\hspace{1pt}'').   Let us talk about $\brecurrence$ first, as all other branching operations are definable in terms of it.

What is common for the members of the family of game operations called recurrences  is that, when applied to a game $A$, they turn it into a game playing which means repeatedly playing $A$. In terms of resources, recurrence operations generate multiple ``copies'' of $A$, thus making $A$ a reusable/recyclable resource. In classical logic, recurrence-style operations would be meaningless, because classical logic is resource-blind and thus sees no difference between one and multiple copies of $A$. In the resource-conscious CoL, however, recurrence operations are not only meaningful, but also necessary to achieve a satisfactory level of expressiveness and realize its potential and ambitions. Hardly any computer program is used only once; rather, it is run over and over again. Loops within such programs also assume multiple repetitions of the same subroutine. In general, the tasks performed in real life by computers, robots or humans are typically recurring ones or involve recurring subtasks.

There is more than one naturally emerging recurrence operation. The differences between various recurrence operations are related to how ``repetition'' or ``reusage''  is exactly understood. Imagine a computer with a chess-playing  program. The resource that such a computer provides is obviously something stronger than just our old friend $\chess$ (as long as it always wins), for it permits to play $\chess$ as many times as the user wishes, while $\chess$, as such, only assumes one play. The simplest operating system would allow to start a session of $\chess$, then---after finishing or abandoning and destroying it---start a new play again, and so on. The game that such a system plays---i.e. the resource that it supports/provides---is $\srecurrence \chess$,  which assumes an unbounded number of plays of $\chess$ in a sequential fashion. A formal definition of the operation $\srecurrence$, called {\em sequential recurrence}, will be given later is Section \ref{ss}.

A more advanced operating system, however, would not require to destroy the old sessions before starting new ones; rather, it would allow to run as many parallel sessions as the user wants. This is what is captured by $\precurrence \chess$, meaning nothing but the infinite parallel conjunction $\chess\pand  \chess\pand  \chess\pand\cdots$. As we remember from Section \ref{sp},  $\precurrence$ is called {\em parallel recurrence}. 

Yet a really good operating system would not only allow the user to start new sessions of $\chess$ without destroying old ones; it would also make it possible to branch/replicate any particular stage of any particular session, i.e., create any number of ``copies'' of any already reached position of the multiple parallel plays of $\chess$, thus giving the user the possibility to try different continuations from the same position. What corresponds to this intuition is the branching recurrence $\brecurrence \chess$ of $\chess$.

At the intuitive level, the difference between $\brecurrence$ and $\precurrence$ is that in $\brecurrence A$, unlike $\precurrence A$,  the environment does not have to restart $A$ from the very beginning every time it wants to reuse it (as a resource); rather, it is allowed to backtrack to any  of the previous---not necessarily starting---positions and try a new continuation from there, thus depriving the adversary of the possibility to reconsider the moves it has already made in that position. This is in fact the type of reusage every purely software resource allows or would allow in the presence of an advanced operating system and unlimited memory: one can start running a process (task, game); then fork it at any stage thus creating two threads that have a common past but possibly diverging futures  (with the possibility to treat one of the threads as a ``backup copy'' and preserve it for backtracking purposes); then further fork any of the branches at any time; and so on.

The less flexible type of reusage of $A$ assumed by $\precurrence A$, on the other hand, is closer to what infinitely many autonomous physical resources would naturally offer, such as an unlimited number of independently acting robots each performing task $A$, or an unlimited number of computers with limited memories, each one only capable of and responsible for running a single thread of process $A$. Here the effect of forking an advanced stage of $A$ cannot be achieved unless, by good luck, there are two identical copies of the stage, meaning that the corresponding two robots or computers have so far acted in precisely the same ways.

In early papers \cite{Jap03, Japfin} on CoL,   the formal definitions of $\brecurrence$ and its dual $\cobrecurrence$    were direct formalizations of the above intuitions, with an explicit presence of ``replicative'' moves used by players to fork a given thread of $A$ and create two threads out of one. Later, in \cite{face},  another definition was found which was proven to be equivalent to the old one in the sense of mutual reducibility of the old and the new versions of $\brecurrence A$. The new definition less directly corresponds to the above intuitions, but is technically simpler, and we choose it as our ``canonical'' definition of branching (co)recurrence. To be able to state it, we agree on the following:

\begin{notation} Where $\Gamma$ is a run and $w$ is a {\bf bitstring} (finite or infinite sequence of $0$s and $1$s), $\Gamma^{\preceq w}$ means the result of deleting from $\Gamma$  all moves except those that look like $u.\alpha$ for some initial segment $u$ of $w$, and then further deleting the prefix ``$u.$'' from such moves. E.g., $\seq{\bot 00.77,\top 01.88, \top 0.66}^{\preceq 00}=\seq{\bot 77, \top 66}$.
\end{notation}

\begin{definition} 
Assume $A = (Dm, Dn, Vr, A)$ is a gameframe.
\begin{description}
  \item[(a)] 
$\brecurrence A $ is  defined as the gameframe $G=(Dm, Dn, Vr, G)$ such that:
\begin{itemize}
\item 
$\Phi\in\lp{G}{e}$ iff every move of $\Phi$ has the prefix ``$u.$'' for some finite bitstring $u$ and, for every infinite bitstring $w$,  $\Phi^{\preceq w}\in \lp{A}{e}$;
\item $\win{G}{e}\seq{\Gamma}=\top$  iff, for every infinite bitstring $w$, $\win{A}{e}\seq{\Gamma^{\preceq w}}=\top$.
\end{itemize}
\item[(b)]
$\cobrecurrence A $ is  dual to $\brecurrence A $. 
\end{description}
\end{definition}

The direct intuitions underlying this definition are as follows. To play $\brecurrence A$ or $\cobrecurrence A$ means to simultaneously play in multiple parallel copies/threads of $A$.  Each infinite bitstring $w$ denotes one such thread    (so, there are in fact uncountably many threads, even if some of them coincide). Every legal move by either player looks like $u.\alpha$ for some finite bitstring $u$, and the effect/meaning of such a move is simultaneously making the move $\alpha$ in all threads $w$ such that $u$ is an initial segment of $w$. So, where $\Gamma$ is the overall run of $\brecurrence A$ or $\cobrecurrence A$, the run in a given thread $w$ of $A$ is $\Gamma^{\preceq w}$. In order to win $\brecurrence A$, the machine needs to win $A$ in all threads, while for winning $\cobrecurrence  A$ it is sufficient to win in just one thread.

$\brecurrence$ can be shown to be stronger than its parallel counterpart $\precurrence$, in the sense that the principle  $\brecurrence P\pimplication \precurrence P$ is valid while $\precurrence P\pimplication \brecurrence P$ is not. The two   operators, in isolation from each other, also validate different principles. For instance, $P\pand \precurrence(P\pimplication Q\pand P)\pimplication  \precurrence Q$ is valid while  $P\pand \brecurrence(P\pimplication Q\pand P)\pimplication  \brecurrence Q$ is not;  
$\brecurrence(P\chor Q)\pimplication \brecurrence P\chor\brecurrence Q $ is valid while $\precurrence(P\chor Q)\pimplication \precurrence P\chor\precurrence Q $ is not.  
In its overall spirit, the earlier mentioned Blass's repetition operator $\mathfrak{R}$ is much closer to $\brecurrence$ than $\precurrence$, yet Blass's semantics validates a different set of principles with $\mathfrak{R}$ than CoL does with $\brecurrence$. For instance, 
the following formula is invalid  \cite{separating} in CoL but valid in Blass's semantics with $\mathfrak{R}$ in the role of $\brecurrence$: $P\pand \brecurrence(P\pimplication Q\pand P)\pand \brecurrence(R\por Q\pimplication R)\pimplication \brecurrence R$. 

The branching sorts of rimplication and repudiation  are defined in terms of   $\neg,\pimplication$ and  $\brecurrence,\cobrecurrence$  the same way as the parallel sorts of rimplication  and  repudiation  are defined in terms of $\neg,\pimplication$ and  $\precurrence,\coprecurrence$:

\begin{definition} {\bf (a)}  $A\brimplication B\ =_{def}\ \brecurrence A\pimplication B$. \ \ 
{\bf (b)} $\brepudiation A\ =_{def}\ \cobrecurrence \neg A$.

\end{definition}  

Similarly to the earlier defined pimplicative reducibility, for games $A,B$  we say that $B$ is {\bf brimplicatively} (resp. primplicatively, etc.) {\bf reducible} to $A$ iff $A\brimplication B$ (resp. $A\primplication B$, etc.) is computable.  

\begin{exercise}\label{kolmogorov}
 The {\em Kolmogorov complexity} $k(x)$ of a natural number $x$ is the size of a smallest Turing machine that outputs $x$ on input $0$. The Kolmogorov complexity problem $\chall x\chexists y\bigl(y=k(x)\bigr)$ has no algorithmic solution. Nor is it pimplicatively reducible to the halting problem. It, however, is reducible to the halting problem in the weaker sense of brimplicative reducibility, meaning that $\top$ has a winning strategy for  $\chall x\chall y\bigl(\neg\mbox{\em Halts}(x,y)\chor\mbox{\em Halts}(x,y)\bigr) \brimplication \chall x\chexists y\bigl(y=k(x)\bigr)$. Describe such a strategy, informally.
\end{exercise}    

Both brimplicative and primplicative reducibilities are  conservative generalizations of Turing reducibility: for any predicates $p(x)$ and $q(x)$,   $\chall x\bigl(\neg p(x)\chor  p(x)\bigr)\brimplication \chall x\bigl(\neg q(x) \chor q(x)\bigr)$ is computable iff $q(x)$ is Turing reducible to $p(x)$ iff  $\chall x\bigl(\neg p(x)\chor  p(x)\bigr)\primplication \chall x\bigl(\neg q(x)\chor  q(x)\bigr)$ is computable. Generally, when restricted to traditional sorts of problems such as problems of deciding a predicate or computing a function as in Exercise \ref{kolmogorov},   $\brimplication$  and $\primplication$  are extensionally indistinguishable. This, however, stops being the case when these operators are applied to problems with higher degrees of interactivity. For instance, the following problem is computable, but becomes incomputable with $\primplication$ instead of $\brimplication$:
\[\chexists y\chall x\bigl(\neg\mbox{\em Halts}(x,y)\chor\mbox{\em Halts}(x,y)\bigr)\brimplication \chexists y\Bigl(\chall x\bigl(\neg\mbox{\em Halts}(x,y)\chor\mbox{\em Halts}(x,y)\bigr)\pand \chall x\bigl(\neg\mbox{\em Halts}(x,y)\chor\mbox{\em Halts}(x,y)\bigr)\Bigr).\]
Generally, $(P\primplication Q)\pimplication (P\brimplication Q)$ is valid but $(P\brimplication Q)\pimplication (P\primplication Q)$ is not.

While both  $\primplication$ and $\brimplication$  are weaker than $\pimplication$ and hence more general than the latter, $\brimplication$ is still a more interesting operation of weak reduction than $\primplication$. What makes it special is the  belief stated in Thesis \ref{thesis} below. The latter, in turn, is based on the belief that $\brecurrence$ (and by no means $\precurrence$)  is the operation allowing to reuse its argument in the strongest algorithmic sense possible.

\begin{thesis}\label{thesis}
  Brimplicative  reducibility is an adequate mathematical counterpart of our intuition of reducibility in the weakest---and thus most general---algorithmic sense possible. Specifically:

   (a) Whenever a problem $B$ is brimplicatively reducible to a problem $A$, $B$ is also algorithmically reducible to $A$ according to anyone's reasonable intuition.

   (b) Whenever a problem $B$ is algorithmically reducible to a problem $A$ according to anyone's reasonable intuition, $B$ is also brimplicatively reducible to $A$.
\end{thesis}

The above is pretty much in the same sense as, by the Church-Turing thesis, a function $f$ is computable by a Turing machine iff  $f$  has an algorithmic solution according to anyone's reasonable intuition.

Understanding the intuitionistic negation, implication, conjunction, disjunction   and   quantifiers as $\brepudiation$,  $\brimplication$, $\chand$, $\chor$, $\chall$, $\chexists$, respectively, Heyting's system for intuitionistic logic has been shown \cite{int1} to be sound with respect to the semantics of CoL. It is also ``almost complete'', as the following formula of an imposing length is among the shortest known propositional formulas  valid in CoL but unprovable in Heyting's calculus:
\[(\brepudiation P\brimplication Q\chor R)\chand(\brepudiation \brepudiation P\brimplication  Q\chor R) \brimplication
 (\brepudiation P\brimplication Q)\chor (\brepudiation P\brimplication  R) \chor (\brepudiation \brepudiation P\brimplication Q)\chor(\brepudiation \brepudiation P\brimplication R).\]


\subsection{Sequential operations}\label{ss}
This group consists of $\sand$ ({\bf sequential conjunction}, read as ``{\em sand}\hspace{1pt}''), $\sor$ ({\bf sequential disjunction}, read as ``{\em sor}\hspace{1pt}''), $\simplication$ ({\bf sequential implication}, read as ``{\em simplication}\hspace{1pt}''), $\sall$ ({\bf sequential universal quantifier}, read as ``{\em sall}\hspace{1pt}''), $\sexists$ ({\bf sequential existential quantifier}, read as ``{\em sexists}\hspace{1pt}''), 
$\srecurrence$ ({\bf sequential recurrence}, read as ``{\em srecurrence}\hspace{1pt}''), $\cosrecurrence$ ({\bf sequential corecurrence}, read as ``{\em cosrecurrence}\hspace{1pt}''),  
$\srimplication$ ({\bf sequential rimplication}, read as ``{\em srimplication}\hspace{1pt}'') and $\srepudiation$ ({\bf sequential repudiation}, read as ``{\em srepudiation}\hspace{1pt}'').   

The game $A \sand B$ starts and proceeds as $A$. It will also end as $A$ unless, at some point, the environment decides to switch to the next component, in which case $A$ is abandoned, and the game restarts, continues and ends as $B$.  $A\sor B$ is similar, with the difference that here it is the machine who decides whether and when to switch from $A$ to $B$.

The original formal definition of $A \sand B$ and $A \sor B$ found in \cite{Japseq} was a direct formalization of the above description. Definition \ref{srach} given below, while less direct, still faithfully formalizes the above intuitions, and we opt for it because it is technically simpler. Specifically, Definition  \ref{srach} allows either player to continue making moves in $A$ even after a switch takes place; such moves are meaningless but harmless. Similarly, it allows either player to make moves in $B$ without waiting for a switch to take place, even though a smart player would only start making such moves if and when a switch happens.  

\begin{definition} \label{srach} 
Assume $A_0=(Dm, Dn, Vr_0, A_0)$ and $A_1 =(Dm, Dn, Vr_1, A_1)$ are gameframes.
\begin{description}
  \item[(a)] 
$A_0\sand A_1$ is  defined as the gameframe $G=(Dm, Dn, Vr_0\cup Vr_1, G)$ such that:
\begin{itemize}
  \item $\Phi\in\lp{G}{e}$ iff $\Phi$  has the form $\seq{\Psi,\Theta}$ or $\seq{\Psi,\bot 1,\Theta}$, where every move of $\seq{\Psi,\Theta}$ has the prefix ``$0.$'' or ``$1.$'' and, for both $i\in\{0,1\}$, $\seq{\Psi,\Theta}^{i.}\in \lp{A_i}{e}$.
  \item If $\Gamma$ does not contain a (``{\em switch}'') move $\bot 1$, $\win{G}{e}\seq{\Gamma}=\win{A_0}{e}\seq{\Gamma^{0.}}$; otherwise   $\win{G}{e}\seq{\Gamma}=\win{A_1}{e}\seq{\Gamma^{1.}}$
\end{itemize}
\item[(b)]
 $A_0\sor A_1$ is dual to $A_0\sand A_1$.
\item[(c)] $A_0\simplication A_1\ =_{def}\ \neg A_0\sor A_1$.
\end{description}
\end{definition}

Recall that, for a predicate $p(x)$,  $\chall x\bigl(p(x) \chimplication p(x)\bigr)$  is the problem of deciding $p(x)$. The 
similar-looking $\chall x\bigl(p(x) \simplication p(x)\bigr)$, on the other hand, can be seen to be the problem of {\em semideciding} $p(x)$: the machine has a winning strategy in this game if and only if $p(x)$ is semidecidable, i.e., recursively enumerable. Indeed, if $p(x)$ is recursively enumerable, a winning strategy   by $\top$  is to wait until $\bot$ brings the game down to $ p(n) \simplication p(n) $, i.e., $\neg p(n)\sor p(n)$, for some particular $n$.  After that, $\top$ starts looking for a certificate of $p(n)$'s  being true. If and when such a certificate is found (meaning that $p(n)$ is indeed true), $\top$ makes a switch move turning  $\neg p(n)\sor p(n)$ into the true $p(n)$; and if no certificate exists (meaning that $p(n)$ is false), then $\top$ keeps looking for a non-existent certificate forever and thus never makes any moves, so the game ends as $\neg p(n)$, which, again, is true. And vice versa: any effective winning strategy for $\chall x\bigl(\neg p(x) \sor p(x)\bigr)$ can obviously be seen as a semidecision procedure for $p(x)$, which accepts an input $n$ iff the strategy ever makes a switch move in the scenario where $\bot$'s initial choice of a value for $x$ is $n$.

As we remember from Section \ref{sreduc}, Turing reducibility of a predicate $p(x)$ to a predicate $q(x)$ means nothing but computability of 
$\chall x\bigl(q(x)\chimplication q(x)\bigr)\primplication \chall x\bigl(p(x)\chimplication p(x)\bigr)$ (the same holds with $\brimplication$ instead of $\primplication$). One can show that changing  $\chimplication$ to $\simplication$ here yields another known concept of reducibility, called  {\em enumeration reducibility} (cf. \cite{soskova}). That is,  $p(x)$ is enumeration reducible to $q(x)$ iff $\chall x\bigl(q(x)\simplication q(x)\bigr)\primplication \chall x\bigl(p(x)\simplication p(x)\bigr)$ is computable in our sense. Similarly, the formula $\chall x\bigl(q(x)\chimplication q(x)\bigr)\primplication \chall x\bigl(p(x)\simplication p(x)\bigr)$ captures {\em relative computable enumerability} (again, cf. \cite{soskova}). And so on and so forth.
 
Existence of effective winning strategies for games is known \cite{Jap03} to be closed under  
``{\em from $A\pimplication B$ and $A$ conclude $B$}'',  ``{\em from $A$ and $B$ conclude $A\pand B$}'', ``{\em from $A$ conclude 
$\chall xA$}'', ``{\em from $A$ conclude $\brecurrence A$}'' and similar rules. In view of such closures, the validity of the principles 
discussed below implies certain known facts from the theory of computation. Those examples once again 
demonstrate how CoL can be used as a systematic tool for defining new interesting properties and relations between 
computational problems, and not only reproducing already known theorems but also discovering an infinite variety of new facts. 

The valid formula $\chall x\bigl(p(x)\simplication  p(x)\bigr)\pand \chall x\bigl(\neg p(x)\simplication \neg p(x)\bigr)\pimplication  \chall x\bigl(p(x)\chimplication p(x)\bigr)$ ``expresses''  the well known fact that, if  a predicate $p(x)$ and its complement $\neg p(x)$ are both recursively enumerable, then $p(x)$ is decidable. Actually, the validity of this formula means something more: it means that the problem of deciding $p(x)$ is reducible to (the $\pand$-conjunction of) the problems of semideciding $p(x)$ and $\neg p(x)$. In fact, reducibility in an even stronger sense---a sense that has no name---holds, expressed by the formula 
$\chall x\Bigl(\bigl(p(x)\simplication  p(x)\bigr)\pand \bigl(\neg p(x) \simplication \neg p(x)\bigr)\pimplication \bigl(p(x)\chimplication   p(x)\bigr)\Bigr)$.

The formula $\chall x\chexists y \bigl(q(x) \leftrightarrow p(y)\bigr)\pand   \chall x\bigl(p(x)\simplication p(x)\bigr)\pimplication 
\chall x\bigl(q(x)\simplication q(x)\bigr)$
is also valid, which implies the known fact that, if a predicate $q(x)$ is mapping reducible to a predicate $p(x)$ and $p(x)$ is recursively enumerable, then  $q(x)$ is also recursively enumerable. Again, the validity of this formula, in fact, means something even more: it means that the problem of semideciding $q(x)$ is reducible to the problems of mapping reducing $q(x)$ to $p(x)$ and semideciding $p(x)$.

Certain other reducibilities hold only in the sense of rimplications rather than implications. Here is an example. Two Turing machines are said to be equivalent iff they accept exactly the same inputs.  Let $\mbox{\em Neq}(x,y)$ be the predicate ``Turing machines $x$ and $y$ are not equivalent''. This predicate is neither semidecidable nor co-semidecidable. However, the problem of its semideciding primplicatively (and hence also brimplicatively) reduces to the halting problem.  Specifically, $\top$ has an effective winning strategy for the game
\[\chall z\chall t\bigl(\neg \mbox{\em Halts}(z,t)\chor  \mbox{\em Halts}(z,t)\bigr)\primplication \chall x\chall y\bigl(\neg 
\mbox{\em Neq}(x,y)\sor  \mbox{\em Neq}(x,y)\bigr),\]
in terms of  \cite{soskova} meaning that   $\mbox{\em Neq}(x,y)$ is computably enumerable relative to $\mbox{\em Halts}(z,t)$.
The strategy is to wait till the environment specifies some values $m$ and $n$ for $x$ and $y$. Then, create a variable $i$, initialize it to $1$ and do the following. Specify $z$ and $t$ as $m$ and $i$ in one yet-unused copy of the antecedent, and as $n$ and $i$ in another yet-unused copy. That is, ask the environment whether $m$ halts on input $i$ and whether $n$ halts on the same input. The environment will have to provide the correct pair of answers, or else it loses. (1) If the answers are ``No,No'', increment $i$ to $i+1$ and repeat the step. (2) If the answers are ``Yes,Yes'', simulate both $m$ and $n$ on input $i$ until they halt. If both machines accept or both reject, increment $i$ to $i+1$ and repeat the step. Otherwise, if one accepts and one rejects, make a switch move in the consequent and celebrate victory. (3) If the answers are ``Yes,No'', simulate $m$ on $i$ until it halts. If $m$ rejects $i$, increment $i$ to $i+1$ and repeat the step. Otherwise, if $m$ accepts $i$, make a switch move in the consequent and you win. (4) Finally, if the answers are ``No,Yes'', simulate $n$ on $i$ until it halts. If $n$ rejects $i$, increment $i$ to $i+1$ and repeat the step. Otherwise, if $n$ accepts $i$, make a switch move in the consequent and you win.

As expected, $\sall xA(x)$ is essentially  the infinite sequential conjunction $A(0)\sand  A(1)\sand  A(2)\sand \cdots$, \ $\sexists xA(x)$ is $A(0)\sor  A(1)\sor  A(2)\sor \cdots$, $\srecurrence A$ is $A\sand  A\sand  A\sand\cdots$ and $\cosrecurrence A$ is 
$A\sor  A\sor  A\sor\cdots$. Formally, we have:

\begin{definition}
Assume $A(x)=(Dm, Dn, Vr, A)$ is a gameframe.
\begin{description}
  \item[(a)] 
$\sall xA(x)$ is  defined as the gameframe $G=(Dm, Dn, Vr-\{x\}, G)$ such that:
\begin{itemize}
  \item $\Phi\in\lp{G}{e}$ iff $\Phi$  has the form $\seq{\Psi_0,\bot 1,\Psi_1, \cdots,\bot n,\Psi_n}$ ($n\geq 0$), where every move of 
$\seq{\Psi_0,\cdots,\Psi_n}$ has the prefix ``$c.$'' for some $c\in\mbox{\em Constants}$  and, for every such $c$, $\seq{\Psi_0,\cdots,\Psi_n}^{c.}\in \lp{A(c)}{e}$.
  \item Call  $\bot 1,\bot 2,\cdots$ {\em switch moves}. If $\Gamma$ does not contain a switch move, then
$\win{G}{e}\seq{\Gamma}=\win{A(0)}{e}\seq{\Gamma^{0.}}$; if $\Gamma$ contains infinitely many switch moves, then $\win{G}{e}\seq{\Gamma}=\top$;   otherwise, where $\bot n$ is the last switch move of  $\Gamma$, $\win{G}{e}\seq{\Gamma}=\win{A(n)}{e}\seq{\Gamma^{n.}}$. 
\end{itemize}
\item[(b)] $\sexists xA(x)$ is dual to $\sall xA(x)$.
\end{description}
\end{definition}
 
\begin{definition}  Assume $A = (Dm, Dn, Vr, A)$ is a gameframe.
\begin{description}
  \item[(a)] 
$\srecurrence A$ is  defined as the gameframe $G=(Dm, Dn, Vr, G)$ such that:
\begin{itemize}
  \item $\Phi\in\lp{G}{e}$ iff $\Phi$  has the form $\seq{\Psi_0,\bot 1,\Psi_1,\cdots,\bot n,\Psi_n}$ ($n\geq 0$), where every move of 
$\seq{\Psi_0,\cdots,\Psi_n}$ has the prefix ``$c.$'' for some $c\in\mbox{\em Constants}$ and, for every such $c$, $\seq{\Psi_0,\cdots,\Psi_n}^{c.}\in \lp{A}{e}$.
  \item Call  $\bot 1,\bot 2,\cdots$ {\em switch moves}. If $\Gamma$ does not contain a switch move, then
$\win{G}{e}\seq{\Gamma}=\win{A}{e}\seq{\Gamma^{0.}}$; if $\Gamma$ contains infinitely many switch moves, then $\win{G}{e}\seq{\Gamma}=\top$;   otherwise, where $\bot n$ is the last switch move of  $\Gamma$, $\win{G}{e}\seq{\Gamma}=\win{A}{e}\seq{\Gamma^{n.}}$. 
\end{itemize}
\item[(b)] $\cosrecurrence A$ is dual to $\srecurrence A$.
\end{description}
\end{definition}

For insights into the above-defined operations, remember the Kolmogorov complexity function $k(x)$ from Exercise \ref{kolmogorov}. It is known that the value of $k(x)$ is always smaller than $x$ (in fact, logarithmically smaller).   While $\chall x\chexists y\bigl(y=k(x)\bigr)$ is not computable, $\top$ does have an algorithmic winning strategy for the problem $\chall x\cosrecurrence \chexists y\bigl(y=k(x)\bigr)$. It goes like this: Wait till $\bot$ specifies a value $m$ for $x$, thus asking ``what is the Kolmogorov complexity of $m$?'' and bringing the game down to $\cosrecurrence \chexists y\bigl(y=k(m)\bigr)$. Answer (generously) that the complexity is $m$, i.e. specify $y$ as $m$. After that, start simulating, in parallel, all machines $n$ of sizes smaller than $m$ on input $0$. Whenever you find a machine $n$ that returns $m$ on input $0$ and is smaller than any of the previously found such machines, make a switch move and, in the new copy of  $\chexists y\bigl(y=k(m)\bigr)$, specify $y$ as the size $|n|$ of $n$.   This obviously guarantees success: sooner or later the real Kolmogorov complexity $c$ of $m$ will be reached and named; and, even though the strategy will never be sure that $k(m)$ is not something yet smaller than $c$, it will never really find a reason to further reconsider its latest claim that $c=k(m)$. 

\begin{exercise} Describe a winning strategy for $\chall x\sexists y\bigl(k(x)=x-y\bigr)$.\end{exercise}

\begin{definition} {\bf (a)}  $ A\srimplication  B\  =_{def}\  \srecurrence A\pimplication  B$. \ \  {\bf (b)} $ \srepudiation A\  =_{def}\  \cosrecurrence \neg A$. 
\end{definition}

\subsection{Toggling operations}
This group consists of $\tand$ ({\bf toggling conjunction}, read as ``{\em tand}\hspace{1pt}''), $\tor$ ({\bf  toggling disjunction}, read as \mbox{``{\em tor}\hspace{1pt}''),} \mbox{$\timplication$} ({\bf  toggling implication}, read as ``{\em timplication}\hspace{1pt}''), \mbox{$\tall$} ({\bf  toggling universal quantifier}, read as ``{\em tall}\hspace{1pt}''), $\texists$ ({\bf  toggling existential quantifier}, read as ``{\em texists}\hspace{1pt}''), $\trecurrence$ ({\bf  toggling recurrence}, read as \mbox{``{\em trecurrence}\hspace{1pt}''),} $\cotrecurrence$ ({\bf  toggling corecurrence}, read as \mbox{``{\em cotrecurrence}\hspace{1pt}''),} $\trimplication$ ({\bf  toggling rimplication}, read as \mbox{``{\em trimplication}\hspace{1pt}'')} and $\trepudiation$ ({\bf  toggling repudiation}, read as ``{\em trepudiation}\hspace{1pt}'').        

Let us for now focus on $\tor$.
One of the ways to characterize $A\tor B$ is the following. This game starts and proceeds as a play of $A$. It will also end as an ordinary play of $A$ unless, at some point, $\top$ decides to switch to $B$, after which the game becomes $B$ and continues as such. It will also end as $B$ unless, at some point, $\top$ ``changes its mind'' and switches back to $A$. In such a case the game again becomes $A$, where $A$ resumes from the position in which it was abandoned (rather than from its start position, as would be the case, e.g., in $A\sor B\sor A$). Later $\top$ may again switch to  the abandoned position of $B$, and so on. $\top$ wins the overall play iff it switches from one component to another 
at most finitely many times and wins in its final choice, i.e., in the component which was chosen last to switch to.

An alternative  characterization $A\tor B$, on which our formal definition of $\tor$ is directly based, is to say that it is played just like $A\chor B$, with the only difference that $\top$ is allowed to make a ``choose $A$''  or ``choose $B$'' move any number of times. If infinitely many choices are made, $\top$ loses. Otherwise, the winner in the play will be the player who wins in the component that was chosen last (``the eventual choice''). The case of $\top$ having made no choices at all is treated as if it had chosen $A$. Thus, as in $A\sor B$, the left component is the ``default'', or ``automatically made'', initial choice.
It is important to note that $\top$'s adversary---or perhaps even $\top$ itself---never knows whether a given choice of a component of $A\tor B$ is the last choice or not.

What would happen if we did not require that $\top$ can change its mind only finitely many times? There would be no ``final choice'' in this case. So, the only natural winning condition in the case of infinitely many choices would be to say that $\top$ wins iff it simply wins in one of the components. But then the resulting operation would be essentially the same as $\por$, as a smart $\top$  would always opt for keeping switching between components forever. That is, allowing infinitely many choices would amount to not requiring any choices at all, as is the case with $A\por B$.

The very weak sort of choice captured by  $\tor$ is the kind of choice that, in real life, one would ordinarily call choice after trial and error. Indeed, a problem is generally considered to be solved after trial and error (a correct choice/solution/answer found) if, after perhaps coming up with several wrong solutions, a true solution is eventually found. That is, mistakes are tolerated and forgotten as long as they are eventually corrected. It is however necessary that new solutions stop coming at some point, so that there is a last solution whose correctness determines the success of the effort. Otherwise, if answers have kept changing all the time, no answer has really been given after all.    

As we remember, for a predicate $p(x)$,  $\chall x\bigl(p(x) \chimplication p(x)\bigr)$ is the problem of deciding $p(x)$, and $\chall x\bigl(p(x) \simplication p(x)\bigr)$ is the weaker (easier to solve) problem of semideciding $p(x)$. Not surprisingly, $\chall x\bigl(p(x) \timplication p(x)\bigr)$---which abbreviates $\chall x\bigl(\neg p(x) \tor p(x)\bigr)$---is also a decision-style problem, but still weaker than the problem of semideciding $p(x)$. This problem has been studied in the literature under several names, the most common of which is {\em recursively approximating}  $p(x)$. It means telling whether $p(x)$ is true or not, but doing so in the same style as semideciding does in negative cases: by correctly saying ``Yes'' or ``No'' at some point (after perhaps taking back previous answers several times) and never reconsidering this answer afterwards.  In similar terms, semideciding $p(x)$ can be seen as always saying (the default) ``No'' at the beginning and then, if this answer is incorrect, changing it to ``Yes'' at some later time; so, when the answer is negative, this will be expressed by saying ``No'' and never taking back this answer, yet without ever indicating that the answer is final and will not change. Thus, the difference between semideciding and recursively approximating is that, unlike a semidecision procedure, a recursive approximation procedure can reconsider both negative and positive answers, and  do so several times rather than only once.

As an example of a predicate which is recursively approximable but neither semidecidable nor co-semidecidable, consider the predicate $k(x)\hspace{-3pt}<\hspace{-3pt}k(y)$, saying that number $x$ is simpler than number $y$ in the sense of Kolmogorov complexity. As noted earlier, $k(z)$ (the Kolmogorov complexity of $z$) is always smaller than $z$. Here is an algorithm that recursively approximates the predicate $k(x)\hspace{-3pt}<\hspace{-3pt}k(y)$, i.e., solves the problem $\chall x\chall y\bigl(k(x)\hspace{-3pt}\geq\hspace{-3pt} k(y)\tor  k(x)\hspace{-3pt}< \hspace{-3pt} k(y)\bigr)$. Wait till the environment brings the game down to $ k(m)\hspace{-3pt}\geq\hspace{-3pt}k(n)\tor  k(m)\hspace{-3pt}<\hspace{-3pt} k(n)$ for some $m$ and $n$. Then start simulating, in parallel, all Turing machines $t$ of sizes less than $max(m,n)$ on input $0$. Whenever you see that a machine $t$ returns $m$  and the size of $t$ is smaller than that of any other previously found machine that returns $m$ or $n$ on input $0$, choose $k(m)\hspace{-3pt}<\hspace{-3pt} k(n)$. Quite similarly, whenever you see that a machine $t$ returns $n$ and the size of $t$ is smaller than that of any other previously found machine that returns $n$ on input $0$, as well as smaller or equal to the size of any other previously found machine that returns $m$ on input $0$, choose $k(m)\hspace{-3pt}\geq\hspace{-3pt}k(n)$. Obviously, the correct choice between $k(m)\hspace{-3pt}\geq \hspace{-3pt}k(n)$ and $k(m)\hspace{-3pt}<\hspace{-3pt}k(n)$ will be made sooner or later and never reconsidered afterwards. This will happen when the procedure hits a smallest-size machine $t$ that returns either $m$ or $n$ on input $0$.

Anyway, here is our formal definition of $\tand$, $\tor$ and $\timplication$:

\begin{definition} \label{tand} 
Assume $A_0=(Dm, Dn, Vr_0, A_0)$ and $A_1 =(Dm, Dn, Vr_1, A_1)$ are gameframes.
\begin{description}
  \item[(a)] 
$A_0\tand A_1$ is  defined as the gameframe $G=(Dm, Dn, Vr_0\cup Vr_1, G)$ such that:
\begin{itemize}
  \item $\Phi\in\lp{G}{e}$ iff $\Phi$  has the form $\seq{\Psi_0,\bot i_1,\Psi_1,\cdots,\bot i_n,\Psi_n}$ ($n\geq 0$), where 
$i_1,\cdots, i_n\in\{0,1\}$, every move of 
$\seq{\Psi_0,\cdots,\Psi_n}$ has the prefix ``$0.$'' or ``$1.$'' and, for both $i\in\{0,1\}$, $\seq{\Psi_0,\cdots,\Psi_n}^{i.}\in \lp{A_i}{e}$.
  \item Call $\bot 0$ and $\bot 1$ {\em switch moves}. If $\Gamma$ does not contain a switch move, then
$\win{G}{e}\seq{\Gamma}=\win{A_0}{e}\seq{\Gamma^{0.}}$; if $\Gamma$ contains infinitely many switch moves, then $\win{G}{e}\seq{\Gamma}=\top$;   otherwise, where $\bot i$ is the last switch move of  $\Gamma$, $\win{G}{e}\seq{\Gamma}=\win{A_{i}}{e}\seq{\Gamma^{i.}}$. 
\end{itemize}
\item[(b)]
 $A_0\tor A_1$ is dual to $A_0\tand A_1$.
\item[(c)] $A_0\timplication A_1\ =_{def}\ \neg A_0\tor A_1$.
\end{description}
\end{definition}

From the formal definitions that follow one can see that, as expected, $\tall xA(x)$ is essentially   $A(0)\tand A(1)\tand$ $A(2)\tand\cdots$, $\texists xA(x)$ is  $A(0)\tor A(1)\tor A(2)\tor \cdots$, \ $\trecurrence A$ is $A \tand A \tand A \tand\cdots$ \ and $\cotrecurrence  xA(x)$ is  $A \tor A \tor A \tor \cdots$.

\begin{definition}
Assume $A(x)=(Dm, Dn, Vr, A)$ is a gameframe.
\begin{description}
  \item[(a)] 

$\tall xA(x)$ is  defined as the gameframe $G=(Dm, Dn, Vr-\{x\}, G)$ such that:
\begin{itemize}
  \item $\Phi\in\lp{G}{e}$ iff $\Phi$  has the form $\seq{\Psi_0,\bot c_1,\Psi_1,\cdots,\bot c_n,\Psi_n}$ ($n\geq 0$), where $c_1,\cdots,c_n\in\mbox{\em Constants}$, every move of 
$\seq{\Psi_0,\cdots,\Psi_n}$ has the prefix ``$c.$'' for some $c\in\mbox{\em Constants}$ and, for every such $c$, $\seq{\Psi_0,\cdots,\Psi_n}^{c.}\in \lp{A(c)}{e}$.
  \item Call the above  $\bot c_1,\bot c_2,\cdots$ {\em switch moves}. If $\Gamma$ does not contain a switch move, then
$\win{G}{e}\seq{\Gamma}=\win{A(0)}{e}\seq{\Gamma^{0.}}$; if $\Gamma$ contains infinitely many switch moves, then $\win{G}{e}\seq{\Gamma}=\top$;   otherwise, where $\bot c$ is the last switch move of  $\Gamma$, $\win{G}{e}\seq{\Gamma}=\win{A(c)}{e}\seq{\Gamma^{c.}}$. 
\end{itemize}
\item[(b)] $\texists xA(x)$ is dual to $\tall xA(x)$.
\end{description}
\end{definition}
 
\begin{definition}  Assume $A = (Dm, Dn, Vr, A)$ is a gameframe.
\begin{description}
  \item[(a)] 
$\trecurrence A$ is  defined as the gameframe $G=(Dm, Dn, Vr, G)$ such that:
\begin{itemize}
  \item $\Phi\in\lp{G}{e}$ iff $\Phi$  has the form $\seq{\Psi_0,\bot c_1,\Psi_1,\cdots,\bot c_n,\Psi_n}$ ($n\geq 0$), where  $c_1,\cdots,c_n\in\mbox{\em Constants}$, every move of 
$\seq{\Psi_0,\cdots,\Psi_n}$ has the prefix ``$c.$'' for some $c\in\mbox{\em Constants}$ and, for every such $c$,  $\seq{\Psi_0,\cdots,\Psi_n}^{c.}\in \lp{A}{e}$.
  \item Call the above  $\bot c_1,\bot c_2,\cdots$ {\em switch moves}. If $\Gamma$ does not contain a switch move, then
$\win{G}{e}\seq{\Gamma}=\win{A}{e}\seq{\Gamma^{0.}}$; if $\Gamma$ contains infinitely many switch moves, then $\win{G}{e}\seq{\Gamma}=\top$;   otherwise, where $\bot c$ is the last switch move of  $\Gamma$, $\win{G}{e}\seq{\Gamma}=\win{A}{e}\seq{\Gamma^{c.}}$. 
\end{itemize}
\item[(b)] $\cotrecurrence A$ is dual to $\trecurrence A$.
\end{description}
\end{definition}

To see toggling quantifiers at work, remember that Kolmogorov complexity $k(x)$ is not a computable function, i.e., the problem $\chall x\chexists y\bigl(y=k(x)\bigr)$ has no algorithmic solution. However, replacing $\chexists y$ with $\texists y$ in it yields an algorithmically solvable problem. A solution for $\chall x\texists y\bigl(y=k(x)\bigr)$ goes like this. Wait till the environment chooses a number $m$ for $x$, thus bringing the game down to $\texists y\bigl(y=k(m)\bigr)$, which is essentially nothing but $0=k(m)\tor 1=k(m)\tor  2=k(m) \tor\cdots$.  Create a variable $i$ initialized to $m$, and perform the following routine: Switch to the disjunct $i=k(m)$ of $0=k(m)\tor 1=k(m)\tor  2=k(m) \tor\cdots$ and then start simulating on input $0$, in parallel, all Turing machines whose sizes are smaller than $i$; if and when you see that one of such machines returns $m$, update $i$ to the size of that machine, and repeat the present routine.

\begin{definition} {\bf (a)}  $ A\trimplication  B\  =_{def}\  \trecurrence A\pimplication  B$. \ \  {\bf (b)} $ \trepudiation A\  =_{def}\  \cotrecurrence \neg A$. 
\end{definition}

\subsection{Cirquents}
The constructs called {\bf cirquents} take the expressive power of CoL to a qualitatively higher level, allowing us to form, in a systematic way, an infinite variety of game operations. Each cirquent is---or can be seen as---an independent operation on games, generally not expressible via composing operations taken from some fixed finite pool of primitives, such as the operations seen in the preceding subsections of the present section.

Cirquents come in a variety of versions, but common to all them is having mechanisms for explicitly accounting for possible {\em sharing} of subcomponents between different components. Sharing is the main distinguishing feature of cirquents from more traditional means of expression such as formulas, sequents,  hypersequents \cite{Avr87}, or structures of the calculus of structures \cite{Gug07}.  While the latter can be drawn as (their parse) trees, cirquents more naturally call for circuit- or graph-style constructs.   The earliest cirquents \cite{Cirq} were intuitively conceived as collections of one-sided sequents (sequences of formulas) that could share some formulas and, as such, could be drawn like circuits rather than linear expressions. This explains the etimology of the word: CIRcuit+seQUENT. All Boolean circuits are cirquents, but not all cirquents are Boolean circuits. Firstly, because cirquents may have various additional sorts of gates ($\chand$-gates,  $\sand$-gates,  $\tand$-gates, etc.). Secondly, because cirquents may often have more evolved sharing mechanisms than just child- (input-) sharing between different gates. For instance,  a ``cluster'' \cite{LMCS2011} of $\chor$-gates may share choices associated with  $\chor$ in game-playing: if the machine chooses the left or the right child for one gate of the cluster, then the same left or right choice automatically extends to all gates of the cluster regardless of whether they share children or not.

We are not going to introduce cirquents and their semantics in full generality or formal detail here. For intuitive insights, let us only focus on cirquents that look like Boolean circuits with $\pand$- and $\por$-gates. Every such cirquent $C$ can be seen as an $n$-ary parallel operation on games, where $n$ is the number of inputs of $C$. 

\begin{center} \begin{picture}(235,87)

\put(36,57){\line(-2,1){29}}
\put(36,57){\line(2,1){29}}
\put(7,57){\line(0,1){15}}
\put(7,57){\line(2,1){29}}

\put(36,31){\line(2,1){29}}
\put(36,31){\line(-2,1){29}}
\put(36,31){\line(0,1){14}}

\put(65,57){\line(-2,1){29}}
\put(65,57){\line(0,1){15}}

\put(3,75){$P$}
\put(32,75){$Q$}
\put(61,75){$R$}

\put(33,48){$\pand$}
\put(36,51){\circle{11}}

\put(4,48){$\pand$}
\put(7,51){\circle{11}}

\put(62,48){$\pand$}
\put(65,51){\circle{11}}

\put(33,23){$\por$}
\put(36,26){\circle{11}}
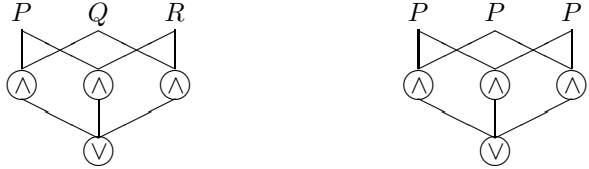
\put(-14,8){{\bf Figure 3:} The two-out-of-three combination of resources}

\put(186,57){\line(-2,1){29}}
\put(186,57){\line(2,1){29}}
\put(157,57){\line(0,1){15}}
\put(157,57){\line(2,1){29}}

\put(186,31){\line(2,1){29}}
\put(186,31){\line(-2,1){29}}
\put(186,31){\line(0,1){14}}

\put(215,57){\line(-2,1){29}}
\put(215,57){\line(0,1){15}}

\put(153,75){$P$}
\put(182,75){$P$}
\put(211,75){$P$}

\put(183,48){$\pand$}
\put(186,51){\circle{11}}

\put(154,48){$\pand$}
\put(157,51){\circle{11}}

\put(212,48){$\pand$}
\put(215,51){\circle{11}}

\put(183,23){$\por$}
\put(186,26){\circle{11}}

\end{picture}
\end{center}

The left cirquent of Figure 3  
represents the 3-ary game operation $\heartsuit$ informally defined as follows. Playing $\heartsuit(P,Q,R)$, as is the case with all parallel operations, means playing simultaneously in all  components of it. In order to win, $\top$ needs to win in at least two out of the three components. Any attempt to express this operation in terms of $\pand,\por$  or other already defined operations is going to fail. For instance, the natural candidate $(P\pand Q)\por(P\pand R)\por(Q\pand R)$ is very far from being adequate. The latter is a game on six rather than three boards, with $P$ played on boards \#1 and \#3, $Q$ on boards \#2 and \#5, and $R$ on boards \#4 and \#6. Similarly, the formula $(P\pand P)\por(P\pand P)\por(P\pand P)$ is not an adequate representation of the right cirquent of Figure 3. It fails to indicate for instance that the 1st and the 3rd occurrences of $P$ stand for the same copy of $P$ while the 2nd occurrence for a different copy in which a different run can be generated. 

Cirquents are thus properly more expressive than formulas even at the most basic $(\pand,\por)$ level. It is this added expressiveness and flexibility that, for some fragments of CoL, makes a difference between axiomatizability and unaxiomatizability: even if one is only trying to set up a deductive system for proving valid formulas, intermediate steps in proofs of such formulas still inherently require using cirquents that cannot be written as formulas. An example is the system {\bf CL15} found in Section \ref{scl15}.

The present article is exclusively focused on the formula-based version of CoL, seeing cirquents (in Section \ref{scl15}) only as technical servants to formulas. This explains why we do not attempt to define the semantics of cirquents formally.  It should however be noted that cirquents are naturally called for not only within the specific formal framework of CoL, but also in the framework of all resource-sensitive approaches in logic, like linear logic. Such approaches may intrinsically require the ability to account for the  ubiquitous phenomenon of resource sharing. 
The insufficient expressiveness of linear logic is due to the inability of formulas to explicitly show (sub)resource sharing or the absence thereof. The right cirquent  of Figure 3 stands for a multiplicative-style disjunction of three resources, with each disjunct, in turn, being a conjunction of two subresources of type $P$. However, altogether there are three rather than six  such  subresources, each one being shared between two different disjuncts of the main resource.
From the abstract resource-philosophical point of view of cirquent-based CoL, classical logic and linear logic are two imperfect extremes.  In the former, all occurrences of a same subformula mean the same (represent the same resource), i.e., {\em everything is shared  that can be shared}; and in the latter, each occurrence stands for a separate resource, i.e., {\em nothing is shared  at all}. Neither approach does thus permit to account for mixed cases where certain occurrences are meant to represent the same resource while some other occurrences stand for different resources of the same type.

\section{Static games}

While games in the sense of Definition \ref{game} are apparently general enough to model anything one would call an interactive computational problem, they are a little bit too general. Consider a game {\em Bad} whose only nonempty legal runs are $\seq{\top \alpha}$, won by $\top$,  and $\seq{\bot\alpha}$, won by $\bot$. Whichever player is fast enough to make the move $\alpha$ first will  thus be the winner. Since there are no natural, robust assumptions regarding the relative speeds of the players, obviously {\em Bad} is not something that could qualify as a meaningful computational problem. For such reasons, CoL limits its focus on a natural proper subclass of all games   in the sense of Definition \ref{game} called {\em static}. Intuitively, static games are games where speed is irrelevant because, using Blass's words, ``it never hurts a player to postpone making moves''. 

In order to define static games, recall that, for a player $\wp$, $\overline{\wp}$ means ``the other player''. Further recall 
the concepts of a $\wp$-legal and $\wp$-won runs from Section \ref{sgames}.  Given a run $\Gamma$, we let $\Gamma^\top$ denote  the subsequence of (all and only) $\top$-labeled moves of $\Gamma$; similarly for $\Gamma^\bot$. 
We say that a run $\Omega$ is a {\bf $\wp$-delay} of a run $\Gamma$ iff the following two conditions are satisfied:\vspace{-5pt}
\begin{itemize}
  \item $\Omega^\top=\Gamma^\top$ and $\Omega^\bot=\Gamma^\bot$;\vspace{-5pt}
  \item For any $n,k \geq 1$, if the $k$th $\overline{\wp}$-labeled move is  made earlier than the $n$th $\wp$-labeled  move in $\Gamma$, then so is it in $\Omega$.\vspace{-5pt}
\end{itemize} 
The above $\Omega$, in other words, is the result of possibly shifting to the right (``delaying'') some $\wp$-labeled moves in $\Gamma$ without otherwise violating the order of moves by either player.     

\begin{definition} \label{static} We say that a game $G$ is {\bf static} iff, for either player $\wp\in\{\top,\bot\}$ and for any runs $\Gamma,\Omega$ where $\Omega$ is a $\wp$-delay of $\Gamma$, the following conditions are satisfied:\vspace{-5pt}
\begin{enumerate}
  \item If $\Gamma$ is a $\wp$-legal run of $G$, then so is $\Omega$.\vspace{-5pt}
  \item If $\Gamma$ is a $\wp$-won  run of $G$, then so is $\Omega$.\vspace{-5pt}
\end{enumerate}
\noindent A  gameframe is said to be {\bf static} iff so are all of its instances.
\end{definition}

\begin{exercise} Verify that the game of Figure 1 is static.\end{exercise}

 The class of static games or gameframes is very broad. Suffice it to say that all elementary gameframes are static, and that all operations defined in the preceding long section preserve the static property of gameframes. Thus, the closure of elementary gameframes under those operations is one natural subclass of the class of all static games.

\section{The formal language of computability logic and its semantics}\label{s6}

It is not quite accurate to say ``the language'' of CoL because, as pointed out earlier, CoL has an open-ended formalism. Yet, in the present article, by ``the language of CoL'' we will mean the particular language defined below. It extends the language of first-order classical logic by adding to it all operators defined in Section \ref{zoo}, and differentiating between two---elementary and general---sorts of  atoms.

The set {\em Variables} of {\bf variables} and the set {\em Constants} of {\bf constants} of the language are those fixed in Section \ref{frames}.   Per each natural number $n$, we also have infinitely many $n$-ary {\em extralogical} {\bf function letters}, {\bf elementary gameframe letters} and {\bf general gameframe  letters}. We usually use $f,g,h,\cdots$ as metavariables for function letters, $p,q,r,\cdots$ for elementary gameframe letters, and $P,Q,R,\cdots$ for general gameframe letters.  Other than  these extralogical  letters, there are three {\em logical} gameframe letters,  all  elementary:  $\top$ (nullary), $\bot$ (nullary) and $=$ (binary).\vspace{3pt} 

{\bf Terms} are defined inductively as follows:\vspace{-5pt} 
\begin{itemize}
  \item All variables and constants are terms.\vspace{-5pt} 
  \item If $t_1,\cdots,t_n$  are terms ($n\geq 0$) and  $f$  is an $n$-ary function letter, then $f(t_1,\cdots,t_n)$ is a term.\vspace{-2pt}    
\end{itemize}   

{\bf Atoms} are defined by:\vspace{-5pt} 
\begin{itemize}
  \item $\top$ and $\bot$ are 
atoms. These two atoms are said to be {\em logical}, and all other atoms {\em extralogical}.\vspace{-5pt} 
  \item     If $t_1$ and $t_2$ are terms, then $t_1=t_2$ is an atom.\vspace{-5pt} 
  \item If $t_1,\cdots,t_n$ are terms ($n\geq 0$)  and $L$ is an extralogical $n$-ary 
game letter, then $L(t_1,\cdots,t_n)$ is an atom. Such an extralogical atom is said to be {\em elementary} or {\em general} iff $L$ is so.\vspace{-2pt} 
\end{itemize}

Finally, {\bf formulas} are defined by:\vspace{-5pt} 
\begin{itemize}
  \item All atoms are formulas.\vspace{-5pt} 
  \item If $E$ is a formula, then so are  $\neg(E)$,  $\brecurrence (E)$,  $\cobrecurrence (E)$,  $\precurrence (E)$,  $\coprecurrence (E)$, $\srecurrence (E)$,  $\cosrecurrence (E)$,  $\trecurrence (E)$,  $\cotrecurrence (E)$,   $\brepudiation (E)$,  $\prepudiation (E)$,  $\srepudiation (E)$,  $\trepudiation (E)$.\vspace{-5pt} 
  \item If $E$ and $F$ are formulas, then so are $(E)\pand(F)$,  $(E)\por (F)$,  $(E)\chand(F)$,  $(E)\chor (F)$,  $(E)\sand(F)$,  $(E)\sor (F)$, $(E)\tand(F)$,  $(E)\tor (F)$, $(E)\pimplication (F)$, $(E)\chimplication (F)$, $(E)\simplication (F)$, $(E)\timplication (F)$,   $(E)\brimplication (F)$, $(E)\primplication (F)$, $(E)\srimplication (F)$, $(E)\trimplication (F)$.\vspace{-5pt} 
  \item  If $E$ is a formula and $x$ is a variable, then $\blall x(E)$, $\blexists x(E)$,  $\pall x(E)$,  $\pexists x(E)$, $\chall x(E)$,  $\chexists x(E)$, $\sall x(E)$,  $\sexists x(E)$, $\tall x(E)$,  $\texists x(E)$  are formulas.\vspace{-2pt} 
\end{itemize}

Unnecessary parentheses will be usually omitted in formulas according to the standard conventions, with partial precedence order as agreed upon earlier for the corresponding game operations. The notions of {\em free} and {\em bound} occurrences of variables are also standard, with the only adjustment that now we have eight rather than two quantifiers.  A {\bf sentence}, or a {\bf closed formula},   is a formula with no free occurrences of variables. While officially $\pand$ is a binary operator, we may still write $E_1\pand\cdots\pand E_n$ for a possibly unspecified $n\geq 0$. This should be understood as $E_1\pand(E_2\pand\cdots (E_{n-1}\pand E_n)\cdots)$ when $n>2$, as just $E_1$ when $n=1$, and as $\top$ when $n=0$. Similarly for all other sorts of conjunctions. And similarly for all disjunctions, with the difference that an empty disjunction of whatever sort is understood as $\bot$ rather than $\top$.

For the following definitions, recall Conventions \ref{cuxu} and \ref{cuxuu}. Also recall that $var_1,\cdots,var_n$ are the first $n$ variables from the lexicographic list of all variables.

\begin{definition} An {\bf interpretation} is a mapping $^*$ such that, for some fixed universe $U$ called the {\bf universe of $^*$}, we have:\vspace{-2pt} 
\begin{itemize}
  \item     $^*$ sends every $n$-ary function letter $f$ to an $n$-ary function $f^*(var_1,\cdots,var_n)$ whose universe is $U$ and whose variables are the first $n$ variables of \mbox{\em Variables}.\vspace{-5pt} 
  \item 
    $^*$ sends every $n$-ary extralogical game letter $L$ to an $n$-ary static gameframe $L^*(var_1,\cdots,var_n)$  whose universe is $U$ and whose variables are the first $n$ variables of \mbox{\em Variables}; besides, if the letter $L$ is elementary, then so is the gameframe $L^*(var_1,\cdots,var_n)$.\vspace{-2pt}
\end{itemize}
Such a $^*$ is said to be {\bf admissible for} a formula $E$ (or {\bf $E$-admissible}) iff, whenever $E$ has an occurrence of a general atom $P(t_1,\cdots,t_n)$ in the scope of  $\blall x$ or $\blexists x$ and one of the terms $t_i$ ($1\leq i\leq n$) contains the variable $x$, $P^*$ is unistructural in $var_i$. We uniquely extend $^*$ to a mapping that sends each term $t$ to a function $t^*$, and each formula $E$ for which it is admissible to a game $E^*$, by stipulating the following:\vspace{-2pt}
\begin{itemize}
  \item Where $c$ is a constant, $c^*$ is (the nullary function) $c^{\mbox{\tiny{\it U}}}$.\vspace{-5pt}
  \item Where $x$ is a variable, $x^*$ is (the unary function) $x^{\mbox{\tiny{\it U}}}$.\vspace{-5pt}
  \item     Where $f$ is an $n$-ary function letter and $t_1,\cdots,t_n$ are terms, $\bigl(f(t_1,\cdots,t_n)\bigr)^*$  is $f^*(t_{1}^{*},\cdots,t_{n}^{*})$.\vspace{-5pt}
  \item $\top^*$  is $\top$ and $\bot^*$ is $\bot$.\vspace{-5pt}
  \item  Where $t_1$ and $t_2$ are terms, $(t_1=t_2)^*$ is $t_{1}^{*}=t_{2}^{*}$.\vspace{-5pt} 
  \item     Where $L$ is an $n$-ary gameframe letter and $t_1,\cdots,t_n$ are terms, $\bigl(L(t_1,\cdots,t_n)\bigr)^*$ is $L^*(t_{1}^{*},\cdots,t_{n}^{*})$.\vspace{-5pt}
  \item  $^*$ commutes with all logical operators, seeing them as the corresponding game operations: $(\neg E)^*$  is $\neg(E^*)$, $(\brecurrence E)^*$ is $\brecurrence (E^*)$,  $(E\pand F)^*$ is $(E^*)\pand (F^*)$, $(\chall xE)^*$ is $\chall x(E^*)$, etc.\vspace{-2pt}
\end{itemize}
When $O$ is a function letter, gameframe letter, term or formula and $O^*=W$, we 
 refer to $W$ as ``$O$ {\bf under} interpretation $^*$''.
\end{definition}

\begin{definition}
 For a sentence $S$ we say that:

1.    $S$ is {\bf logically} {\bf valid} iff there is an HPM $\cal M$ such that, for every $S$-admissible interpretation $^*$, $\cal M$ computes $S^*$. Such an $\cal M$ is said to be a {\bf logical}  {\bf solution} of $S$.

2.    $S$ is {\bf extralogically}  {\bf valid} iff for every $S$-admissible interpretation $^*$ there is an HPM $\cal M$ such that $\cal M$ computes $S^*$.
\end{definition}

\begin{convention} When $S$ is a formula but not a sentence, its validity is understood as that of  the $\chall$-closure of $S$, i.e., of the sentence  $\chall x_1\cdots\chall x_n S$, where  $x_1,\cdots,  x_n$  are all free variables of $S$ listed lexicographically.
\end{convention}

Every logically valid formula is, of course, also extralogically valid. But some extralogically valid formulas may not  necessarily be also logically valid. For instance, where $p$ is a $0$-ary elementary gameframe letter, the formula $\neg p\chor  p$ is valid extralogically but not logically. It is extralogically valid for a trivial reason: given an interpretation $^*$, either $\neg p$ or $p$ is true under $^*$. If $\neg p$ is true, then the strategy that chooses the first disjunct wins; and if $p$ is true, then the strategy that chooses the second disjunct wins. The trouble is that, even though we know that one of these two strategies succeeds, generally
we have no way to tell which one does. And this is why $\neg p\chor  p$ fails to be logically valid.

Extralogical validity  is not only a non-constructive, but also  a fragile sort of validity: this property, unlike logical validity, is not closed under substitution of extralogical atoms. For instance, where $p$ is as before and $q$ is a unary extralogical elementary gameframe letter, the formula $ \neg q(x)\chor q(x)$, while having the same form as $\neg p\chor  p$, is no longer extralogically valid. The papers on CoL written prior to 2016 had a more relaxed understanding of interpretations than our present understanding. Namely, there was no requirement that an interpretation should respect the arity of a gameframe letter. In such a case, as it turns out, the extensional difference between logical and extralogical validity disappears: while the class of logically valid principles remains the same, the class of extralogically valid principles shrinks down to that of logically valid ones. 

Intuitively, a logical solution $\cal M$ for a sentence $S$ is an interpretation-independent winning strategy: since the  intended  interpretation   is not known to the machine,  $\cal M$  has to play in some standard, uniform way that would be successful for any possible interpretation of $S$. It is logical rather than extralogical validity that is of interest in all applied systems based on CoL (cf. Section \ref{sappl}).  In such applications we want a logic that could be built into a universal problem-solving machine.  Such a machine  should be able to solve problems represented by  logical formulas without any specific knowledge of the meanings  of their atoms, other than the knowledge explicitly provided in the knowledgebase (extralogical axioms) of the system. Otherwise the machine would be  special-purpose rather than universal.  
For such reasons, in the subsequent sections we will only be focused on the logical sort of validity, which will be the default meaning of the word ``valid''.   

\begin{definition}\label{LC}
We say that a  sentence $F$  is a {\bf logical consequence} of a set $\mathbb{B}$ of sentences iff, for some $E_1,\cdots,E_n\in\mathbb{B}$, 
the sentence  $E_1\pand\cdots\pand E_n\brimplication F$ is logically valid.  
\end{definition}

Remember the symmetry between computational resources and computational problems: a problem for one player is a resource for the other.
Having a problem $A$ as a computational resource intuitively means having the (perhaps externally provided) ability to successfully solve/win $A$. For instance, as a resource, $\chall x\chexists y(y=x^2)$ means the ability to tell the square of any number.   According to Thesis \ref{scopy} below,      
the relation of logical consequence lives up to its name. The main utility of this thesis, as will be illustrated in Section \ref{scl12}, is that it allows us to rely on informal, intuitive arguments instead of formal proofs when reasoning within CoL-based applied theories.

\begin{thesis}\label{scopy}
Consider sentences $E_1,\cdots,E_n,F$ ($n\geq 0$) and an admissible interpretation $^*$ for them. Assume 
there is a    winning strategy for $F^*$ that relies on  availability and ``recyclability''---in the strongest sense possible---of $E_{1}^{*},\cdots,E_{n}^{*}$ as computational resources but no other knowledge or assumptions about $^*$ (see  Example \ref{stra} for an instance of such a strategy).   Then $F$ is a logical consequence of $E_1,\cdots,E_n$.
\end{thesis}

 \section{Axiomatizations} 
 While a semantical setup for CoL in the  language of Section \ref{s6} is complete, a corresponding proof theory is still at earlier stages of development. Due to the inordinate expressive power of the language, successful axiomatization attempts have only been made for various fragments of CoL obtained by moderating its language in one way or another. It should be pointed out that every conceivable application of CoL will only need some fragment of CoL rather than the ``whole'' CoL anyway.

As of 2020 there are seventeen  deductive systems for various fragments of CoL, named {\bf CL1} through {\bf CL17}. Based on their languages, these systems can be divided into three groups: {\bf elementary-base}, {\bf general-base} and {\bf mixed-base}. Of the extralogical gameframe letters, the languages of elementary-base systems only allow elementary ones, the languages of  general-base systems only allow  general ones, and the languages of  mixed-base system allow  both sorts of letters. Based on the style of the underlying proof theory, the systems can be further subdivided into two groups: {\bf cirquent calculus systems} and {\bf brute force systems}.  Either sort is rather unusual, not seen elsewhere in proof 
theory. The cirquent calculus systems operate with cirquents rather than formulas, with formulas understood as special cases of cirquents. The brute force systems operate with formulas (sometimes referred to as sequents for technical reasons), but in an unusual way, with their inference rules being relatively directly derived from the underlying game semantics and hence somewhat resembling games themselves.  

\begin{theorem} \label{adeq}
Each of the above-mentioned systems ${\bf S}\in\{{\bf CL1}, \cdots, {\bf CL17}\}$ is {\bf adequate} in the   sense that, for any sentence $F$ of the language of {\bf S},  we have:
\begin{description}
  \item[(a) Soundness:] If $F$ is provable in ${\bf S}$, then it is logically valid and, furthermore, a logical solution for $F$ can be automatically extracted from a proof of $F$.
  \item[(b) Completeness:]   If $F$ is logically valid, then it is provable in {\bf S}.
\end{description}
\end{theorem}

In this article we shall take a look at only three of the systems: the general-base cirquent calculus system {\bf CL15} in the logical signature $\{\neg,\pand,\por,\brecurrence, \cobrecurrence\}$, the mixed-base brute force system {\bf CL13} in the signature $\{\neg,\pand,\por,\chand,\chor,\sand,\sor,\tand,\tor\}$, and the elementary-base brute force system {\bf CL12} in the signature 
$\{\neg,\pand,\por,\chand,\chor,\blall,\blexists,\chall,\chexists,\brimplication\}$ (with  $\brimplication$ only allowed to be applied externally).
Their adequacy proofs can respectively be found in \cite{taming1,taming}, \cite{Japtoggling} and \cite{cl12}.

\subsection{The cirquent calculus system {\bf CL15}}\label{scl15}
{\bf CL15-formulas}---or just {\em formulas} in this subsection---are formulas of the language of CoL that do not contain any function letters, do not contain any gameframe letters  other than 
$0$-ary general gameframe letters, and do not contain any operators other than $\neg,\pand,\por,\brecurrence,\cobrecurrence$.  
 Besides, $\neg$ is only allowed to be applied to  atoms. Shall we still write $\neg E$ for a nonatomic $E$,  it is to be understood as the standard {\em DeMorgan abbreviation} defined by $\neg\neg F=F$, $\neg(F\pand G)=\neg F\por\neg G$, $\neg(F\por G)=\neg F\pand\neg G$, 
$\neg\brecurrence F=\cobrecurrence \neg F$, $\neg\cobrecurrence F=\brecurrence \neg F$. Similarly,   $F\pimplication G$, $F\brimplication G$ and $\brepudiation F$ should be understood as  $\neg F\por G$, $\cobrecurrence \neg F\por G$ and $\cobrecurrence\neg F$, respectively.

\begin{definition}\label{cirqdef}
 A {\bf CL15-cirquent} (henceforth simply ``cirquent'') is a triple $C=(\vec{F},\vec{U},\vec{O})$ where: 

1. $\vec{F}$ is a nonempty finite sequence of {\bf CL15}-formulas, whose elements are said to be the {\bf oformulas} of $C$.  Here the prefix ``o''  is for ``occurrence'', and is used to mean a formula together with a particular occurrence of it in $\vec{F}$. So, for instance, if $\vec{F}=\seq{E,G,E}$, then the cirquent has three oformulas even if  only two formulas. 

2.  Both $\vec{U}$ and $\vec{O}$ are nonempty finite sequences of nonempty sets of oformulas of $C$. The elements of $\vec{U}$ are said to be the {\bf undergroups} of $C$, and the elements of $\vec{O}$ are said to be the {\bf overgroups} of $C$. As in the case of oformulas, it is possible that two undergroups or two overgroups are identical as sets (have identical {\bf contents}), yet they count as different undergroups or overgroups because they occur at different places in the sequence $\vec{U}$ or $\vec{O}$. Simply ``group'' will be used as a common name for undergroups and overgroups. 

3. Additionally, every oformula is required to be in at least one undergroup and  at least one overgroup. 
\end{definition}

While oformulas are not the same as formulas, we may often identify an oformula with the corresponding formula and, for instance, say ``the oformula $E$'' if it is clear from the context which of the possibly many occurrences of $E$ is meant. Similarly, we may not always be very careful about differentiating between groups and their contents.

We represent cirquents using three-level diagrams such as the one shown below:

 
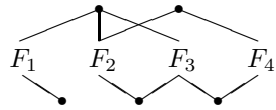
\begin{center} \begin{picture}(98,35)
\put(34,35){\circle*{3}}
\put(64,35){\circle*{3}}
\put(34,35){\line(0,-1){12}}
\put(34,35){\line(5,-2){30}}
\put(34,35){\line(-5,-2){30}}
\put(64,35){\line(5,-2){30}}
\put(64,35){\line(-5,-2){30}}
\put(0,13){$F_1$}
\put(30,13){$F_2$}
\put(60,13){$F_3$}
\put(90,13){$F_4$}

\put(20,0){\line(-3,2){15}}
\put(49,0){\line(-3,2){15}}
\put(49,0){\line(3,2){15}}
\put(79,0){\line(-3,2){15}}
\put(79,0){\line(3,2){15}}
\put(20,0){\circle*{3}}
\put(49,0){\circle*{3}}
\put(79,0){\circle*{3}}
\end{picture}
\end{center}

\noindent This diagram represents the cirquent with four oformulas $F_1, F_2, F_3, F_4$, three undergroups $\{F_1\},$ $ \{F_2,F_3\},$ $ \{F_3,F_4\}$ and two overgroups $\{F_1,F_2,F_3\}$, $\{F_2,F_4\}$.  
Each (under- or over-) group is  represented by a \raisebox{0.2mm}{ \mbox{\scriptsize $\bullet$}}, where the {\bf arcs} (lines connecting the \raisebox{0.2mm}{ \mbox{\scriptsize $\bullet$}}'s with oformulas) are pointing to the oformulas that the given group contains.

{\bf CL15} has ten rules of inference. The first one takes no premises, which qualifies it as an axiom. All other rules take a single premise. Below we explain them in a relaxed fashion, in terms of deleting arcs, swapping oformulas, etc. Such explanations are rather clear, and translating them into rigorous formulations in the style and terms of Definition \ref{cirqdef}, while possible, is hardly necessary.\vspace{5pt}

{\bf Axiom (A):} The conclusion of this premiseless rule looks like an array of  $n$ ($n\geq 1$)  ``diamonds'' as seen below for the case of $n=3$,  where the oformulas within each diamond are $\neg F$ and $F$ for some formula $F$. 

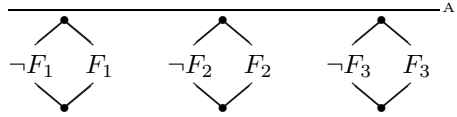
\begin{center} \begin{picture}(156,40)

\put(164,36){\tiny A}
\put(0,37){\line(1,0){163}}

\put(0,13){$\neg F_1$}
\put(29,13){$F_1$}
\put(21,0){\line(-6,5){11}}
\put(21,0){\line(6,5){11}}
\put(21,0){\circle*{3}}
\put(21,33){\line(-6,-5){11}}
\put(21,33){\line(6,-5){11}}
\put(21,33){\circle*{3}}

\put(60,13){$\neg F_2$}
\put(89,13){$F_2$}
\put(81,0){\line(-6,5){11}}
\put(81,0){\line(6,5){11}}
\put(81,0){\circle*{3}}
\put(81,33){\line(-6,-5){11}}
\put(81,33){\line(6,-5){11}}
\put(81,33){\circle*{3}}

\put(120,13){$\neg F_3$}
\put(149,13){$F_3$}
\put(141,0){\line(-6,5){11}}
\put(141,0){\line(6,5){11}}
\put(141,0){\circle*{3}}
\put(141,33){\line(-6,-5){11}}
\put(141,33){\line(6,-5){11}}
\put(141,33){\circle*{3}}

\end{picture}
\end{center}

{\bf Exchange (E):} This rule comes in three flavors: Undergroup Exchange, Oformula Exchange and Overgroup Exchange. Each one allows us to swap any two adjacent objects (undergroups, oformulas or overgroups) of a cirquent, otherwise preserving all oformulas, groups and arcs. Below we see three examples, one per each sort of Exchange.   In all cases, of course,  the upper cirquent is the premise and the lower cirquent is the conclusion of an application of the rule.  

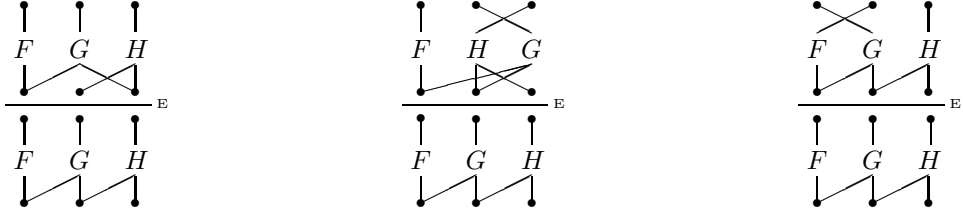
\begin{center} \begin{picture}(373,76)

\put(16,75){\circle*{3}}
\put(16,75){\line(0,-1){10}}
\put(37,75){\circle*{3}}
\put(37,75){\line(0,-1){10}}
\put(58,75){\circle*{3}}
\put(58,75){\line(0,-1){10}}

\put(12,55){$F$}
\put(33,55){$G$}
\put(54,55){$H$}

\put(16,42){\line(0,1){10}}
\put(16,42){\line(2,1){21}}
\put(16,42){\circle*{3}}
\put(37,42){\circle*{3}}
\put(58,42){\circle*{3}}
\put(37,42){\line(2,1){21}}
\put(58,42){\line(-2,1){21}}
\put(58,42){\line(0,1){10}}

\put(66,36){\tiny E}
\put(9,37){\line(1,0){55}}

\put(16,32){\circle*{3}}
\put(16,32){\line(0,-1){10}}
\put(37,32){\circle*{3}}
\put(37,32){\line(0,-1){10}}
\put(58,32){\circle*{3}}
\put(58,32){\line(0,-1){10}}

\put(12,13){$F$}
\put(33,13){$G$}
\put(54,13){$H$}

\put(16,0){\line(0,1){10}}
\put(16,0){\line(2,1){21}}
\put(16,0){\circle*{3}}
\put(37,0){\circle*{3}}
\put(58,0){\circle*{3}}
\put(37,0){\line(0,1){10}}
\put(37,0){\line(2,1){21}}
\put(58,0){\line(0,1){10}}


\put(216,36){\tiny E}
\put(159,37){\line(1,0){55}}

\put(166,75){\circle*{3}}
\put(166,75){\line(0,-1){10}}
\put(187,75){\circle*{3}}
\put(187,75){\line(2,-1){21}}
\put(208,75){\circle*{3}}
\put(208,75){\line(-2,-1){21}}

\put(162,55){$F$}
\put(183,55){$H$}
\put(204,55){$G$}

\put(166,42){\line(0,1){10}}
\put(166,42){\line(4,1){42}}
\put(166,42){\circle*{3}}
\put(187,42){\circle*{3}}
\put(208,42){\circle*{3}}
\put(187,42){\line(0,1){10}}
\put(187,42){\line(2,1){21}}
\put(208,42){\line(-2,1){21}}

\put(166,32){\circle*{3}}
\put(166,32){\line(0,-1){10}}
\put(187,32){\circle*{3}}
\put(187,32){\line(0,-1){10}}
\put(208,32){\circle*{3}}
\put(208,32){\line(0,-1){10}}

\put(162,13){$F$}
\put(183,13){$G$}
\put(204,13){$H$}

\put(166,0){\line(0,1){10}}
\put(166,0){\line(2,1){21}}
\put(166,0){\circle*{3}}
\put(187,0){\circle*{3}}
\put(208,0){\circle*{3}}
\put(187,0){\line(0,1){10}}
\put(187,0){\line(2,1){21}}
\put(208,0){\line(0,1){10}}


\put(366,36){\tiny E}
\put(309,37){\line(1,0){55}}

\put(316,75){\circle*{3}}
\put(316,75){\line(2,-1){21}}
\put(337,75){\circle*{3}}
\put(337,75){\line(-2,-1){21}}
\put(358,75){\circle*{3}}
\put(358,75){\line(0,-1){10}}

\put(312,55){$F$}
\put(333,55){$G$}
\put(354,55){$H$}

\put(316,42){\line(0,1){10}}
\put(316,42){\line(2,1){21}}
\put(316,42){\circle*{3}}
\put(337,42){\circle*{3}}
\put(358,42){\circle*{3}}
\put(337,42){\line(0,1){10}}
\put(337,42){\line(2,1){21}}
\put(358,42){\line(0,1){10}}

\put(316,32){\circle*{3}}
\put(316,32){\line(0,-1){10}}
\put(337,32){\circle*{3}}
\put(337,32){\line(0,-1){10}}
\put(359,32){\circle*{3}}
\put(359,32){\line(0,-1){10}}

\put(312,13){$F$}
\put(333,13){$G$}
\put(354,13){$H$}

\put(316,0){\line(0,1){10}}
\put(316,0){\line(2,1){21}}
\put(316,0){\circle*{3}}
\put(337,0){\circle*{3}}
\put(358,0){\circle*{3}}
\put(337,0){\line(0,1){10}}
\put(337,0){\line(2,1){21}}
\put(358,0){\line(0,1){10}}

\end{picture}
\end{center} 
The presence of Exchange essentially allows us to treat all three components $(\vec{F},\vec{U},\vec{O})$  of a cirquent as multisets rather than  sequences.\vspace{5pt}  

{\bf Weakening (W):} The premise of this rule is obtained from the conclusion by deleting an arc between some undergroup $U$ with $\geq 2$ elements and some oformula $F$; if $U$ was the only undergroup containing $F$, then $F$ should also be deleted (to satisfy condition 3 of Definition \ref{cirqdef}), together with all arcs between $F$ and overgroups; if such a deletion makes some overgroups empty, then they should also be deleted (to satisfy condition 2 of Definition  \ref{cirqdef}).   Below are three examples:

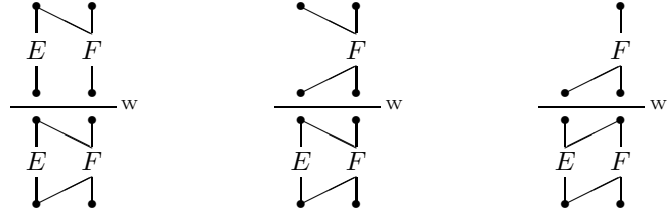
\begin{center} \begin{picture}(295,76)
\put(24,75){\line(0,-1){10}}
\put(24,75){\line(2,-1){21}}
\put(24,75){\circle*{3}}
\put(45,75){\circle*{3}}
\put(45,75){\line(0,-1){10}}

\put(20,55){$E$}
\put(41,55){$F$}
\put(24,42){\line(0,1){10}}
\put(24,42){\circle*{3}}
\put(45,42){\circle*{3}}
\put(45,42){\line(0,1){10}}

\put(56,36){\tiny W}
\put(14,37){\line(1,0){40}}

\put(24,32){\line(0,-1){10}}
\put(24,32){\line(2,-1){21}}
\put(24,32){\circle*{3}}
\put(45,32){\circle*{3}}
\put(45,32){\line(0,-1){10}}

\put(20,13){$E$}
\put(41,13){$F$}
\put(24,0){\line(0,1){10}}
\put(24,0){\line(2,1){21}}
\put(24,0){\circle*{3}}
\put(45,0){\circle*{3}}
\put(45,0){\line(0,1){10}}


\put(124,75){\line(2,-1){21}}
\put(124,75){\circle*{3}}
\put(145,75){\circle*{3}}
\put(145,75){\line(0,-1){10}}

\put(141,55){$F$}
\put(124,42){\line(2,1){21}}
\put(124,42){\circle*{3}}
\put(145,42){\circle*{3}}
\put(145,42){\line(0,1){10}}

\put(156,36){\tiny W}
\put(114,37){\line(1,0){40}}

\put(124,32){\line(0,-1){10}}
\put(124,32){\line(2,-1){21}}
\put(124,32){\circle*{3}}
\put(145,32){\circle*{3}}
\put(145,32){\line(0,-1){10}}

\put(120,13){$E$}
\put(141,13){$F$}
\put(124,0){\line(0,1){10}}
\put(124,0){\line(2,1){21}}
\put(124,0){\circle*{3}}
\put(145,0){\circle*{3}}
\put(145,0){\line(0,1){10}}


\put(245,75){\circle*{3}}
\put(245,75){\line(0,-1){10}}

\put(241,55){$F$}
\put(224,42){\line(2,1){21}}
\put(224,42){\circle*{3}}
\put(245,42){\circle*{3}}
\put(245,42){\line(0,1){10}}

\put(256,36){\tiny W}
\put(214,37){\line(1,0){40}}
 
\put(224,32){\line(0,-1){10}}
\put(245,32){\line(-2,-1){21}}
\put(224,32){\circle*{3}}
\put(245,32){\circle*{3}}
\put(245,32){\line(0,-1){10}}

\put(220,13){$E$}
\put(241,13){$F$}
\put(224,0){\line(0,1){10}}
\put(224,0){\line(2,1){21}}
\put(224,0){\circle*{3}}
\put(245,0){\circle*{3}}
\put(245,0){\line(0,1){10}}

\end{picture}
\end{center}

{\bf Contraction (C):}  The premise of this rule is obtained from the conclusion through replacing an oformula $\cobrecurrence F$ by two adjacent oformulas $\cobrecurrence F,\cobrecurrence F$, and including them in exactly the same undergroups and overgroups in which the original oformula was contained. Example:

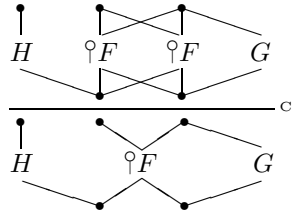
\begin{center} \begin{picture}(106,76)

\put(4,75){\circle*{3}}
\put(4,75){\line(0,-1){10}}
\put(34,75){\circle*{3}}
\put(34,75){\line(0,-1){10}}
\put(34,75){\line(3,-1){30}}
\put(65,75){\circle*{3}}
\put(65,75){\line(0,-1){10}}
\put(65,75){\line(3,-1){30}}
\put(65,75){\line(-3,-1){30}}
\put(0,55){$H$}
\put(28,55){$\cobrecurrence  F$}
\put(59,55){$\cobrecurrence  F$}
\put(91,55){$G$}
\put(34,42){\circle*{3}}
\put(65,42){\circle*{3}}
\put(34,42){\line(-3,1){30}}
\put(34,42){\line(3,1){30}}
\put(65,42){\line(3,1){30}}
\put(65,42){\line(-3,1){30}}
\put(34,42){\line(0,1){10}}
\put(65,42){\line(0,1){10}}

\put(102,36){\tiny C}
\put(0,37){\line(1,0){100}}

\put(4,32){\circle*{3}}
\put(4,32){\line(0,-1){10}}
\put(34,32){\circle*{3}}
\put(34,32){\line(3,-2){16}}
\put(66,32){\circle*{3}}
\put(66,32){\line(3,-1){29}}
\put(66,32){\line(-3,-2){16}}
\put(0,13){$H$}
\put(43,13){$\cobrecurrence  F$}
\put(92,13){$G$}
\put(34,0){\circle*{3}}
\put(66,0){\circle*{3}}
\put(34,0){\line(-3,1){29}}
\put(66,0){\line(3,1){29}}
\put(34,0){\line(3,2){16}}
\put(66,0){\line(-3,2){16}}

\end{picture}
\end{center}

{\bf Duplication (D):} This rule comes in two versions: Undergroup Duplication and Overgroup Duplication. The conclusion of Undergroup Duplication is the result of replacing, in the premise, some undergroup $U$ with two adjacent undergroups whose contents are identical to that of $U$. Similarly for Overgroup Duplication. Examples:

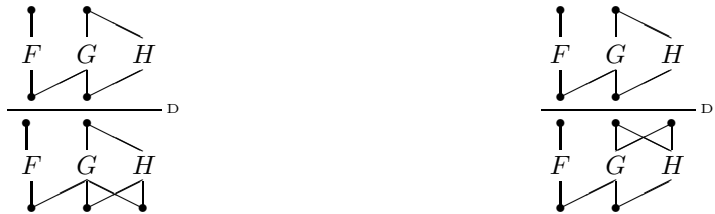
\begin{center} \begin{picture}(283,76)

\put(26,75){\circle*{3}}
\put(26,75){\line(0,-1){10}}
\put(47,75){\circle*{3}}
\put(47,75){\line(0,-1){10}}
\put(47,75){\line(2,-1){21}}

\put(22,55){$F$}
\put(43,55){$G$}
\put(64,55){$H$}

\put(26,42){\line(0,1){10}}
\put(26,42){\line(2,1){21}}
\put(26,42){\circle*{3}}
\put(47,42){\circle*{3}}
\put(47,42){\line(2,1){21}}
\put(47,42){\line(0,1){10}}

\put(77,36){\tiny D}
\put(17,37){\line(1,0){58}}

\put(24,32){\circle*{3}}
\put(24,32){\line(0,-1){10}}
\put(47,32){\circle*{3}}
\put(47,32){\line(0,-1){10}}
\put(47,32){\line(2,-1){21}}

\put(22,13){$F$}
\put(43,13){$G$}
\put(64,13){$H$}

\put(26, 0){\line(0,1){10}}
\put(26, 0){\line(2,1){21}}
\put(26, 0){\circle*{3}}
\put(47, 0){\circle*{3}}
\put(68, 0){\circle*{3}}
\put(47, 0){\line(0,1){10}}
\put(47, 0){\line(2,1){21}}
\put(68, 0){\line(0,1){10}}
\put(68, 0){\line(-2,1){21}}

\put(226,75){\circle*{3}}
\put(226,75){\line(0,-1){10}}
\put(247,75){\circle*{3}}
\put(247,75){\line(0,-1){10}}
\put(247,75){\line(2,-1){21}}

\put(222,55){$F$}
\put(243,55){$G$}
\put(264,55){$H$}

\put(226,42){\line(0,1){10}}
\put(226,42){\line(2,1){21}}
\put(226,42){\circle*{3}}
\put(247,42){\circle*{3}}
\put(247,42){\line(2,1){21}}
\put(247,42){\line(0,1){10}}

\put(279,36){\tiny D}
\put(219,37){\line(1,0){58}}

\put(226,32){\circle*{3}}
\put(226,32){\line(0,-1){10}}
\put(247,32){\circle*{3}}
\put(247,32){\line(0,-1){10}}
\put(247,32){\line(2,-1){21}}
\put(268,32){\circle*{3}}
\put(268,32){\line(-2,-1){21}}
\put(268,32){\line(0,-1){10}}

\put(222,13){$F$}
\put(243,13){$G$}
\put(264,13){$H$}

\put(226,0){\line(0,1){10}}
\put(226,0){\line(2,1){21}}
\put(226,0){\circle*{3}}
\put(247,0){\circle*{3}}
\put(247,0){\line(0,1){10}}
\put(247,0){\line(2,1){21}}

\end{picture}
\end{center}

{\bf Merging (M):} In the top-down view, this rule merges any two adjacent overgroups, as illustrated below. 

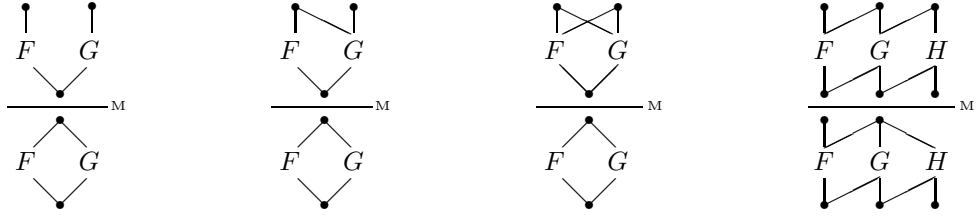
\begin{center} \begin{picture}(370,76)

\put(14,75){\circle*{3}}
\put(14,75){\line(0,-1){10}}
\put(39,75){\circle*{3}}
\put(39,75){\line(0,-1){10}}

\put(10,55){$F$}
\put(34,55){$G$}

\put(46,36){\tiny M}
\put(7,37){\line(1,0){38}}

\put(27,32){\circle*{3}}
\put(27,32){\line(-1,-1){10}}
\put(27,32){\line(1,-1){10}}

\put(10,13){$F$}
\put(34,13){$G$}

\put(27,0){\line(-1,1){10}}
\put(27,0){\line(1,1){10}}
\put(27,0){\circle*{3}}

\put(116,75){\circle*{3}}
\put(116,75){\line(0,-1){10}}
\put(116,75){\line(2,-1){21}}
\put(138,75){\circle*{3}}
\put(138,75){\line(0,-1){10}}

\put(110,55){$F$}
\put(134,55){$G$}

\put(127,42){\circle*{3}}
\put(127,42){\line(-1,1){10}}
\put(127,42){\line(1,1){10}}
\put(27,42){\circle*{3}}
\put(27,42){\line(-1,1){10}}
\put(27,42){\line(1,1){10}}

\put(146,36){\tiny M}
\put(107,37){\line(1,0){38}}

\put(127,32){\circle*{3}}
\put(127,32){\line(-1,-1){10}}
\put(127,32){\line(1,-1){10}}

\put(110,13){$F$}
\put(134,13){$G$}

\put(127,0){\line(-1,1){10}}
\put(127,0){\line(1,1){10}}
\put(127,0){\circle*{3}}

\put(215,75){\circle*{3}}
\put(215,75){\line(0,-1){10}}
\put(215,75){\line(2,-1){21}}
\put(238,75){\line(-2,-1){21}}
\put(238,75){\circle*{3}}
\put(238,75){\line(0,-1){10}}

\put(210,55){$F$}
\put(234,55){$G$}

\put(227,42){\circle*{3}}
\put(227,42){\line(-1,1){11}}
\put(227,42){\line(1,1){11}}

\put(249,36){\tiny M}
\put(207,37){\line(1,0){40}}

\put(227,32){\circle*{3}}
\put(227,32){\line(-1,-1){10}}
\put(227,32){\line(1,-1){10}}

\put(210,13){$F$}
\put(234,13){$G$}

\put(227,0){\line(-1,1){10}}
\put(227,0){\line(1,1){10}}
\put(227,0){\circle*{3}}

\put(316,75){\circle*{3}}
\put(316,75){\line(0,-1){10}}

\put(337,75){\circle*{3}}
\put(337,75){\line(0,-1){10}}
\put(337,75){\line(-2,-1){21}}
\put(358,75){\circle*{3}}
\put(358,75){\line(0,-1){10}}
\put(358,75){\line(-2,-1){21}}

\put(312,55){$F$}
\put(333,55){$G$}
\put(354,55){$H$}

\put(316,42 ){\line(0,1){10}}
\put(316,42){\line(2,1){21}}
\put(316,42){\circle*{3}}
\put(337,42){\circle*{3}}
\put(358,42){\circle*{3}}
\put(337,42){\line(0,1){10}}
\put(337,42){\line(2,1){21}}
\put(358,42){\line(0,1){10}}

\put(367,36){\tiny M}
\put(310,37){\line(1,0){55}}

\put(316,32){\circle*{3}}
\put(316,32){\line(0,-1){10}}
\put(337,32){\circle*{3}}
\put(337,32){\line(-2,-1){21}}
\put(337,32){\line(0,-1){10}}
\put(337,32){\line(2,-1){21}}

\put(312,13){$F$}
\put(333,13){$G$}
\put(354,13){$H$}

\put(316,0){\line(0,1){10}}
\put(316,0){\line(2,1){21}}
\put(316,0){\circle*{3}}
\put(337,0){\circle*{3}}
\put(358,0){\circle*{3}}
\put(337,0){\line(0,1){10}}
\put(337,0){\line(2,1){21}}
\put(358,0){\line(0,1){10}}

\end{picture}
\end{center}

{\bf Disjunction Introduction ($\por$):} The premise of this rule is obtained from the conclusion through replacing an oformula $F\por G$ by two adjacent oformulas $F,G$, and including both of them in exactly the same undergroups and overgroups in which the original oformula was contained, as illustrated below:

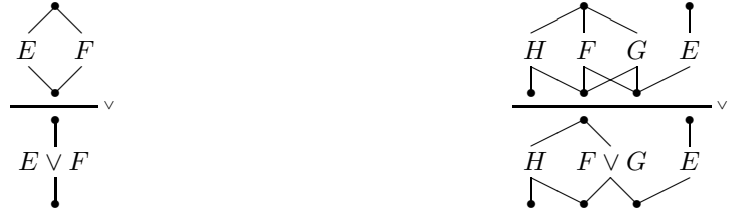
\begin{center} \begin{picture}(275,76)

\put(23,75){\circle*{3}}
\put(23,75){\line(1,-1){10}}
\put(23,75){\line(-1,-1){10}}

\put(8,55){$E$}
\put(30,55){$F$}
\put(23,42){\circle*{3}}
\put(23,42){\line(1,1){10}}
\put(23,42){\line(-1,1){10}}

\put(41,36){\tiny $\por$}
\put(6,37){\line(1,0){33}}

\put(23,32){\circle*{3}}
\put(23,32){\line(0,-1){10}}

\put(9,13){$E\por F$}

\put(23,0){\circle*{3}}

\put(23,0){\line(0,1){10}}

\put(223,75){\circle*{3}}
\put(223,75){\line(2,-1){20}}
\put(223,75){\line(-2,-1){20}}
\put(223,75){\line(0,-1){10}}
\put(263,75){\circle*{3}}
\put(263,75){\line(0,-1){10}}

\put(200,55){$H$}
\put(220,55){$F$}
\put(239,55){$G$}
\put(259,55){$E$}
\put(203,42){\circle*{3}}
\put(223,42){\circle*{3}}
\put(243,42){\circle*{3}}
\put(203,42){\line(0,1){10}}
\put(223,42){\line(2,1){20}}
\put(223,42){\line(0,1){10}}
\put(223,42){\line(-2,1){20}}
\put(243,42){\line(-2,1){20}}
\put(243,42){\line(2,1){20}}
\put(243,42){\line(0,1){10}}

\put(273,36){\tiny $\por$}
\put(196,37){\line(1,0){75}}

\put(223,32){\circle*{3}}
\put(223,32){\line(-2,-1){20}}
\put(223,32){\line(1,-1){10}}
\put(263,32){\circle*{3}}
\put(263,32){\line(0,-1){10}}

\put(200,13){$H$}
\put(220,13){$F\por G$}
\put(259,13){$E$}
\put(203,0){\circle*{3}}
\put(223,0){\circle*{3}}
\put(243,0){\circle*{3}}
\put(203,0){\line(0,1){10}}
\put(223,0){\line(1,1){10}}
\put(223,0){\line(-2,1){20}}
\put(243,0){\line(-1,1){10}}
\put(243,0){\line(2,1){20}}

\end{picture}
\end{center}

{\bf Conjunction Introduction ($\pand$):} The premise of this rule is obtained from the conclusion by picking an arbitrary oformula $F\pand G$ and applying the following two steps:
\begin{itemize}
\item Replace  $F\pand G$ by two adjacent oformulas $F,G$, and include both of them in exactly the same undergroups and overgroups in which the original oformula was contained.
\item Replace each undergroup $U$ originally containing  $F\pand G$ (and now containing $F,G$ instead) by the two adjacent undergroups $U-\{G\}$ and $U-\{F\}$. 
\end{itemize}
Below we see three examples.

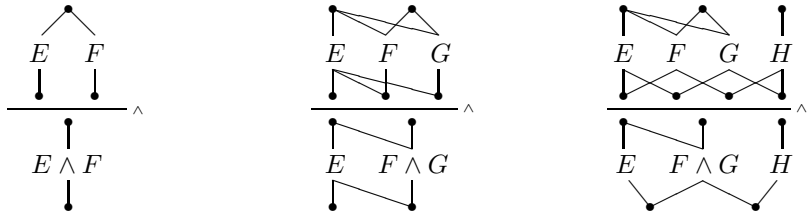
\begin{center} \begin{picture}(287,76)

\put(23,75){\circle*{3}}
\put(23,75){\line(1,-1){10}}
\put(23,75){\line(-1,-1){10}}

\put(8,55){$E$}
\put(29,55){$F$}
\put(12,42){\circle*{3}}
\put(12,42){\line(0,1){10}}
\put(33,42){\circle*{3}}
\put(33,42){\line(0,1){10}}

\put(47,35){\tiny $\pand$}
\put(0,37){\line(1,0){45}}

\put(23,32){\circle*{3}}
\put(23,32){\line(0,-1){10}}

\put(9,13){$E\pand F$}
\put(23,0){\circle*{3}}
\put(23,0){\line(0,1){10}}


\put(153,75){\circle*{3}}
\put(153,75){\line(1,-1){10}}
\put(153,75){\line(-1,-1){10}}
\put(123,75){\circle*{3}}
\put(123,75){\line(0,-1){10}}
\put(123,75){\line(2,-1){20}}
\put(123,75){\line(4,-1){40}}

\put(120,55){$E$}
\put(140,55){$F$}
\put(160,55){$G$}
\put(123,42){\circle*{3}}
\put(123,42){\line(0,1){10}}
\put(143,42){\circle*{3}}
\put(163,42){\circle*{3}}
\put(163,42){\line(0,1){10}}
\put(143,42){\line(0,1){10}}
\put(143,42){\line(-2,1){20}}
\put(163,42){\line(-4,1){40}}

\put(172,36){\tiny $\pand$}
\put(115,37){\line(1,0){55}}

\put(153,32){\circle*{3}}
\put(153,32){\line(0,-1){10}}
\put(123,32){\circle*{3}}
\put(123,32){\line(0,-1){10}}
\put(123,32){\line(3,-1){30}}

\put(120,13){$E$}
\put(140,13){$F\pand G$}
\put(123,0){\circle*{3}}
\put(123,0){\line(0,1){10}}
\put(153,0){\circle*{3}}
\put(153,0){\line(0,1){10}}
\put(153,0){\line(-3,1){30}}


\put(263,75){\circle*{3}}
\put(263,75){\line(1,-1){10}}
\put(263,75){\line(-1,-1){10}}

\put(233,75){\circle*{3}}
\put(233,75){\line(0,-1){10}}
\put(233,75){\line(4,-1){40}}
\put(233,75){\line(2,-1){20}}

\put(293,75){\circle*{3}}
\put(293,75){\line(0,-1){10}}

\put(229,55){$E$}
\put(249,55){$F$}
\put(269,55){$G$}
\put(288,55){$H$}
\put(233,42){\circle*{3}}
\put(233,42){\line(0,1){10}}
\put(233,42){\line(2,1){20}}
\put(253,42){\circle*{3}}
\put(273,42){\circle*{3}}
\put(273,42){\line(2,1){20}}
\put(253,42){\line(2,1){20}}
\put(253,42){\line(-2,1){20}}
\put(273,42){\line(-2,1){20}}
\put(293,42){\circle*{3}}
\put(293,42){\line(0,1){10}}
\put(293,42){\line(-2,1){20}}

\put(298,36){\tiny $\pand$}
\put(227,37){\line(1,0){69}}

\put(263,32){\circle*{3}}
\put(263,32){\line(0,-1){10}}
\put(233,32){\circle*{3}}
\put(233,32){\line(0,-1){10}}
\put(233,32){\line(3,-1){30}}
\put(293,32){\circle*{3}}
\put(293,32){\line(0,-1){10}}

\put(230,13){$E$}
\put(250,13){$F\pand G$}
\put(288,13){$H$}
\put(243,0){\circle*{3}}
\put(243,0){\line(-4,5){8}}
\put(243,0){\line(2,1){20}}
\put(283,0){\circle*{3}}
\put(283,0){\line(4,5){8}}
\put(283,0){\line(-2,1){20}}

\end{picture}
\end{center}

{\bf Recurrence Introduction ($\brecurrence$):} The premise of this rule is obtained from the conclusion through replacing an oformula $\brecurrence F$ by $F$ (while preserving all arcs), and inserting, anywhere in the cirquent, a new overgroup that contains $F$ as its only oformula. Examples:

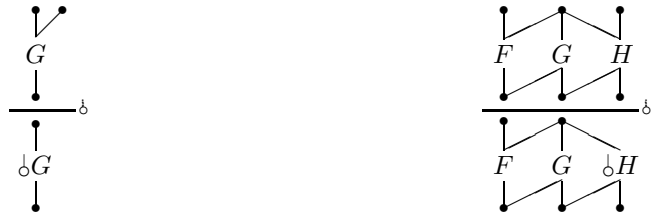
\begin{center} \begin{picture}(273,76)

\put(37,75){\circle*{3}}
\put(37,75){\line(0,-1){10}}
\put(47,75){\circle*{3}}
\put(47,75){\line(-1,-1){10}}

\put(33,55){$G$}

\put(37,42){\circle*{3}}
\put(37,42){\line(0,1){10}}

\put(53,36){$\sti$}
\put(27,37){\line(1,0){25}}

\put(37,32){\circle*{3}}
\put(37,32){\line(0,-1){10}}

\put(30,13){$\brecurrence G$}

\put(37,0){\circle*{3}}
\put(37,0){\line(0,1){10}}

\put(214,75){\circle*{3}}
\put(214,75){\line(0,-1){11}}
\put(236,75){\circle*{3}}
\put(236,75){\line(-2,-1){22}}
\put(236,75){\line(0,-1){11}}
\put(236,75){\line(2,-1){22}}
\put(258,75){\circle*{3}}
\put(258,75){\line(0,-1){11}}

\put(210,55){$F$}
\put(232,55){$G$}
\put(254,55){$H$}

\put(214,42){\line(0,1){11}}
\put(214,42){\line(2,1){22}}
\put(214,42){\circle*{3}}
\put(236,42){\circle*{3}}
\put(258,42){\circle*{3}}
\put(236,42){\line(0,1){11}}
\put(236,42){\line(2,1){22}}
\put(258,42){\line(0,1){11}}

\put(266,36){$\sti$}
\put(206,37){\line(1,0){59}}

\put(214,33){\circle*{3}}
\put(214,33){\line(0,-1){11}}
\put(236,33){\circle*{3}}
\put(236,33){\line(-2,-1){22}}
\put(236,33){\line(0,-1){11}}
\put(236,33){\line(2,-1){22}}

\put(210,13){$F$}
\put(232,13){$G$}
\put(251,13){$\brecurrence  H$}

\put(214,0){\line(0,1){10}}
\put(214,0){\line(2,1){22}}
\put(214,0){\circle*{3}}
\put(236,0){\circle*{3}}
\put(258,0){\circle*{3}}
\put(236,0){\line(0,1){10}}
\put(236,0){\line(2,1){22}}
\put(258,0){\line(0,1){10}}

\end{picture}
\end{center}

{\bf Corecurrence Introduction ($\cobrecurrence$):} The premise of this rule is obtained from the conclusion through replacing an oformula $\cobrecurrence F$ by $F$, and including $F$ in any (possibly zero) number of the already existing overgroups in addition to those in which the original oformula $\cobrecurrence F$ was already present. Examples:

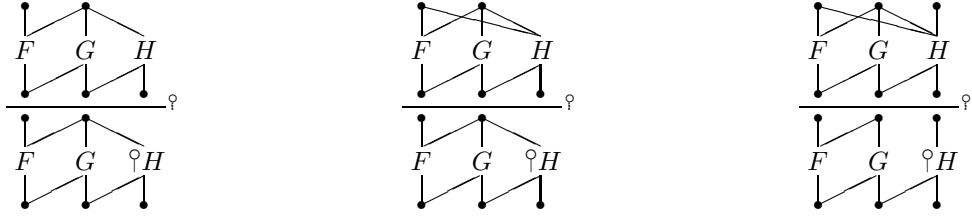
\begin{center} \begin{picture}(373,76)

\put(14,75){\circle*{3}}
\put(14,75){\line(0,-1){11}}
\put(37,75){\circle*{3}}
\put(37,75){\line(-2,-1){22}}
\put(37,75){\line(0,-1){11}}
\put(37,75){\line(2,-1){22}}

\put(10,55){$F$}
\put(33,55){$G$}
\put(55,55){$H$}

\put(14,42){\line(0,1){11}}
\put(14,42){\line(2,1){22}}
\put(14,42){\circle*{3}}
\put(37,42){\circle*{3}}
\put(59,42){\circle*{3}}
\put(37,42){\line(0,1){11}}
\put(37,42){\line(2,1){22}}
\put(59,42){\line(0,1){11}}

\put(68,36){$\costi$}
\put(7,37){\line(1,0){60}}

\put(14,33){\circle*{3}}
\put(14,33){\line(0,-1){11}}
\put(37,33){\circle*{3}}
\put(37,33){\line(-2,-1){22}}
\put(37,33){\line(0,-1){11}}
\put(37,33){\line(2,-1){22}}

\put(10,13){$F$}
\put(33,13){$G$}
\put(53,13){$\cobrecurrence  H$}

\put(14,0){\line(0,1){11}}
\put(14,0){\line(2,1){22}}
\put(14,0){\circle*{3}}
\put(37,0){\circle*{3}}
\put(59,0){\circle*{3}}
\put(37,0){\line(0,1){11}}
\put(37,0){\line(2,1){22}}
\put(59,0){\line(0,1){11}}

\put(164,75){\circle*{3}}
\put(164,75){\line(0,-1){11}}
\put(164,75){\line(4,-1){45}}
\put(187,75){\circle*{3}}
\put(187,75){\line(-2,-1){22}}
\put(187,75){\line(0,-1){11}}
\put(187,75){\line(2,-1){22}}

\put(160,55){$F$}
\put(183,55){$G$}
\put(205,55){$H$}

\put(164,42){\line(0,1){11}}
\put(164,42){\line(2,1){22}}
\put(164,42){\circle*{3}}
\put(187,42){\circle*{3}}
\put(209,42){\circle*{3}}
\put(187,42){\line(0,1){11}}
\put(187,42){\line(2,1){22}}
\put(209,42){\line(0,1){11}}

\put(218,36){$\costi$}
\put(157,37){\line(1,0){60}}

\put(164,33){\circle*{3}}
\put(164,33){\line(0,-1){11}}
\put(187,33){\circle*{3}}
\put(187,33){\line(-2,-1){22}}
\put(187,33){\line(0,-1){11}}
\put(187,33){\line(2,-1){22}}

\put(160,13){$F$}
\put(183,13){$G$}
\put(203,13){$\cobrecurrence  H$}

\put(164,0){\line(0,1){11}}
\put(164,0){\line(2,1){22}}
\put(164,0){\circle*{3}}
\put(187,0){\circle*{3}}
\put(209,0){\circle*{3}}
\put(187,0){\line(0,1){11}}
\put(187,0){\line(2,1){22}}
\put(209,0){\line(0,1){11}}

\put(314,75){\circle*{3}}
\put(314,75){\line(0,-1){11}}
\put(314,75){\line(4,-1){45}}
\put(337,75){\circle*{3}}
\put(337,75){\line(-2,-1){22}}
\put(337,75){\line(0,-1){11}}
\put(337,75){\line(2,-1){22}}
\put(359,75){\circle*{3}}
\put(359,75){\line(0,-1){11}}

\put(310,55){$F$}
\put(333,55){$G$}
\put(355,55){$H$}

\put(314,42){\line(0,1){11}}
\put(314,42){\line(2,1){22}}
\put(314,42){\circle*{3}}
\put(337,42){\circle*{3}}
\put(359,42){\circle*{3}}
\put(337,42){\line(0,1){11}}
\put(337,42){\line(2,1){22}}
\put(359,42){\line(0,1){11}}

\put(368,36){$\costi$}
\put(307,37){\line(1,0){60}}

\put(314,33){\circle*{3}}
\put(314,33){\line(0,-1){11}}
\put(337,33){\circle*{3}}
\put(337,33){\line(-2,-1){22}}
\put(337,33){\line(0,-1){11}}
\put(359,33){\circle*{3}}
\put(359,33){\line(0,-1){11}}

\put(310,13){$F$}
\put(333,13){$G$}
\put(353,13){$\cobrecurrence  H$}

\put(314,0){\line(0,1){11}}
\put(314,0){\line(2,1){22}}
\put(314,0){\circle*{3}}
\put(337,0){\circle*{3}}
\put(359,0){\circle*{3}}
\put(337,0){\line(0,1){11}}
\put(337,0){\line(2,1){22}}
\put(359,0){\line(0,1){11}}

\end{picture}
\end{center}

A {\bf proof} (in {\bf CL15})   of a cirquent $C$ is a sequence of cirquents ending in $C$ such that the first cirquent is  the conclusion of (an instance of) Axiom, and every subsequent cirquent follows from the immediately preceding cirquent by one of the rules of {\bf CL15}. A  proof of a formula $F$ is understood as a proof of the cirquent $(\seq{F}, \{F\},\{F\})$.

As an example, below is a proof of the formula $\cobrecurrence\brecurrence F\pimplication \brecurrence \cobrecurrence F$, i.e., $\brecurrence\cobrecurrence \neg F\por \brecurrence \cobrecurrence F$. 
To save space, the cirquents in it have been arranged horizontally, separated with $\Longrightarrow$'s together with the symbolic names of the rules used; if such a name is duplicated as in \raisebox{0.3mm}{\tiny DD}, it means that the rule was applied twice rather than once.

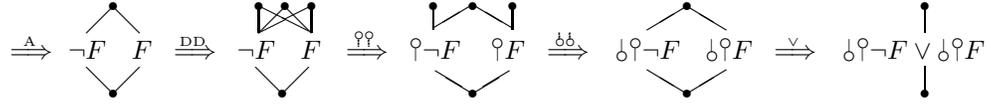
\begin{center}\begin{picture}(370,36)
\put(2,18){\tiny A}
\put(-3,13){$\Longrightarrow$}

\put(36,33){\line(1,-1){10}}
\put(36,33){\line(-1,-1){10}}
\put(36,33){\circle*{3}}
\put(19,13){$\neg F$}
\put(43 ,13){$ F$}
\put(36,0){\circle*{3}}
\put(36,0){\line(1,1){10}}
\put(36,0){\line(-1,1){10}}

\put(61,18){\tiny DD}
\put(59,13){$\Longrightarrow$}

\put(101,33){\line(1,-1){10}}
\put(111,33){\line(-2,-1){20}}
\put(91,33){\circle*{3}}
\put(91,33){\line(0,-1){10}}
\put(111,33){\circle*{3}}
\put(111,33){\line(0,-1){10}}
\put(91,33){\line(2,-1){19}}
\put(101,33){\line(-1,-1){10}}
\put(101,33){\circle*{3}}
\put(83,13){$\neg F$}
\put(107,13){$ F$}
\put(100,0){\circle*{3}}
\put(100,0){\line(1,1){10}}
\put(100,0){\line(-1,1){10}}

\put(127,18){$\costi\costi$}
\put(124,13){$\Longrightarrow$}

\put(157,33){\circle*{3}}
\put(157,33){\line(0,-1){9}}
\put(187,33){\circle*{3}}
\put(187,33){\line(0,-1){9}}
\put(173,33){\line(5,-3){14}}
\put(171,33){\line(-5,-3){14}}
\put(172,33){\circle*{3}}
\put(148,13){$\cobrecurrence \neg F$}
\put(179,13){$\cobrecurrence F$}
\put(172,0){\circle*{3}}
\put(172,0){\line(5,3){14}}
\put(172,0){\line(-5,3){14}}

\put(203,18){$\sti\sti$}
\put(200,13){$\Longrightarrow$}

\put(253,33){\line(5,-3){15}}
\put(253,33){\line(-5,-3){15}}
\put(253,33){\circle*{3}}
\put(226,13){$\brecurrence \cobrecurrence  \neg F$}
\put(260,13){$\brecurrence \cobrecurrence  F$}
\put(253,0){\circle*{3}}
\put(253,0){\line(5,3){15}}
\put(253,0){\line(-5,3){15}}

\put(291,18){{\tiny $\por$}}
\put(286,13){$\Longrightarrow$}

\put(343,33){\line(0,-1){10}}
\put(343,33){\circle*{3}}
\put(312,13){$\brecurrence \cobrecurrence  \neg F\por \brecurrence \cobrecurrence  F$}
\put(343,0){\line(0,1){10}}
\put(343,0){\circle*{3}}
\end{picture}
\end{center}

\begin{exercise}
Prove the following formulas in {\bf CL15}:\vspace{-5pt}
\begin{itemize}
  \item $F\brimplication F$ (i.e. $\brepudiation F\por F$, i.e. $\cobrecurrence \neg F\por F$).\vspace{-5pt}
  \item $F\pand F\pimplication F$.\vspace{-5pt}
  \item $F\brimplication \brecurrence F\pand\brecurrence F$.\vspace{-5pt}
  \item $F\brimplication \brecurrence \brecurrence F$.\vspace{-5pt}
  \item $\brecurrence E\por\brecurrence F\pimplication\brecurrence (E\por F)$. \vspace{-5pt}
  \item $(E\pand F)\por (G\pand H)\pimplication (E\por G)\pand(F\por H)$. 
\end{itemize}
\end{exercise}

W. Xu and S. Liu \cite{xuND} showed that {\bf CL15} remains sound with $\precurrence,\coprecurrence$ instead of $\brecurrence,\cobrecurrence$. Completeness, however, is lost in this case because, for instance, as shown in \cite{separating}, the formula $F\pand \precurrence(F\pimplication F \pand F)\pimplication \precurrence F$ is logically valid while $F\pand \brecurrence(F\pimplication F \pand F)\pimplication \brecurrence F$ is not.

\begin{openproblem} \ \vspace{-5pt}
\begin{enumerate}
\item Is (the problem of provability in) {\bf CL15} decidable?\vspace{-5pt} 
\item Extend the language of {\bf CL15} by including $\chand,\chor$ and axiomatize (if possible) the set of logically valid formulas in this extended language.\vspace{-5pt}
\item Replace $\brecurrence,\cobrecurrence$ with $\precurrence,\coprecurrence$ in the language of {\bf CL15}. Is the set of logically valid formulas in this new language axiomatizable and, if yes, how?\vspace{-5pt}
\item Does {\bf CL15} remain complete with respect to extralogical (as opposed to logical) validity? 
\end{enumerate}
\end{openproblem}

\subsection{The brute force system {\bf CL13}}

{\bf CL13-formulas}---or just {\em formulas} in this subsection---are formulas of the language of CoL that do not contain any function letters or non-nullary gameframe letters, and do not contain any operators other than $\neg, \pand,\por, \chand,\chor,  \sand,\sor, \tand,\tor$. As in the case of {\bf CL15}, officially $\neg$ is only allowed to be applied to extralogical atoms, otherwise understood as the corresponding DeMorgan abbreviation,  including understanding $\neg\top$ as $\bot$ and $\neg \bot$ as $\top$. Each of the implication operators $\pimplication,\chimplication,\simplication,\timplication$  should also be understood as an abbreviation of its standard meaning in terms of negation and the corresponding sort of disjunction. To define the system axiomatically, we need certain terminological conventions.\vspace{-5pt}

\begin{itemize}
  \item  A {\bf literal} means an atom $A$ with or without negation $\neg$. Such a literal is said to be elementary or general iff $A$ is so. \vspace{-5pt} 
  \item As in Section \ref{scl15}, we often need to differentiate between {\em subformulas} as such, and particular {\em occurrences} of subformulas. We will be using the term {\bf osubformula} to mean a subformula together with a particular occurrence.  The prefix ``o'' will be used with a similar meaning in terms such as {\bf oatom}, {\bf oliteral}, etc. \vspace{-5pt}
  \item An osubformula is {\bf positive} iff it is not in the scope of $\neg$. Otherwise it is {\bf negative}.  According to our conventions regarding the usage of $\neg$, only oatoms may be negative. \vspace{-5pt}
  \item   A {\bf politeral} is a positive oliteral.\vspace{-5pt}
  \item A {$\pand$-(sub)formula} is a (sub)formula of the form $E\pand F$. Similarly for the other connectives.\vspace{-5pt}
  \item A {\bf sequential (sub)formula} is one of the form $E\sand F$ or $E\sor F$. We say that $E$ is the {\bf head} of such a (sub)formula, and $F$ is its {\bf tail}.\vspace{-5pt}
  \item Similarly, a {\bf parallel (sub)formula} is one of the form $E\pand F$ or $E\por F$, a {\bf choice (sub)formula} is one of the form $E\chand F$ or $E\chor F$, and a  {\bf toggling (sub)formula} is one of the form $E\tand F$ or $E\tor F$. \vspace{-5pt}
  \item   A formula is said to be {\bf quasielementary} iff it contains no general atoms and no operators other than $\neg,\pand,\por,\tand,\tor$.   \vspace{-5pt}
  \item A formula is said to be {\bf elementary} iff it is a formula of classical propositional logic, i.e., contains no general atoms and no operators other than $\neg,\pand,\por$. \vspace{-5pt}
  \item A {\bf semisurface osubformula} (or {occurrence}) is an osubformula (or occurrence) which is not in the scope of a choice connective. \vspace{-5pt}
  \item A {\bf surface osubformula} (or { occurrence}) is an osubformula (or occurrence) which is not in the scope of any connectives other than $\neg,\pand,\por$.\vspace{-5pt}
  \item The {\bf quasielementarization} of a formula $F$, denoted by $|F|$, is the result of replacing in $F$ every sequential osubformula by its head, every $\chand$-osubformula by $\top$, every  $\chor$-osubformula by $\bot$, and every general politeral by $\bot $ (the order of these replacements does not matter). For instance,  the quasielementarization of $\bigl((P\tor q)\por\bigl((p\pand \neg P)\sand(Q\pand R)\bigr)\bigr)\tand  \bigl(q\chand (r\chor s)\bigr)$   is $\bigl((\bot\tor q)\por (p\pand \bot)\bigr)\tand\top$.\vspace{-5pt}
  \item The {\bf elementarization} of a quasielementary formula $F$,  denoted by $\|F\|$, is the result of replacing in  $F$ every $\tand$-osubformula  by $\top$ and every  $\tor$-osubformula by $\bot$ (again, the order of these replacements does not matter). For instance,  $\|\bigl(s\pand \bigl(p\tand(q\tor r)\bigr)\bigr)\por\bigl(\neg s\por (p\tor r)\bigr)\| = (s\pand \top)\por(\neg s\por \bot)$.\vspace{-5pt}
  \item A quasielementary formula $F$ is said to be {\bf stable} iff its elementarization $\|F\|$ is a tautology of classical logic.  
\end{itemize}

We now define {\bf CL13} by the following six rules of inference, where $\vec{H}\Longrightarrow F$ means ``from premise(s) $\vec{H}$ conclude $F$''.  Axioms are not explicitly stated, but the set of premises of the $(\tand)$ rule can be empty, in which case (the conclusion of) this rule acts like an axiom. 

\begin{description}
  \item[Rule $(\tand )$:]   $\vec{H}\Longrightarrow   F$, where $F$ is a stable quasielementary formula, and $\vec{H}$ is the smallest set of formulas satisfying the following condition:\vspace{-5pt}
\begin{itemize}  
  \item Whenever $F$ has a surface osubformula $E_0\tand E_1$, for both $i\in\{0,1\}$, $\vec{H}$ contains the result of replacing in $F$  that osubformula by $E_i$.\vspace{-2pt}
\end{itemize}  
\item[Rule $(\tor)$:]  $H\Longrightarrow F$, where $F$    is a quasielementary formula, and $H$ is the result of replacing in $F$ a surface osubformula $E\tor G$  by $E$ or $G$.
\item[Rule $(\sand\chand)$:] $|F|,\vec{ H}\Longrightarrow   F$, where $F$ is a non-quasielementary formula (note that otherwise $F=|F|$), and  $\vec{H}$ is the smallest set of formulas satisfying the following two conditions:\vspace{-5pt}
\begin{itemize}
  \item   Whenever $F$ has a semisurface osubformula $G_0\chand G_1$, for both $i\in\{0,1\}$, $\vec{H}$  contains the result of replacing in $F$ that osubformula by $G_i$.\vspace{-5pt}
  \item Whenever $F$ has a semisurface osubformula $E\sand G$, $\vec{H}$ contains the result of replacing in $F$ that osubformula by $G$.\vspace{-2pt}
\end{itemize}  
\item[Rule $(\chor)$:] $H\Longrightarrow F$, where $H$ is the result of replacing in $F$ a semisurface osubformula $E\chor G$ by $E$ or $G$.
\item[Rule $(\sor)$:] $H\Longrightarrow F$, where $H$ is the result of replacing in $F$ a semisurface osubformula $E\sor G$ by $G$.
\item[Rule $(\mbox{M})$:]   $H\Longrightarrow F$, where $H$ is the result of replacing in $F$ two---one positive and one negative---semisurface occurrences of some general atom $P$ by an extralogical elementary atom $p$ which does not occur in $F$.
\end{description}

A {\bf proof} (in {\bf CL13}) of a formula $F$ is a sequence of formulas ending in $F$ such that every formula follows from some (possibly empty) set of earlier formulas by one of the rules of the system.

\begin{example}  
Pick any two distinct connectives $\&_1$ and $\&_2$ from the list $\pand,\tand,\sand,\chand$. Then {\bf CL13} proves the formula $P\&_1Q \pimplication P\&_2Q$ if  and  only if  $\&_1$ is to the left of $\&_2$ in the list. Similarly for the list $\chor,\sor,\tor,\por$. Here we verify this fact only for the case $\{\&_1,\&_2\}=\{\tand,\sand\}$. The reader may want to try some other combinations as exercises.
Below  is a proof of $P\tand Q \pimplication P\sand Q$ together with step justifications:\vspace{-5pt}
\begin{enumerate}
  \item $\neg p\por p$\hspace{30pt}  From no premises by $(\tand)$.\vspace{-5pt}                    
  \item  $(\neg p\tor \bot)\por  p$  \hspace{30pt}   From 1 by           $(\tor)$\vspace{-5pt}
  \item $\neg q\por q $\hspace{30pt}    From no premises by $(\tand)$\vspace{-5pt}                     
  \item $(\neg p\tor\neg q)\por  q $\hspace{30pt} From 3 by $(\tor)$ \vspace{-5pt}             
  \item $(\neg p\tor \neg Q)\por  Q$\hspace{30pt}    From 4 by (M) \vspace{-5pt}          
  \item $(\neg p\tor \neg Q)\por (p\sand Q)$\hspace{30pt}    From 2,5 by $(\sand\chand)$\vspace{-5pt}    
  \item     $(\neg P\tor \neg Q)\por (P\sand Q)$\hspace{30pt}   From 6 by (M)\vspace{-5pt}    
\end{enumerate}
On the other hand, the formula $P\sand Q\pimplication  P\tand Q$, i.e. $(\neg P\sor \neg Q)\por  (P\tand Q)$, has no proof in {\bf CL13}. This can be shown through attempting and failing to construct, bottom-up, a purported proof of the formula. Here we explore one of the branches of a proof-search tree.  $(\neg P\sor \neg Q)\por  (P\tand Q)$ is not quasielementary, so it could not be derived by (be the conclusion of) the $(\tor)$ or $(\tand)$ rule.  The $(\chor )$ rule does not apply either, as there is no $\chor$ in the formula. This leaves us with one of the rules $(\sor)$, $(\sand\chand)$ and (M). Let us see what happens if our target formula is derived by $(\sor)$. In this case the premise should be  $\neg Q\por  (P\tand Q)$. The latter can be derived only by $(\sand\chand)$  or (M). Again, let us try (M). The premise in this case should be $\neg q \por (P\tand q)$ for some elementary atom $q$. But the only way $\neg q \por (P\tand q)$  can be derived is by $(\sand\chand)$ from the premise $\neg q\por  (\bot\tand q)$. This formula, in turn, could only be derived by $(\tand)$, in which case $\neg q\por\bot$  is one of the premises. Now we are obviously stuck, as $\neg q\por\bot$  is not the conclusion of any of the rules of the system. We thus hit a dead end.  All remaining possibilities can be checked in a similar routine/analytic way, and the outcome in each case will be a dead end.
\end{example}

\begin{exercise}  \

1. Construct a  proof of $(P\simplication P) \pand (\neg P\simplication \neg P) \pimplication  P\chimplication P $.

2. For which of the four disjunctions $\cup\in\{\por,\chor,\sor,\tor\}$   are the following formulas  provable and for which are not?  \ (a) $\neg P \cup P$; \ (b)  $P\cup Q\pimplication Q\cup P$; \ (c) $P\cup P\pimplication P$; \ (d)  $p\cup p\pimplication p$.
\end{exercise}

\begin{openproblem} \

 1.   Consider the first-order version of the language of {\bf CL13} with choice (to start with) quantifiers. Adequately axiomatize the set of logically valid formulas in this language.

2. Consider the set of the theorems of {\bf CL13} that do not contain extralogical elementary letters. Does this set  remain complete with respect to extralogical   validity?
\end{openproblem}

\subsection{The brute force system {\bf CL12}}\label{scl12}

{\bf CL12-formulas}---or just {\em formulas} in this subsection---are formulas of the language of CoL that do not contain any general gameframe letters,  and do not contain any operators other than $\neg, \pand,\por , \chand,\chor, \chall,\chexists, \blall,\blexists$.   As in the preceding two sections,  $\neg $ applied to formulas other than extralogical atoms is understood  as the corresponding DeMorgan abbreviation, $E\pimplication F$ is understood as an abbreviation of $\neg E\por F$, and $E\chimplication F$ as an abbreviation of $\neg E\chor F$.  

{\bf CL12-sequents}---or just {\em sequents} in this subsection---are expressions of the form $E_1,\cdots,E_n\brimplication  F$, where $E_1,\cdots,E_n$ ($n\geq 0$) and $F$ are {\bf CL12}-formulas; for simplicity and safety, we require that no variable has both free and bound occurrences in the (not necessarily the same) formulas of the sequent. The sequence $E_1,\cdots,E_n$ is said to be the {\em antecedent} of the sequent, and $F$ its {\em succedent}. Semantically, 
such a sequent is identified with the (non-{\bf CL12}) formula $E_1\pand\cdots\pand E_n\brimplication F$.   So, for instance, when we say that the former is logically valid, we mean that so is the latter, and a logical solution of the former means a logical solution of the latter. Each {\bf CL12}-formula $F$, in turn, can be identified with the empty-antecedent sequent $\brimplication F$.  A {\bf CL12}-sequent is {\bf closed} iff so is its succedent as well as all formulas of the antecedent.   When applied to {\bf CL12}, the word ``sentence'' in Theorem \ref{adeq} should be interpreted as ``closed {\bf CL12}-sequent'' rather than (merely) ``closed {\bf CL12}-formula''.

Note that the language of {\bf CL12} is an extension of  the full  language of classical first-order logic.
Due to this fact, together with the presence of $\chand,\chor,\chall,\chexists,\brimplication$ in the language,   {\bf CL12} is a very powerful tool for constructing CoL-based applied theories (see Section \ref{sappl}), and  has been repeatedly \cite{Japtowards,cla4,cla8,cl12,cla5,cla11a,cla11b,Jap18a} used as such with significant advantages over the less expressive and computationally less meaningful classical logic.  Below is some terminology employed in our axiomatization of {\bf CL12}.\vspace{-5pt}   

\begin{itemize}
  \item A {\bf surface occurrence} of a subformula is an occurrence that is not in the scope of any choice operators.\vspace{-5pt}
  \item  A formula not containing choice operators---i.e., a formula of the language of classical first order logic---is said to be {\bf elementary}. A sequent is {\bf elementary} iff all of its formulas are so. The {\bf elementarization} $\|F\|$ of a formula $F$ is the result of replacing in $F$ all surface occurrences of $\chor$- and $\chexists$-subformulas by $\bot$, and all surface occurrences of $\chand$- and $\chall$-subformulas by $\top$. Note that $\|F\|$ is (indeed) an elementary formula. The {\bf elementarization} $\|E_1,\cdots,E_n\brimplication  F\|$ of a sequent $E_1,\cdots,E_n\brimplication  F$  is the elementary formula $\|E_1\|\pand\cdots\pand \|E_n\|\pimplication  \|F\|$.\vspace{-5pt} 
  \item A sequent  is said to be {\bf stable} iff its elementarization is classically valid (i.e., provable in some standard version of classical first-order calculus with constants, function letters and $=$).\vspace{-5pt} 
  \item   We will be using the notation $F[E]$ to mean a formula $F$ together with some  fixed  surface occurrence of a subformula $E$. Using this notation sets a context, in which $F[H]$ will mean the result of replacing in $F[E]$ that occurrence of $E$ by $H$.\vspace{-5pt}
  \item $\vec{G},\vec{K},\vec{L},\vec{M}, $ stand for finite sequences of formulas.\vspace{-5pt}  
\end{itemize}

We now define {\bf CL12} by the following six rules of inference, where $S_1,\cdots,S_m\Longrightarrow S$  means ``from premise(s) $S_1,\cdots,S_n$  conclude $S$''. Axioms are not explicitly stated, but the set of premises of the Wait rule can be empty, in which case (the conclusion of) this rule acts like an axiom. 
In each rule, $i$ is assumed to be either $0$ or $1$, $t$ is  either a constant or a variable with no bound occurrences in the premise, and $y$ is a variable not occurring in the conclusion; $H(t)$ (resp. $H(y)$) is the result of replacing in the formula $H(x)$ all free occurrences of the variable $x$ by $t$ (resp. $y$). 
\begin{description}
  \item[$\chor$-Choose:]  $\vec{G}\brimplication  F[H_i] \ \Longrightarrow \     \vec{G}\brimplication  F[H_0\chor H_1]$,  for either $i$.
  \item[$\chand$-Choose:]  $\vec{G},E[H_i] \brimplication  F    \Longrightarrow \     \vec{G},E [H_0\chand H_1]\brimplication F $,   for either $i$.     
  \item[$\chexists$-Choose:] $\vec{G}\brimplication  F\bigl[H(t)\bigr]\ \Longrightarrow\     \vec{G}\brimplication F\bigl[\chexists x  H(x)\bigr]$, for any $t$.
  \item[$\chall$-Choose:] $\vec{G}, E\bigl[H(t)\bigr], \vec{K}  \brimplication  F   \ \Longrightarrow\    \vec{G}, E\bigl[\chall xH(x)\bigr], \vec{K} \brimplication   F$, for any $t$.
  \item[Replicate:]  $\vec{G}, E, \vec{K}, E \brimplication    F   \ \Longrightarrow\     \vec{G}, E, \vec{K} \brimplication   F$. 
  \item[Wait:]  $S_1,\cdots,S_n\Longrightarrow S$   ($n\geq 0$),  where $S$ is   stable   and the following four conditions are satisfied:\vspace{-2pt}
\begin{itemize}       
  \item Whenever $S$ has the form $\vec{K} \brimplication E[H_0\chand H_1]$, both $\vec{K}\brimplication E[H_0]$ and $\vec{K}\brimplication E[H_1]$ are among $S_1,\cdots,S_n$. 
  \item Whenever $S$ has the form $\vec{L}, J[H_0\chor H_1], \vec{M}\brimplication  E$,   both  $\vec{L}, J[H_0], \vec{M} \brimplication E$  and  $\vec{L}, J[H_1], \vec{M} \brimplication  E $ are among $S_1,\cdots,S_n$. 
  \item Whenever $S$ has the form $\vec{K}\brimplication E\bigl[\chall xH(x)\bigr]$, for some  $y$, \  $\vec{K}\brimplication E\bigl[H(y)\bigr]$ is among $S_1,\cdots,S_n$. 
  \item Whenever $S$ has the form $\vec{L}, J\bigl[\chexists xH(x)\bigr], \vec{M}  \brimplication E$,  for some  $y$, \  $\vec{L}, J\bigl[H(y)\bigr],\vec{ M} \brimplication E$  is among $S_1,\cdots,S_n$. 
\end{itemize}
\end{description}

Each rule---seen bottom-up---encodes an action that a winning strategy should take in a corresponding situation, and the name of each rule is suggestive of that action. For instance, Wait (indeed) prescribes the strategy to wait till the adversary moves. This explains why we use the name  ``Replicate'' for one of the rules rather than the more standard ``Contraction''.  

A {\bf proof} (in {\bf CL12}) of a sequent $S$ is a sequence $S_1,\cdots,S_n$  of sequents, with $S_n =S$, such that each $S_i$ follows  by one of the rules of {\bf CL12} from some (possibly empty in the case of Wait, and certainly empty in the case of $i=1$) set $\vec{P}$ of premises such that $\vec{P}\subseteq\{S_1,\cdots,S_{i-1}\}$. A proof of a formula $F$ is understood as a proof of the empty-antecedent sequent $\brimplication F$. 

\begin{example}\label{mor}
 Here $\times$   is a binary function letter and $^3$ is a unary function letter. We write $x\times y$ and $x^3$ instead of $\times(x,y)$ and $^3(x)$. The following sequence is a proof of its last sequent.

1.   $\blall x \bigl(x^3=(x\times x)\times x\bigr),   t=s\times s,  r=t\times s \brimplication     r=s^3 $                         \hspace{15pt}                  Wait: (no premises)

2.   $\blall x \bigl(x^3=(x\times x)\times x\bigr),   t=s\times s,  r=t\times s \brimplication    \chexists y(y=s^3)$                       \hspace{15pt}              $\chexists$-Choose: 1

3.   $\blall x \bigl(x^3=(x\times x)\times x\bigr),   t=s\times s,  \chexists z(z=t\times s)\brimplication    \chexists y(y=s^3)$                \hspace{15pt}           Wait: 2

4.   $\blall x \bigl(x^3=(x\times x)\times x\bigr),   t=s\times s,  \chall y\chexists z(z=t\times y)\brimplication    \chexists y(y=s^3)$               \hspace{15pt}           $\chall$-Choose: 3

5.   $\blall x \bigl(x^3=(x\times x)\times x\bigr),   t=s\times s, \chall x\chall y\chexists z(z=x\times y)\brimplication    \chexists y(y=s^3)$            \hspace{15pt}           $\chall$-Choose: 4

6.   $\blall x \bigl(x^3=(x\times x)\times x\bigr),   \chexists z(z=s\times s), \chall x\chall y\chexists z(z=x\times y)\brimplication    \chexists y(y=s^3)$         \hspace{15pt}       Wait: 5

7.   $\blall x \bigl(x^3=(x\times x)\times x\bigr),  \chall y\chexists z(z=s\times y), \chall x\chall y\chexists z(z=x\times y)\brimplication    \chexists y(y=s^3)$       \hspace{15pt}    $\chall$-Choose: 6

8.   $\blall x \bigl(x^3=(x\times x)\times x\bigr), \chall x\chall y\chexists z(z=x\times y), \chall x\chall y\chexists z(z=x\times y)\brimplication    \chexists y(y=s^3)$     \hspace{15pt}  $\chall$-Choose: 7

9.   $\blall x \bigl(x^3=(x\times x)\times x\bigr),  \chall x\chall y\chexists z(z=x\times y)\brimplication    \chexists y(y=s^3)$                   \hspace{15pt}               Replicate: 8

10. $\blall x \bigl(x^3=(x\times x)\times x\bigr),  \chall x\chall y\chexists z(z=x\times y)\brimplication   \chall x\chexists y(y=x^3)$        \hspace{15pt}                  Wait: 9
\end{example}

\begin{exercise} To see the resource-consciousness of {\bf CL12}, show that it does not prove $p\chand q\pimplication (p\chand q)\pand (p\chand q)$, even though this formula has the form $F\pimplication F\pand F$ of a classical tautology. Then show that, in contrast, {\bf CL12} proves the sequent $p\chand q\brimplication (p\chand q)\pand (p\chand q)$ because, unlike the antecedent of a pimplication, the antecedent of a brimplication is reusable (trough Replicate).
\end{exercise} 

For any closed sequent  $E_1,\cdots, E_n\brimplication F$,  following holds due to
the adequacy  Theorem \ref{adeq}: 
\begin{equation}\label{eqq} \mbox{\em {\bf CL12} proves   
 $E_1,\cdots, E_n\brimplication F$ if and only if $F$ is a logical consequence  of $E_1,\cdots, E_n$.}\end{equation} 

This explains why we call the following rule of inference {\bf Logical Consequence}:

\[\mbox{\em $E_1,\cdots,E_n \ \Longrightarrow\   F$,   where {\bf CL12} proves the sequent  $E_1,\cdots,E_n\brimplication F$.}\]

Logical Consequence is the only logical rule of inference in all {\bf CL12}-based applied theories briefly discussed in Section \ref{sappl}. To appreciate the convenience that Thesis \ref{scopy} offers when reasoning in such theories, let us look at the following example. 

\begin{example}\label{stra}
  Imagine a {\bf CL12}-based applied formal theory, in which we have  proven or postulated  $\blall x\bigl(x^3=(x\times x)\times x\bigr)$ (the meaning of  ``cube'' in terms of multiplication) and $\chall x\chall y\chexists z(z=x\times y)$ (the computability of multiplication), and now we want to derive $\chall x\chexists y(y=x^3)$ (the computability of ``cube''). This is how we can reason to justify $\chall x\chexists y(y=x^3)$: 
\begin{quote}{\em 
Consider any $s$ (selected by the environment for $x$ in $\chall x\chexists y(y=x^3)$). We need to find $s^3$. Using the resource $\chall x\chall y\chexists z(z=x\times y)$ twice, we first find the value $t$ of $s\times s$, and then  the value $r$ of $t\times s$. According to $\blall x\bigl(x^3=(x\times x)\times x\bigr)$, such an $r$ is the sought $s^3$.}\end{quote}

\noindent Thesis \ref{scopy}, in view of (\ref{eqq}), promises that the above intuitive argument will be translatable into a proof of
\[\blall x\bigl(x^3=(x\times x)\times x\bigr), \chall x\chall y\chexists z(z=x\times y)\brimplication  \chall x\chexists y(y=x^3)\]
in {\bf CL12}, and hence the succedent $\chall x\chexists y(y=x^3)$ will be derivable in the  theory by Logical Consequence  as the formulas of the antecedent are already proven. Such a proof indeed exists---see Example \ref{mor}.
\end{example}

\begin{openproblem} \ 

1. Add the sequential connectives $\sand,\sor$ to the language of {\bf CL12} and adequately axiomatize the corresponding logic.

2. Axiomatize (if possible) the set of extralogically valid {\bf CL12}-sequents.

3.    Along with elementary gameframe letters, allow also general gameframe letters in the language of {\bf CL12}, and axiomatize (if possible) the set of valid sequents of this extended language.
\end{openproblem}

\section{Applied systems based on computability logic}\label{sappl}
The main utility of CoL, actual or potential, is related to the benefits of using it as a logical basis for applied systems,  such as axiomatic theories or knowledgebase systems. 

The most common logical basis for applied systems is classical first-order logic (CFOL). This is due to the fact that CFOL is universal: its language allows one to say anything one could say, and its proof system allows one to justify anything one could justify logically. But when it comes to expressing {\em tasks} (as opposed to {\em facts}) and reasoning about them, such as the task/problem expressed by $\chall x\chexists y(y=x^2)$, using CFOL can be an extremely circuitous and awkward way. Asking why we need  CoL if everything can be done with CFOL is akin to asking,   for instance,  why  we   study modal logics if anything one can express or reason about in modal logic can just as well be expressed or reasoned about using CFOL. 

For specificity, let us imagine what a typical applied system $\mathbb{S}$ based on the already axiomatized fragment {\bf CL12} of CoL would look like. The construction of such a system 
 would start from building its extralogical basis $\mathbb{B}$,  with some fixed interpretation $^*$ in mind. In what follows, for readability, we    omit explicit references to this $^*$ and, terminologically, identify each sentence $E$ with the game $E^*$.  Depending on the context or traditions, $\mathbb{B}$ would generally be referred to as the {\bf knowledgebase}, or the set of {\bf axioms}, of $\mathbb{S}$. It  would be a collection of  relevant (to the purposes of the system) sentences expressing
computational problems with already known, fixed solutions. Those can be 
atomic elementary sentences expressing  true facts such as $Jane=MotherOf(Bob)$ or $0\not=1$; nonatomic elementary sentences 
expressing general or conceptual knowledge such as $\blall x\blall y\bigl(y=MotherOf(x)\pimplication Female(y)\bigr)$ or $\blall x(x^2=x\times x)$; 
nonatomic nonelementary sentences such as $\chall x\chexists y\bigl(y=DateOfBirth(x)\bigr)$ expressing the ability to tell    any person's date of birth (perhaps due to having access to an external database), $\chall x\chall y\chexists z(z=x\times y)$ expressing the ability to compute multiplication, or 
$\chall x\chall y\bigl(\neg\mbox{\em Halts}(x,y)\chor\mbox{\em Halts}(x,y)\bigr) \pimplication \chall x\chall y\bigl(\neg\mbox{\em Accepts}(x,y)\chor\mbox{\em Accepts}(x,y)\bigr) $ expressing the ability to reduce the acceptance problem to the halting problem. The only logical rule of inference in $\mathbb{S}$ would be Logical Consequence defined in Section \ref{scl12}. 
 
A {\bf proof} $\cal P$ of a sentence $F$ in such a system $\mathbb{S}$ will be defined in a standard way, with the elements of $\mathbb{B}$ acting as axioms.  The rule of Logical Consequence preserves computability in the sense that,   as long as all sentences of the antecedent of a {\bf CL12}-sequent are computable       under a given interpretation, so is the succedent and, furthermore, a solution of the latter  can be extracted from solutions of the sentences of the antecedent.   
Since solutions of all axioms are already known,  $\cal P$ thus automatically translates into  a solution $\cal H$ of $F$. Think of $\mathbb{S}$ as a   declarative programming language, $\cal P$ as a program written in that language, the sentence $F$ as a specification of (the goal of) such a program,  the mechanism extracting solutions from proofs  as a compiler, and the above $\cal H$ as a machine-language-level translation of the high-level program $\cal P$.  Note that the notoriously hard problem of program verification is fully neutralized in this paradigm:   being a proof of the sentence $F$, $\cal P$ automatically also serves as a formal verification of the fact that the  program $\cal P$ meets its specification $F$. Further, $\cal P$ is a program commented in an extreme sense, with every line/sentence in this program being its own, best possible, comment.  

A relevant question, of course, is how efficient the above solution $\cal H$ would be in terms of computational complexity. Here come more pieces of positive news. The traditional complexity-theoretic concepts such as time or space complexities find in CoL natural and conservative generalizations from the traditional sorts of problems to all games (cf. \cite{cl12}).  
The Logical Consequence rule is   complexity-theoretically well behaved, with the time (resp. space) complexity of its conclusion  guaranteed to be at most linearly (resp. logarithmically) different from the time (resp. space) complexities of the  solutions of the premises. So, how efficient the solutions extracted from proofs in $\mathbb{S}$ are, is eventually determined by how efficient the solutions of the  axioms comprising $\mathbb{B}$ are. If, for instance, all axioms have linear time and/or logarithmic space solutions, then so do all theorems of $\mathbb{S}$ as well.  

In some cases, the extralogical postulates of $\mathbb{S}$ would consist of not only the axioms $\mathbb{B}$, but also certain extralogical rules of inference such as some versions of induction or comprehension. Depending on the version, such a rule may less closely preserve computational complexity than Logical Consequence  does. For instance, each application of  induction may increase the time complexity quadratically rather than linearly. By limiting the number of such applications or imposing certain other restrictions, we can still get a system all of whose theorems are problems with low-order polynomial time complexities as long as all \mbox{axioms are so.}    

By now CoL has found applications in a series {\bf CLA1-CLA11} of formal number theories dubbed ``{\bf clarithmetics}''. All of these theories are based on  {\bf CL12}, and differ between each other only in their extralogical postulates. The language of each  theory has the same extralogical vocabulary $\{0, +, \times, '\}$ (where $x'$ means ``the successor of $x$'')  as the language of first-order Peano arithmetic. Unlike other approaches with similar aspirations such as that of bounded arithmetic \cite{Buss}, this approach avoids a need for adding more extralogical primitives to the language as, 
due to extending rather than restricting traditional Peano arithmetic, all arithmetical functions or predicates remain expressible in standard ways. The extralogical postulates of clarithmetical systems are also remarkably simple, with their sets of axioms consisting of all axioms of Peano arithmetic plus the single nonelementary sentence $\chall x\chexists y(y=x')$ or just a few similarly innocuous-looking axioms, and the set of extralogical inference rules consisting of induction and perhaps one more rule such as comprehension. Different clarithmetics serve  different computational complexity classes, which explains their  multiplicity. Each  system has been proven to be sound and complete 
with respect to its target complexity class $C$. Sound in the sense that every theorem $T$ of the system expresses an arithmetical 
problem $A$ with a $C$ complexity solution and, furthermore, such a solution can be automatically obtained from the proof of $T$.
 And complete in the sense that every  arithmetical problem $A$ with a $C$ complexity solution is expressed by some theorem $T$ of the system. Furthermore, if one adds all true sentences of Peano arithmetic to the set of axioms, then this {\em extensional} completeness result strengthens to {\em intensional} completeness, according to which every (rather than just some) sentence $F$ expressing such an 
$A$ is a theorem of the system.  

 Among {\bf CLA1-CLA11}, the system {\bf CLA11} stands out in that it is a scheme of clarithmetical theories rather than a particular theory, taking three parameters $\mathfrak{a},\mathfrak{s},\mathfrak{t}$ and correspondingly written as {\bf CLA11}$(\mathfrak{a},\mathfrak{s},\mathfrak{t})$. These parameters range over sets of terms or pseudoterms used as bounds for $\chall,\chexists$ in certain postulates. $\mathfrak{t}$ determines the time complexity of all theorems of the system, $\mathfrak{s}$ determines space complexity and $\mathfrak{a}$ the so called {\em amplitude complexity} (the complexity measure concerned with the sizes of $\top$'s moves relative to the sizes of $\bot$'s moves). By tuning these three parameters in an essentially mechanical, brute force fashion, one immediately gets a system sound and complete with respect to one or another combination of time, space and amplitude complexities. For instance, for {\em Linear amplitude + Logarithmic space + Polynomial time}, it is sufficient to choose $\mathfrak{a}$ to be the canonical set of terms expressing all linear functions (i.e. terms built from variables, $0$, $+$ and $'$),  $\mathfrak{t}$  the canonical set of terms for all polynomial functions (namely, terms built from variables, $0$, $+$, $\times$ and $'$),  and $\mathfrak{s}$ the set of canonical pseudoterms for all logarithmic functions. This way one can obtain a system for essentially all  natural (whatever this means) combinations of time, space and amplitude complexities. See the introductory section of \cite{cla11a} for a more detailed account. 

\def\DITTO{--}

\end{document}